\documentclass[12pt,reqno,a4paper]{amsart}

\usepackage{a4}
\usepackage{amsmath, amssymb}

\usepackage{graphicx}                      
\newcommand{\color}[2][{}]{}        

\usepackage{bbm}         

\theoremstyle{plain}            
\newtheorem{theorem}{Theorem}[section]
\newtheorem{proposition}[theorem]{Proposition}
\newtheorem{lemma}[theorem]{Lemma}

\theoremstyle{definition}       
\newtheorem{definition}[theorem]{Definition}

\theoremstyle{remark}
\newtheorem{remark}[theorem]{Remark}
\newtheorem{notation}[theorem]{Notation}

\newcommand{\Sec}[1]{Section~\ref{sec:#1}}

\newcommand{\App}[1]{Appendix~\ref{app:#1}}
\newcommand{\Apps}[2]{Appendices~\ref{app:#1} and~\ref{app:#2}}

\newcommand{\Thm}[1]{Theorem~\ref{thm:#1}}

\newcommand{\Thms}[2]{Theorems~\ref{thm:#1} and~\ref{thm:#2}}
\newcommand{\Lem}[1]{Lemma~\ref{lem:#1}}

\newcommand{\Lems}[2]{Lemmas~\ref{lem:#1} and~\ref{lem:#2}}
\newcommand{\LemS}[2]{Lemmas~\ref{lem:#1}--\ref{lem:#2}}

\newcommand{\Rem}[1]{Remark~\ref{rem:#1}}
\newcommand{\Remenum}[2]{Remark~\ref{rem:#1}~(\ref{#2})}
\newcommand{\Def}[1]{Definition~\ref{def:#1}}
\newcommand{\Defs}[2]{Definitions~\ref{def:#1} and~\ref{def:#2}}
\newcommand{\Defenum}[2]{Definition~\ref{def:#1}~(\ref{#2})}

\newcommand{\Not}[1]{Notation~\ref{not:#1}}
\newcommand{\Fig}[1]{Figure~\ref{fig:#1}}

\newcommand{\Footnote}[1]{Footnote~\ref{fn:#1}}

\numberwithin{equation}{section}

\DeclareMathOperator{\dom}    {dom}

\DeclareMathOperator{\vol}    {vol}

\DeclareMathOperator{\Ric}    {Ric}

\DeclareMathOperator{\injrad} {inj\,rad}

\DeclareMathOperator{\id}     {id}  

\renewcommand{\Re}     {\mathrm {Re}\,}
\renewcommand{\Im}     {\mathrm {Im}\,}

\newcommand{\spec}[2][{}]   {\sigma_{\mathrm{#1}}(#2)} 
\newcommand{\essspec}[1]{\spec[ess] {#1}}
\newcommand{\disspec}[1]{\spec[d]{#1}}

\newcommand{\Err}{\mathcal O}

\newlength{\maxbreite}%
\newlength{\maxhoehe}%
\newlength{\maxtiefe}%

\newcommand{\stelldrueber}[3][0pt]{
  \settowidth{\maxbreite}{#3}%
  \settoheight{\maxhoehe}{#3}%
  \settodepth{\maxtiefe}{#2}%
  \addtolength{\maxhoehe}{\maxtiefe}%
  {\makebox[\maxbreite]{\raisebox{\maxhoehe}{\hspace{#1}#2}}%
  \makebox[0pt][r]{#3}}%
}

\newcommand{\overcirc}[1]       
{\stelldrueber[.45ex]{$\scriptscriptstyle\circ$}{${#1}$}}

\newcommand{\R}{\mathbb{R}} 
\newcommand{\C}{\mathbb{C}} 
\newcommand{\N}{\mathbb{N}} 
\newcommand{\Z}{\mathbb{Z}} 

\newcommand{\eps}{\varepsilon} 
\renewcommand{\phi}{\varphi}   
\newcommand{\e}{\mathrm e}  
\newcommand{\im}{\mathrm i} 
\DeclareMathOperator{\dd}    {d\!}     
\newcommand{\de}   {\mathrm d}         
\newcommand{\wt}{\widetilde}           
\newcommand {\qf}[1]{\mathfrak{#1}}    

\newcommand{\HS}{\mathcal H}           

\newcommand{\Sobsymb} {\mathsf H}      
\newcommand{\Contsymb} {\mathsf C}     
\newcommand{\Lsymb}    {\mathsf L}     
\newcommand{\Lsqrsymb} {\mathsf L_2}   
\newcommand{\lsymb}    {\ell}          

\newcommand{\Ci} [2][{}]{\Contsymb^\infty_{#1} ({#2})}
\newcommand{\Cci}[1]{\Ci[\mathrm c]{#1}}
\newcommand{\Cont}[2][{}]{\Contsymb_{#1}({#2})}

\newcommand{\Lp}[2][p]{\Lsymb_{#1}({#2})} 

\newcommand{\Lsqr}[2][{}]{\Lsymb_2^{#1}({#2})} 

\newcommand{\lsqr}[2][{}]{\lsymb_2^{#1}({#2})}   

\newcommand{\Linfty}[2][{}]{\Lsymb_\infty^{#1}({#2})} 

\newcommand{\Sob}[2][1]{\Sobsymb^{#1}({#2})} 

\newcommand{\norm}[2][{}]{\|{#2}\|_{{#1}}}    
\newcommand{\normsqr}[2][{}]{\|{#2}\|^2_{#1}} 
\newcommand{\bignormsqr}[2][{}]{\bigl\|{#2}\bigr\|^2_{#1}}
\newcommand{\Bignorm}[2][{}]{\Bigl\|{#2}\Bigr\|_{#1}}     

\newcommand{\iprod}[3][{}]{\langle{#2},{#3}\rangle_{#1}}  
\newcommand{\bigiprod}[3][{}]{\bigl\langle{#2},{#3}\bigr\rangle_{#1}}

\newcommand{\set}[2]{\{ \, #1 \, | \, #2 \, \} } 
\newcommand{\bigset}[2]{\bigl\{ \, #1 \, \bigl|\bigr. \, #2 \, \bigr\} }
\newcommand{\Bigset}[2]{\Bigl\{ \, #1 \, \Bigl|\Bigr. \, #2 \, \Bigr\} }

\newcommand{\map}[3]{ #1 \colon #2 \longrightarrow #3 } 

\newcommand{\bd}  {\partial}                
\newcommand{\clo}[1]{\overline{{#1}}} 

\newcommand{\compl}[1]{#1^{\mathrm c}}       
\newcommand{\dcup}{\mathrel{\uplus}}               

\newcommand{\disjcup}{\mathrel{\overline{\dcup}}} 
\newcommand{\bigdisjcup}{\operatorname*{\overline{\biguplus}}}

\newcommand{\restr}[1]{{\restriction}_{#1}} 

\newcommand{\conj}[1]{\overline {{#1}}}       
\newcommand{\orth}{\bot}                    
\newcommand{\normder}{\partial_\mathrm{n}}  

\newcommand{\1}{\mathbbm 1}                    
\newcommand{\und}{\qquad\text{and}\qquad}
\newcommand{\Neu}{{\mathrm N}}              
\newcommand{\Dir}{{\mathrm D}}              
\newcommand{\laplacian}[2][{}]{\Delta_{{#2}}^{{#1}}}
\newcommand{\laplacianN}[1]{\laplacian[\Neu]{#1}} 

\newcommand{\EW}[3][{}]{\lambda^{{#1}}_{#2}({#3})}
\newcommand{\EWN}[2]{\EW[\Neu]{#1}{#2}}      

\newcommand{\hiddenfootnote}[1]{} 

\newcommand{\eucl}{{\mathrm {eucl}}}
\newcommand{\rest}{{\mathrm {rest}}}

\newcommand{\wh}{\widehat}

\newcommand{\HSaux}{{\mathcal G}}

\newcommand{\normhat}[2][{}]{\|{#2}\hat\|_{{#1}}}    
\newcommand{\normsqrhat}[2][{}]{\|{#2}\hat\|^2_{#1}} 

\newcommand{\iprodhat}[3][{}]{\langle{#2},{#3}\hat \rangle_{#1}}

\newcommand{\ext}{{\mathrm{ext}}}
\newcommand{\inl}{{\mathrm{int}}}

\newcommand{\vxeps}{{\eps,v}}
\newcommand{\edeps}{{\eps,e}}
\newcommand{\vxed}{{v,e}}
\newcommand{\vxedeps}{{\eps,v,e}}

\begin{document}
\title[Convergence of resonances on quantum wave guides]{Convergence
  of resonances on thin branched quantum wave guides}

\author{Pavel Exner}
\address{Department of Theoretical Physics, NPI, Academy of Sciences,
25068 \v{R}e\v{z} near Prague, and Doppler Institute, Czech
Technical University, B\v{r}ehov\'{a}~7, 11519 Prague, Czechia}
\email{exner@ujf.cas.cz}

\author{Olaf Post} 
\address{Institut f\"ur Mathematik,
         Humboldt-Universit\"at zu Berlin,
         Rudower Chaussee~25,
         12489 Berlin,
         Germany}
\email{post@math.hu-berlin.de}
\date{\today}




\begin{abstract}
  We prove an abstract criterion stating resolvent convergence in the
  case of operators acting in different Hilbert spaces. This result is
  then applied to the case of Laplacians on a family $X_\eps$ of
  branched quantum waveguides. Combining it with an exterior complex
  scaling we show, in particular, that the resonances on $X_\eps$
  approximate those of the Laplacian with ``free'' boundary conditions
  on $X_0$, the skeleton graph of $X_\eps$.
\end{abstract}

\maketitle

%
\section{Introduction}
\label{sec:intro}
%

In a few recent years there was a surge of interest to quantum
mechanics on metric graphs. It is a subject with a long history
reaching back to the paper of Ruedenberg and Scherr
\cite{ruedenberg-scherr:53} on spectra of aromatic carbohydrate
molecules elaborating an idea of L.~Pauling, but a systematic
study motivated by the need to describe semiconductor graph-type
structures began only at the end of the eighties,
cf.~\cite{exner-seba:89}; a survey of the subsequent development
with the appropriate bibliography can be found, e.g., in the
papers~\cite{kostrykin-schrader:99} or~\cite{kuchment:04}.

Since quantum graphs are used in the first place to model various real
graph-like structures whose transverse size is small but non-zero, one
of the most important questions in the theory is how such system
approximate an ideal graph in the limit of zero thickness. This
problem is difficult and the answer is so far known in some cases
only.  In particular, compact ``fat graphs'' with Neumann boundary
conditions has been analyzed, first in~\cite{freidlin-wentzell:93} and
\cite{freidlin:96}, then in~\cite{kuchment-zeng:01} and
\cite{rubinstein-schatzman:01} where the eigenvalue convergence was
demonstrated; an extension of this result to more general Neumann-type
graph-like manifolds can be found in~\cite{exner-post:05}. More
recently, the resolvent convergence on non-compact graph-like
manifolds of this type was dealt with in~\cite{post:06a}. Recall,
however, that the analogous problem in the physically most important
case of tube systems with Dirichlet boundary is more difficult and at
the present moment far from being fully understood, although there are
fresh results in this direction~\cite{post:05a},
\cite{molchanov-vainberg:pre06b}.

Apart of the spectral analysis, one of the most important questions we
study on quantum graphs concerns the resonance scattering. It is
usually easy to find resonances on a graph --- see, e.g.,~\cite{etv:01}
and references therein --- but \emph{a priori} it is not clear how are
these related to possible resonances on an approximating
finite-thickness manifold. This is the main topic of the present
paper.

An efficient way to study resonances understood as poles in the
analytically continued resolvent is to rephrase the question as an
eigenvalue problem. A time-honored trick to achieve this goal is based
on the \emph{complex scaling} --- see, e.g.,~\cite{combes:69,
  aguilar-combes:71, balslev-combes:71, simon:72, combes-thomas:73,
  simon:79, cdks:87, bcdc:89} or~\cite[Sec.~XII.6
and~XIII.10]{reed-simon-1-4} --- which transforms the Hamiltonian by a
non-unitary operator with the aim to rotate the essential spectrum
uncovering a part of the ``second sheet'' while leaving the poles at
place\footnote{While the complex-scaling method was formulated by
  mathematicians it became a practical and often used tool in atomic
  and molecular physics -- see, e.g., the review~\cite{moiseyev:98}.}.
Our aim here is to apply this method to the problem at hand. We will
construct an \emph{exterior} complex-scaling transformation for
Hamiltonians on graph-like manifolds and show that some among its
complex eigenvalues converge to the eigenvalues of the complex-scaled
graph Hamiltonian\footnote{Complex scaling was used to treat
  resonances of thin tubes also in~\cite{nedelec:97, dem:01}, this
  time with Dirichlet boundary conditions. In that case the resonances
  of the limiting zero-thickness problem come from the tube curvature
  rather than the (trivial) graph structure. The complex scaling can
  be also used to demonstrate equivalence of the ``resolvent'' and
  scattering resonances for a wide class of quantum graphs including
  those discussed here~\cite{exner-lipovsky:pre06}.}.  In this way,
resonances of the quantum graph are approximated by those of the
corresponding family of ``fat graphs'' (cf.~\Thms{main}{main2}).
Furthermore, graph Hamiltonians often have embedded eigenvalues, e.g.\
by rational relations between the edges, and these are again
approximated, either by embedded eigenvalues or by resonances (one can
conjecture that the latter case is generic).

As a by-product of our analysis we will prove, using the technique
of~\cite{post:06a}, that a magnetic Laplacian of a family of ``fat
\emph{non-compact} graphs'' converges to the one on the corresponding
graph, this time without any complex scaling (cf.~\Thm{mag.ham}). This
conclusion is rather important because it shows that nice results
about fractal graph spectra such as the one discussed in~\cite{bgp:07}
can be observed in some form with more ``realistic'' systems. Needless
to say, this is a goal which the experimental physicists vigorously
pursue, see e.g.~\cite{albrecht-et-al:01}. The convergence of the
spectrum of the magnetic Laplacian on a \emph{compact} graph was
already established in~\cite{kuchment-zeng:01}.

Let us describe the contents of the paper. To explain our method in a
simple setting first, we analyze in the next section an example of a
``lasso'' graph having one loop and one semi-infinite external link.
After that we describe the two main objects of our approximation,
quantum graphs in \Sec{graph.model} and quantum wave guides in
\Sec{qwg.model}. The following section is devoted to explanation of
the complex-dilation method in our setting, and in \Sec{closeness.mod}
we will state and prove our main results.

Since the arguments are rather technical and demand various auxiliary
material, we collected it in a series of appendices. \App{scale}
contains facts about Hilbert scales associated to sectorial operators,
\App{abstr.crit} provides an abstract convergence theory for
eigenvalues and eigenvectors of non-selfadjoint operators in different
Hilbert spaces.  Finally, we prove in \App{res.est} among other things
the analyticity of the complex dilated operators.

%
\section{A motivating example: a loop with a lead}
\label{sec:loop}
%
Let us start with a slightly informal discussion of a simple example
in order to show the main purpose and to motivate the general analysis
presented in the forthcoming sections. Proofs and more precise
definitions of the operators will also be given there.

\subsection{The graph and its neighbourhood}
\label{sec:loop_graph}
\begin{figure}[h]
  \centering
\begin{picture}(0,0)
  \includegraphics{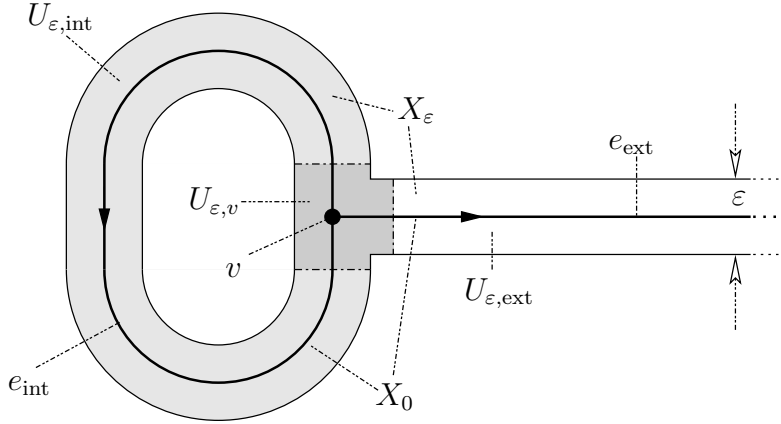}
\end{picture}
\setlength{\unitlength}{4144sp}
\begin{picture}(4522,2481)(631,-2050) 
  \put(546,320){$U_{\eps,\inl}$}
  \put(1720,-1186){$v$} 
  \put(431,-1861){$e_\inl$}
  \put(3986,-421){$e_\ext$}%
  \put(3131,-1321){$U_{\eps,\ext}$}%
  \put(1500,-781){$U_{\eps,v}$}%
  \put(2606,-1951){$X_0$}%
  \put(2736,-241){$X_\eps$}%
  \put(4716,-756){$\eps$}%
\end{picture}%
  \caption{The metric graph $X_0$ consisting of one loop and one
    external line together with the $\eps/2$-neighbourhood $X_\eps$.}
  \label{fig:loop}
\end{figure}

Denote by $X_0$ the metric graph consisting of a loop $e_\inl$ with a
finite length $\ell := \ell_\inl \in (0,\infty)$ and one external
line, i.e., an edge $e_\ext$ of length $\ell_\ext = \infty$ attached
to the loop $e_\inl$ at the vertex $v$; sometimes also called a
\emph{lasso graph}~\cite{exner:97}. For simplicity we assume that the
graph is planar, i.e., $e \subset \R^2$ and $v \in \R^2$ where $e$
denotes either $e_\inl$ or $e_\ext$ (cf.\ \Fig{loop}), and
furthermore, that the edges are straight in a neighbourhood of $v$; we
will simply suppose that the exterior edge $e_\inl$ is embedded as a
straight half-line in $\R^2$.
Denote by $X_\eps$ the open $\eps/2$-neighbourhood of $X_0$. We
decompose $X_\eps$ into three open, mutually disjoint sets
$U_{\eps,\ext}$, $U_\vxeps$ and $U_{\eps,\inl}$ such that the union of
their closures equals $\clo X_\eps$. They are chosen in such a way
that $v \in U_\vxeps$ while $U_\edeps$ is the $\eps$-tubular
neighbourhood of the slightly shortened edge $e$. Since the edges are
straight near $v$ by assumption, $U_\vxeps$ is $\eps$-homothetic to a
fixed set $U_v \subset \R^2$ and there exists an affine transformation
\begin{align}
  \label{eq:trafo.vx}
  \tau_\vxeps \colon U_v & \longrightarrow U_\vxeps\\
                       z & \mapsto         v + \eps z.
\end{align}
The $\eps$-tubular neighbourhood $U_\edeps$ is given by
\begin{align}
  \label{eq:trafo.ed}
  \tau_\edeps \colon e \times F & \longrightarrow U_\edeps\\
   (x,y) & \mapsto \gamma_e \bigl(
                  \phi_\edeps(x)) + \eps y n_e(\phi_\edeps(x)\bigr)
\end{align}
where $\map {\gamma_e} {(0,\ell_e)} {X_\eps \subset \R^2}$ denotes the
path parametrising the edge $e$ by arc-length (according to its
orientation). Furthermore, $\map {n_e} {(0,\ell_e)} {\R^2}$ denotes
one of the two possible unit vector fields along $\gamma_e$ orthogonal
to the tangent vector $\dot \gamma_e$. We can also identify $e$ with
the interval $(0,\ell_e)$ and set $F:=(-1/2,1/2)$. Since the graph is
embedded into $\R^2$, we have to take a slightly smaller part of
$e=e_\inl$ instead of the full edge. This is needed when constructing
the edge neighbourhood $U_\edeps$ in order to make room for the vertex
neighbourhood $U_\vxeps$. We therefore let
\begin{equation*}
  \map {\phi_\edeps} {(0,\ell)}
      {\bigl(\eps \ell/2,(1-\eps/2 )\ell\bigr)}
\end{equation*}
be the affine linear mapping from the full edge $e$ onto the
shortened edge where $\eps \ell/2$ is the amount of $e$ belonging
to the vertex neighbourhoods; for the external edge a simple shift
by $\eps \ell/2$ will do the job.

Since we want to study the (non-relativistic) quantum dynamics on
the graph in presence of external fields we have to introduce the
latter. Denote by $g_\eucl$ the usual Euclidean metric in $\R^2$.
The \emph{vector potential} $\alpha$ in $\R^2$ is given by a
real-valued $1$-form $\alpha = a_1 \de z_1 + a_2 \de z_2$ and we
denote the corresponding vector field by $a=(a_1,a_2)$.
Furthermore, let $q$ be a real-valued function on $\R^2$, the
\emph{electric} potential. Their regularity properties will be
specified below.

In the particular example of this section we could, of course,
perform all the reasoning which follows in the coordinates given
by the embedding. We will, however, employ the $\eps$-independent
sets $U_v$ and $U_e := e \times F$, not only because the argument
is simpler but also because it can be easily be generalized to the
differential geometric setting which we will use in the general
case below. Consequently, let us express the metric, the magnetic
and electric potential in terms of the coordinates given on $U_v$
and $U_e$. We set
\begin{equation}
  \label{eq:trafo}
  \begin{split}
    g_\vxeps &:= \tau_\vxeps^* g_\eucl, \\
    g_\edeps &:= \tau_\edeps^* g_\eucl,
  \end{split} \qquad
  \begin{split}
    \alpha_\vxeps &:= \tau_\vxeps^* \alpha, \\
    \alpha_\edeps &:= \tau_\edeps^* \alpha,
  \end{split} \qquad
  \begin{split}
    q_\vxeps &:= \tau_\vxeps^* q, \\
    q_\edeps &:= \tau_\edeps^* q,
  \end{split}
\end{equation}
where $\tau_\vxeps^* \omega$ denotes the usual pull-back of the
tensor $\omega$ from (a subset of) $\R^2$ to $U_\vxeps$, and the
other map has the analogous meaning. A simple calculation now
shows that quantities at left-hand sides are equal to
\begin{align}
  \label{eq:trafo.met}
  \biggl\{
  \begin{split}
    g_\vxeps(z)   &= \eps^2 g_\eucl,\\
    g_\edeps(x,y) &=
      (1-\eps)^2 \bigl(1-\eps y \kappa_e(\wt x) \bigr)^2 \de x^2 +
      \eps^2 \de y^2,
  \end{split}
  \biggr. \\
  \label{eq:trafo.mag}
  \biggr\{
  \begin{split}
    \alpha_\vxeps(z)   &= \eps \alpha(v+\eps z),\\
    \alpha_\edeps(x,y) &=
      (1-\eps)\bigl(1-\eps y \kappa_e(\wt x)\bigr) a_e^\|(x,y)\, \de x +
      \eps a_e^\orth(x, y)\, \de y,
  \end{split}
  \biggr. \\
  \label{eq:trafo.pot}
  \biggr\{
  \begin{split}
    q_\vxeps(z)   &= q(v+ \eps z),\\
    q_\edeps(x,y) &= q\bigl(\gamma_e(\wt x) + \eps y n_e(\wt x)\bigr),
  \end{split}
  \biggr.
\end{align}
where $\wt x = \phi_\edeps(x)$, $z \in U_v$, $(x,y) \in e \times
F$ and
\begin{equation}
  \label{eq:mag.tan.nor}
  a_e^\|(x,y) := \dot \gamma_e(\wt x) \cdot a(\tau_\edeps(x,y)),
  \qquad
  a_e^\orth(x,y) := n_e(\wt x) \cdot a(\tau_\edeps(x,y))
\end{equation}
denote the \emph{tangential} and \emph{normal} component of the
vector field $a$, respectively, taken at the (shortened) edge $e$
parametrised by $\gamma \circ \phi_\edeps$. Furthermore,
\begin{equation}
  \label{eq:def.curv}
  \kappa_e :=
     \dot \gamma_{e,1} \ddot \gamma_{e,2} -
     \dot \gamma_{e,2} \ddot \gamma_{e,1}
\end{equation}
denotes the (signed) curvature of the curve $\gamma_e =
(\gamma_{e,1},\gamma_{e,2})$ embedded in $\R^2$. As mentioned above we
assume that $\kappa_e=0$ on the external edge $e=e_\ext$, and
therefore
\begin{equation}
\label{eq:met.edext}
   g_{\eps,\ext}(x,y) = \de x^2 + \eps^2 \de y^2
\end{equation}
has a product structure.  In addition, we suppose that the tangential
component of the vector potential vanishes, $a_e^\|=0$; notice that
this can always be achieved by an appropriate gauge transformation
(see~\Sec{gauge} below).  For simplicity, we assume also that there is
no electric potential on the exterior edge $e_\ext$ as well as on its
neighbourhood $U_{\eps,\ext}$.

\subsection{Magnetic Hamiltonians}
\label{sec:mag.ham}
After describing the graph and its neighbourhood we introduce now
the corresponding magnetic Schr\"odinger operator for a vector
potential $a=(a_1,a_2)$ (a vector field) and an electric potential
$q$ (a function). We shall assume that $a_1$, $a_2$, $q$ and their
first derivatives are bounded and, as we have said, that they
vanish on the external edge neighbourhood.

Let us start with the ``fat graph''. The magnetic Hamiltonian
$H_\eps$ in the Hilbert space $\Lsqr {X_\eps}$  is given formally
by the differential expression
\begin{equation}
  \label{eq:def.mag.eps}
  H_\eps := (\nabla - \im a)^* (\nabla - \im a) + q
\end{equation}
acting on $X_\eps$. To define it properly as a self-adjoint operator
one has to specify its domain; namely, we assume \emph{Neumann}
boundary conditions. In terms of coordinates introduced on the edge
and vertex neighbourhoods we have
\begin{multline*}
  H_\edeps = \bigl(-\partial_x + \im a_e^\| + \Err(\eps)\bigr)
             \bigl(\partial_x - \im a_e^\| + \Err(\eps)\bigr)\\ +
     \frac 1 {\eps^2}
             \bigl(-\partial_y + \im \eps a_e^\orth\bigr)
             \bigl(\partial_y - \im \eps a_e^\orth\bigr) + q_\edeps
\end{multline*}
for the internal edge $e=e_\inl$ and
\begin{gather*}
  H_\edeps = -\partial_{xx} - \frac 1 {\eps^2} \partial_{yy},\\
  H_\vxeps = \frac 1 {\eps^2}
             \bigl(-\nabla + \im \eps a_v\bigr)
             \bigl(\nabla  - \im \eps a_v\bigr) + q_\vxeps
\end{gather*}
for the external edge $e=e_\ext$ and the vertex $v$, respectively,
where $a_v$ is the vector field corresponding to the $1$-form
$\alpha_v$. The error term on the internal edge comes from the
curvature and the shortened edge --- cf.~\eqref{eq:trafo.met}
and~\eqref{eq:trafo.mag}.

On the other hand, on the graph we consider the ``limit'' operator
$H_0$ given by
\begin{align*}
  H_{0,\inl} &= (-\partial_x + \im a_e)
        (\partial_x - \im a_e) + q_e, \quad e=e_\inl\\
  H_{0,\ext} &= -\partial_{xx}.
\end{align*}
To fix its domain we have to specify how the functions are related at
the vertex $v$. We suppose that they satisfy the so-called \emph{free}
boundary conditions\footnote{\label{fn:kirchhoff}They are often
  labelled as Kirchhoff boundary conditions, with an allusion to
  classical electrical circuits. The term is unfortunate, however,
  since \emph{every} boundary condition giving rise to a self-adjoint
  graph Hamiltonian must preserve the (probability) current.},
namely
\begin{gather*}
  f_\inl(0) = f_\inl(\ell) = f_\ext(0),\\
  (f' - \im a f)_\inl(0+) -
  (f' - \im a f)_\inl(\ell-) +
  f_\ext'(0+) = 0.
\end{gather*}
More general (self-adjoint) boundary conditions for a magnetic
Hamiltonian on $X_0$ were discussed in~\cite{exner:97}, in particular,
from the point of view of resonances.

The magnetic and electric potential on the internal edge can be
easily found, in particular, one can see from the ``fat graph''
Hamiltonian that
\begin{equation}
  \label{eq:pot.null}
  a_e(x) := \dot \gamma_e(x) \cdot a(\gamma_e(x))
          = a_e^\|(\phi_\edeps^{-1}(x),0), \quad
  q_e(x) := q(\gamma_e(x))
\end{equation}
are the tangent component of $a$ and the value of $q$,
respectively, along the \emph{full} edge $e=e_\inl$.  Indeed, on a
heuristic level the choice of the potentials in the limiting
operator is justified by the relations
\begin{equation}
  \label{eq:est.mag}
  \begin{split}
    |a_e^\| (x,y) + \Err(\eps) -a_e(x)| &\le \eps c_1 \norm[C^1] a,\\
    |\eps a_\vxeps(z)| &\le \eps \norm[\infty] a,
  \end{split}
\end{equation}
where $\norm[C^1] a$ denotes the supremum of $|a|$, $|\nabla a_1|$
and $|\nabla a_2|$ on $X_\eps$, and
\begin{equation}
  \label{eq:est.pot}
  \begin{split}
    |q_\edeps(x,y)-q_e(x)| &\le \eps c_2 \norm[C^1] q,\\
    |q_\vxeps(z)- q(v)|    &\le \eps c_3 \norm[\infty] q,
  \end{split}
\end{equation}
where the constants $c_i>0$ depend only on $\ell$ and
$\norm[\infty] {\kappa_e}$, $0 < \eps \le 1$.

As in the previous work quoted in the introduction our aim is to give
meaning to the intuitive notion that $H_0$ described above is in some
sense a limit of the operators $H_\eps$ as $\eps\to 0$ --- now from
the resonance point of view --- despite the fact they act on different
Hilbert spaces. There is no paradox here, of course, since only the
lowest transverse eigenmode survives, in other words, all functions
which are not constant in the transverse direction $y$ will not
contribute to the limit. We will make this vague observation precise
in \Sec{closeness.mod} and~\App{abstr.crit} below.

Note also that we have a somehow simpler, unitary equivalent
magnetic Hamiltonian $\hat H$ on the graph obtained by the gauge
transformation $\hat f = \Xi f$ where
\begin{equation}
  \label{eq:gauge.ed}
  \Xi_e(x):=\e^{-\im \Phi_e(x)} \qquad \text{and} \qquad
  \Phi_e(x):=\int_0^x a_e(s) \dd s.
\end{equation}
on the loop and $\Xi_e=1$ on the external edge (cf.~\Sec{gauge}),
where $\Phi=\Phi_e(\ell)$ is the \emph{total flux} through the
loop. The free boundary conditions under this unitary
transformation become
\begin{subequations}
  \label{eq:bd.loop}
  \begin{gather}
      \hat f_\inl(0) = \e^{\im \Phi} \hat f_\inl(\ell) =
      \hat f_\ext(0)\,,\\
      \hat f_\inl'(0+) - \e^{\im \Phi} \hat f_\inl'(\ell-) + \hat
      f_\ext'(0+) = 0\,;
  \end{gather}
\end{subequations} %
the price for the simpler expression of the Hamiltonian on an edge
are more complicated boundary conditions, with
\emph{discontinuous} functions at the vertex.

\subsection{Complex dilations and resonances}
\label{sec:complex}
Let us recall briefly the essence of the complex exterior dilation
argument --- for more details we refer, e.g.,
to~\cite[Sec.~XIII.10]{reed-simon-1-4},~\cite{cdks:87}
or~\cite{hislop-sigal:96}. We will do it in our setting, both on
the graph and its neighbourhood, i.e., for a fixed $\eps \ge 0$.
Let us consider the one-parameter unitary group $U_\edeps^\theta$
on the \emph{external} part $X_{\eps,\ext} := U_\edeps$, in
particular $X_{0,\ext} = e_\ext$ for the graph, whose element
characterized by the parameter $\theta \in \R$ acts as
\begin{equation}
  \label{eq:def.unitary}
  \begin{split}
      (U_0^\theta f)_e (x) &:=
                 \e^{\theta/2} f_e (\e^{\theta} x)\\
      (U_\eps^\theta u)_\edeps(x,y) &:=
                 \e^{\theta/2} u(\e^{\theta} x,y)
  \end{split} \quad
\end{equation}
at the external edge $e = e_\ext$; note that $U_\eps^\theta$ is
unitary since the exterior edge $e=e_\ext$ is straight by
assumption and therefore the metric on $U_\edeps$ has the product
structure \eqref{eq:met.edext}. The transformation can be extended
to the whole Hilbert space acting as the identity operator on the
\emph{internal} parts $X_{\eps,\inl}$, in other words,  a function
on the graph or the fat graph is \emph{longitudinally} dilated on
the external edge and remains unchained on the remaining parts. A
simple coordinate transformation shows that for a fixed $\eps \ge
0$, the action of the dilated magnetic Hamiltonian $H_\eps^\theta
:= U_\eps^\theta H_\eps (U_\eps^\theta)^*$ is given by
\begin{equation}
  \label{eq:op.dil}
  \begin{split}
    H_\eps^\theta u &= H_{\eps,\inl} u_\inl + H_{\eps,\ext}^\theta
         u_\ext,\\
    H_0^\theta f &= H_{0,\inl} f_\inl - \e^{-2\theta}
    \partial_{xx} f_\ext,
  \end{split}
\end{equation}
where
\begin{equation*}
   \qquad  H_{\eps,\ext}^\theta :=
     - \e^{-2\theta} \partial_{xx}
           - \frac 1 {\eps^2} \partial_{yy}
\end{equation*}
The domain $\dom H_\eps^\theta$ consists of all functions which are
locally twice weakly $\Lsymb_2$-differentiable and satisfy the
conditions
\begin{equation}
  \label{eq:dom.dil}
  u_\ext  = \e^{\theta/2}  u_\inl  \qquad \text{and} \qquad
  u_\ext' =\e^{3\theta/2} u_\inl'
\end{equation}
on $\Gamma_\eps$, the common boundary of $X_{\eps,\inl}$ and
$X_{\eps,\ext}$, where $u_\ext'=\partial_x u$ and $u_\inl'=\partial_x
u$ denote the (normal) derivatives in the orientation of $x$, i.e.,
the outward normal derivative on $\bd X_{\eps,\inl}$ and the inward
normal derivative on $\bd X_{\eps,\ext}$.\footnote{Here $\bd
  X_{\eps,\bullet} = \bd_{\R^2} X_{\eps,\bullet} \cap X_\eps$ means
  the boundary w.r.t.~the open set $X_\eps$, \emph{not} the boundary
  of $X_\eps$ as subset of $\R^2$}

In the particular case of the graph, $\eps=0$, we have to specify
the boundary conditions. Using the gauge described above we can
write them as
\begin{gather*}
  \hat f_\inl(0) =
  \e^{\im \Phi} \hat f_\inl(\ell) =
  \e^{-\theta/2} \hat f_\ext(0),\\
  \hat f_\inl'(0+) -
      \e^{\im \Phi} \hat f_\inl'(\ell-) +
      \e^{-3\theta/2}\hat f_\ext'(0+) = 0.
\end{gather*}

In the next step we use~\eqref{eq:op.dil} and~\eqref{eq:dom.dil} to
perform the basic trick of the complex-scaling methods by extending
the definition of $H_\eps^\theta$ to complex $\theta$ with $2|\Im
\theta| < \vartheta < \pi$. Note that such a perturbation is very
singular with respect to $\theta$, even for real $\theta$, since not
only the operator domain, but also the form domain depends on $\theta$
as we shall discuss in \App{res.est} below. In the spirit
of~\cite{cdks:87} we are going to show there that
$\{H_\eps^\theta\}_\theta$ defines a self-adjoint family of operators
(i.e., $(H_\eps^\theta)^* = H_\eps^{\conj \theta}$) with spectrum
contained in the common sector $\Sigma_\vartheta$ for $\theta$ in the
strip $S_\vartheta$ where
\begin{equation}
  \label{eq:def.sect.strip}
  \Sigma_\vartheta := \bigset{z \in \C} {|\arg z|\le \vartheta} \und
  S_\vartheta := \bigset{\theta \in \C} {|\Im \theta|< \vartheta/2}.
\end{equation}

Following the usual convention --- see, e.g.,
\cite[Sec.~XII.6]{reed-simon-1-4} --- we define a \emph{resonance} of
$H_\eps$ with $\eps \ge 0$ as the pole of the resolvent analytically
continued over the cut given by the essential spectrum of the
operator. The position of the cut changes once $\theta$ ceases to be
real, in particular, for $\Im \theta>0$ sufficiently large it may
``expose'' the pole which will just become a complex eigenvalue of
$H_\eps^\theta$ in the lower half-plane.  Such eigenvalues will the
main object of our interest.

In the example, the eigenvalues $\lambda = k^2$ of the quantum
graph Hamiltonian $H_0^\theta$ with a magnetic field of total flux
$\Phi$ through the loop (to make things simple we put $q=0$), are
obtained from the condition~\cite{exner:97}
\begin{equation*}
  2(\cos k\ell - \cos\Phi) = \im \sin k\ell\,.
\end{equation*}
If $\Phi\ne 0~(\mathrm{mod}~\pi)$ none of the solution is real,
while for $\Phi=0$ half of the solutions is on the real axis and
the other half in the lower half-plane, explicitly
\begin{equation}
  \label{eq:ex.res}
  \lambda_j = \frac 1 {\ell^2} \, \bigl( 2\pi j \bigr)^2,
         \quad  \text{and} \quad
  \hat \lambda_j =\frac 1 {\ell^2}\bigl( 2\pi j - \im \ln 3 \bigr)^2.
\end{equation}
for $j\in\mathbb{Z}\setminus\{0\}$ and $j\in\mathbb{Z}$,
respectively; as expected the values of $\lambda_j$ and $\hat
\lambda_j$ are independent of the exterior scaling parameter
$\theta$. The real eigenvalues $\lambda_j$ do not turn into
resonances because they correspond to eigenfunctions on the loop
which have a node at the vertex, and therefore do not ``know''
about the presence of the external lead, the half-line part of the
eigenfunction being zero. The remain embedded into the essential
spectrum of $H_0$ coming from the half-line, and naturally become
isolated after the complex scaling whenever $\Im \theta>0$ and
$\essspec {H_0^\theta} = \e^{-2\theta}[0,\infty)$.

On the contrary, the solutions corresponding to $\hat\lambda_j$ are
true resonances. Their half-line component is proportional to
$\exp\bigl((\ln 3 + \im \cdot 2\pi j) x/\ell\bigr)$ and thus not
square integrable, however, after a complex scaling with large enough
$\Im \theta$ it will become (a part of) an $\Lsqrsymb$-eigenfunction
of $H_0^\theta$. Recall that $\essspec {H_0^\theta} =
\e^{-2\theta}[0,\infty)$ is rotated into the lower half-plane by the
angle $2\Im \theta$ and the resonances $\hat \lambda_j$ lie on the
parabola
\begin{equation*}
  \Im \hat \lambda = - \frac{2\ln 3}{\ell}\, \sqrt{
  \Re \hat \lambda + \left(\frac{\ln 3} \ell \right)^2},
\end{equation*}
hence for complex scaling with $\Im \theta$ large enough
\emph{all} resonances are revealed.

On the other hand, for the ``fat graph'' $X_\eps$ one can check
easily that $\essspec {H_\eps^\theta} = \frac 1 {\eps^2} \spec
{\laplacianN F} + \e^{2\theta}[0,\infty)$ consists of an infinite
number of half-lines turned by $2\Im \theta$; each attached to the
base point $(j\pi/2\eps)^2 \in \spec {\laplacianN F}$. All these
base points except the one with $j=0$ tend to $\infty$ as $\eps
\to 0$, so for any bounded set $B \subset \C$ we have
\begin{equation}
  \label{eq:sp.ess}
  \essspec {H_0^\theta} \cap B = \essspec {H_\eps^\theta} \cap B
\end{equation}
provided $\eps>0$ is small enough, in other words, higher sheets
of the Riemann surface associated the resolvent of $H_\eps^\theta$
play no role. The question is whether the complex dilation reveals
resonances of this system --- manifested as complex eigenvalues of
$H_\eps^\theta$ --- and what is their relation to the resonances of
the graph. The answer which are going to demonstrate is the
following.

\begin{theorem}
  \label{thm:main}
  Let $\lambda(0)$ be a resonance of the magnetic Hamiltonian $H_0$
  with a multiplicity $m>0$. Under the stated assumptions, for a
  sufficiently small $\eps>0$ there exist $m$ resonances
  $\lambda_1(\eps), \dots, \lambda_m(\eps)$ of $H_\eps$, satisfying
  $\Im\lambda_j(\eps)<0$ and not necessarily mutually different, which
  all converge to $\lambda(0)$ as $\eps\to 0$. The same is true in the
  case when $\lambda(0)$ is an embedded eigenvalue of $H_0$, except
  that $\Im\lambda_j(\eps)\le 0$ holds in general.
\end{theorem}
In the following sections we will prove this claim in a
considerably more general setting when the loop is replaced by a
finite metric graph to which a finite number half-lines is
attached --- this will be the main result of this paper.

The indicated proof will be divided into several steps. First we
will introduce generally in \Sec{graph.model} and~\Sec{qwg.model},
respectively, the Hamiltonians of the quantum graph and on the
corresponding graph-like waveguide. Next in \Sec{dilation} we
present the exterior scaling argument. Finally, in
\Sec{closeness.mod} we conclude the proof by verifying conditions
of abstract criteria given in \App{abstr.crit}; the aim is to show
the convergence of discrete eigenvalues --- complex in general ---
for non-self-adjoint operators $\wt H^\theta=H_\eps^\theta$ and
$H^\theta=H_0^\theta$ having a ``distance'' which tends to zero.
The difficult part of the argument, in comparison with
\cite{rubinstein-schatzman:01, kuchment-zeng:01, exner-post:05},
is that we cannot use the variational characterization of
eigenvalues because our operators are not self-adjoint, nor even
normal.

%
\section{Quantum graph model}
\label{sec:graph.model}
%
Passing to our main subject we define now the general model in
which we are able to prove the convergence of resonances. We start
with the quantum graph.

\subsection{Metric graphs}
\label{sec:graph}

Suppose $X_0$ is a connected metric graph given by $(V,E,\bd,\ell)$
where $(V,E,\bd)$ is a usual graph, i.e., $V$ denotes the set of
vertices, $E$ denotes the set of edges, $\map \bd E {V \times V}$
associates to each edge $e$ the pair $(\bd_+e,\bd_-e)$ of its terminal
and initial point (and therefore an orientation). That $X_0$ is a
\emph{metric} graph (also called \emph{quantum} graph) means that
there is a \emph{length function} $\map \ell E {(0,\infty]}$
associating to each edge $e$ a length $\ell_e$. We often identify the
edge $e$ with the interval $(0,\ell_e)$. Clearly, the length function
makes $X_0$ into a metric space.

For each vertex $v \in V$ we set
\begin{equation*}
  E_v^\pm := \set {e \in E} {\bd_\pm e = v} \qquad \text{and} \qquad
  E_v := E_v^+ \dcup E_v^-,
\end{equation*}
i.e., $E_v^\pm$ consists of all edges starting ($-$) resp.\ ending
($+$) at $v$ and $E_v$ their \emph{disjoint} union. Note that the
\emph{disjoint} union is necessary in order to allow loops, i.e.,
edges having the same initial and terminal point as in the example in
\Sec{loop}. We adopt the following uniform bounds on the degree $\deg
v := |E_v|$ and the length function $\ell$:
%
\newcommand{\graphass}{H$_0$}
\newcommand{\eqrefgraph}{\textrm{(\graphass){}}} 
\begin{align}
  \label{eq:deg.bd}
  \tag{\graphass1}
  \deg v \le d_0&, \qquad v \in V,\\
  \label{eq:len.bd}
  \tag{\graphass2} \ell_e \ge \ell_0&, \qquad e \in E,
\end{align}
where $0 < d_0 < \infty$ and $0<\ell_0 \le 1$. Of course, both
assumptions are fulfilled if $|E|$ and $|V|$ are finite.

An edge $e$ with $\ell_e = \infty$ will be called \emph{external} and
$E_\ext$ denotes the set of all external edges. Such edges are assumed
to have only an initial point, i.e., $\bd e$ consists only of the
point $\bd_-e$ for $e \in E_\ext$. The remaining edges are called
\emph{internal} and their set will be denoted by $E_\inl := E
\setminus E_\ext$. We call the vertices connecting internal and
external edges \emph{boundary vertices}, denoted by
\begin{equation}
  \label{eq:def.gamma0}
  \Gamma_0 := \set{\bd_- e \in V}{e \in E_\ext}.
\end{equation}
Since we are not aware of reasonable models with an infinite number of
external edges attached, we assume throughout this paper that
$\Gamma_0$ (i.e.,~$E_\ext$) is finite, namely
\begin{equation}
  \tag{\graphass3}
  \label{eq:ext.fin}
  |E_\ext| = |\Gamma_0| < \infty.
\end{equation}

\subsection{Magnetic Hamiltonian on the graph}
\label{sec:mag.graph}
Let $\HS = \Lsqr {X_0} = \bigoplus_{e \in E} \Lsqr e$, and denote
the corresponding norm by $\norm[0] f = \norm f$. Suppose that $a$
and $q$ are bounded, measurable functions on $X_0$, i.e,
\begin{equation}
  \label{eq:pot.bd}
  \tag{\graphass4}
  \norm[\infty] a < \infty, \qquad\text{and}\qquad
  \norm[\infty] q < \infty.
\end{equation}
Without loss of generality, we assume that $q \ge 0$ and that $a_e$ is
a smooth function on each edge (cf.~\Rem{a.smooth}). For simplicity,
we also assume that $q_e$ is smooth on each edge. We set
\begin{equation}
  \label{eq:def.a.0}
  \qf h(f) := \sum_{e \in E} \qf h_e(f_e), \qquad
  \qf h_e(f) :=
     \int_e \bigl[ |D_e f_e|^2 + q_e |f_e|^2 \bigr] \dd x.
\end{equation}
where $D_e f_e := f'_e - \im a_e f_e$. In particular, $\qf h$ is
non-negative, i.e, $\qf h(f) \ge 0$ for all $f$. We specify its
domain below.
\begin{notation}
  Here and in the sequel, the subscript $(\cdot)_e$ refers to the
  restriction onto the edge $e$ (sometimes also identified with the
  interval $(0,\ell_e)$, e.g., $f_e := f \restr e$, $\norm[e] \cdot$
  denotes the norm on $\Lsqr e$, $\qf h_e$ is the restriction of $\qf
  h$ onto $\Lsqr e$ etc. We often omit the index if it is clear from
  the context (e.g.,\ $\qf h_e(f)=\qf h_e(f_e)$).

  Denote by $\Sob[k] e$ the Sobolev space on the interval $e \cong
  (0,\ell_e)$ of $k$-times $\Lsqrsymb$-weakly differentiable
  functions.
\end{notation}
\begin{notation}
  Denote by $\norm[\qf q] \cdot$ the norm associated to a closed,
  non-negative quadratic form $\qf q$ in the Hilbert space $\HS$,
  i.e.,
  \begin{equation}
    \label{eq:norm.qf}
    \normsqr[\qf h] f := \normsqr f + \qf q(f).
  \end{equation}
  This norm turns $\HS^1:=\dom \qf h$ into a complete Hilbert space.
\end{notation}
Denote by $\qf d$ the quadratic form $\qf h$ where $a=0$ and $q=0$.
\begin{lemma}
  \label{lem:h.free}
  Assume that $a, q \in \Linfty {X_0}$. Then $\qf h$ and $\qf d$ are
  closed forms on
  \begin{equation}
    \label{eq:dom.h.free}
    \HS^1 := \Sob {X_0} := \Cont {X_0} \cap \bigoplus_{e \in E} \Sob e.
  \end{equation}
  Furthermore, the norms $\norm[1] \cdot := \norm[\qf d] \cdot$ and
  $\norm[\qf h] \cdot$ are equivalent.
\end{lemma}
\begin{proof}
  It can be quite easily seen that $\qf d$ is a closed form on $\Sob
  {X_0}$ (this is clear by standard arguments for $\qf d_e$, and the
  vertex condition remains true by the continuity of $f_e \mapsto
  f_e(v)$, $v \in \bd e$ (cf.~\eqref{eq:point.vx}). In addition,
  \begin{equation*}
    \qf h_e(f) \le
    2 \qf d_e(f) + (2 \normsqr[\infty] {a_e} +
                            \normsqr[\infty] {q_e})\normsqr[0] {f_e}
  \end{equation*}
  and a similar inequality holds with the roles of $\qf h_e$ and $\qf d_e$
  interchanged, thus the norms $\norm [\qf h] \cdot$ and $\norm[\qf d]
  \cdot$ are equivalent.
\end{proof}
We denote the operators corresponding to $\qf h$ and $\qf d$ by $H$
and $\Delta$, respectively.\footnote{We work in the ``geometric''
  convention in which the Laplacian is a non-negative operator.}
\begin{remark}
  \label{rem:a.smooth}
  We can always assume that $a_e$ is a smooth function on each edge:
  We just have to replace a non-smooth magnetic potential $a_e$ by a
  smooth function $\wt a_e$ having the same values at the endpoints
  and the same integral over $e$. Using the gauge
  transformation~\eqref{eq:gauge.ed} we will see in \Sec{gauge} that
  the operators with magnetic potentials $a$ and $\wt a$ are unitarily
  equivalent.
\end{remark}
Nevertheless the domain $\HS^2:=\dom H$ of $H$ may depends on $a$ in
general, namely, a function $f$ is in $\HS^2$ iff (i)~$f_e \in \Sob[2]
e$ (due to our smoothness assumption of $a_e$), (ii)~$f$, $Hf \in
\Lsqr{X_0}$ and (iii)~the so-called \emph{generalised free boundary
  conditions} (sometimes also labelled as Kirchhoff -- see
\Footnote{kirchhoff})
\begin{subequations}
\label{eq:kirchhoff}
  \begin{gather}
    \label{eq:kirchhoff.cont}
    f_{e_1} (v) = f_{e_2}(v), \qquad e_1, e_2 \in E_v\\
    \label{eq:kirchhoff.der}
    \sum_{e \in E_v} D_e f(v) = 0
  \end{gather}
\end{subequations}
are fulfilled for all $v \in V$ where $D_e f(v):= \vec f_e'(v) - \im
\vec a_e(v) f_e(v)$ and
\begin{equation}
  \label{eq:der.orient}
  \vec f_e'(v) :=
  \begin{cases}
    f_e'(0), & \text{if $v=\bd_-e$,}\\
    -f_e'(\ell_e), & \text{if $v=\bd_+e$}
  \end{cases}
\end{equation}
defines the \emph{outward} derivative of $f_e$ at $v$, and similarly
for $\vec a_e(v)$.  The fact that we need different signs for incoming
and outgoing edges is due to the fact that $f'$ and $a$ formally are
$1$-forms on the quantum graph. Note that $1$-forms \emph{do} see the
orientation (in contrast to the second order operator $H$).
Condition~\eqref{eq:kirchhoff.cont} is the continuity at each vertex
and~\eqref{eq:kirchhoff.der} is the conservation of the current
generated by $D_e$.

If $\sum_{e \in E_v} \vec a_e(v)=0$ for all $v \in V$ then $\HS^2 =
\dom H$, i.e.  \eqref{eq:kirchhoff.der} becomes the usual free
boundary conditions, where $D_ef_e(v)$ is replaced by $\vec f_e'(v)$.

\subsection{Gauge transformations}
\label{sec:gauge}
Without loss of generality, we may assume that $a_e=0$ for
external edges $e \in E_\ext$: Using the simple gauge
transformation $\hat f_e = \Xi_e f_e$ on the external edge
(cf.~eq.~\eqref{eq:gauge.ed}) one easily sees that $\qf h_e(f_e) =
\qf d_e(\Xi_e f_e)$. In addition, $\Xi_e(0)=1$ so that $(\Xi_e
f_e)(v)=f(v)$, i.e., $\Xi_e f_e$ extends to a continuous function
also onto interior edges (where $\Xi_e=1$). In particular, $\hat
f=\Xi f \in \Sob{X_0}$ and the assertion $\qf h (f)= \qf {\hat
h}(\hat f)$ holds where $\qf {\hat h}$ is the quadratic form
without magnetic potential on the external edges, which in turn
implies that the corresponding operators are unitarily equivalent.

Similarly, we can always gauge away the magnetic potential on a
\emph{tree} graph (i.e., a graph without loops). On a general graph,
we can use a gauge transformation to eliminate the vector potential on
each edge; the price for that is a less convenient quadratic form
domain, now consisting of functions generally \emph{discontinuous} at
the vertices. Specifically, the values $\e^{\im \Phi_e(v)}f_e(v)$ have
to be equal for all $e \in E_v$.  Note that
$\Phi_e(\bd_-e)=\Phi_e(0)=0$, but $\Phi_e(\bd_+e)=\Phi_e(\ell_e)
\notin 2\pi \Z$ in general.  Furthermore, the condition for the
operator domain is the free condition with $D_ef(v)$ replaced by
$\e^{\im \Phi_e(v)} f_e'(v)$ (cf.~eq.~\eqref{eq:bd.loop}).

In addition, a magnetic Hamiltonian on a quantum graph is
completely determined (up to a unitary equivalence) by the values
of the magnetic flux $\Phi_L := \Phi_{e_1}(\ell_{e_1}) \cdot
\ldots \cdot \Phi_{e_n}(\ell_{e_n})$ (mod $2\pi$) through all its
primary loops $L=(e_1, \dots, e_n)$ by Stokes theorem. For a
general treatment of magnetic perturbations on quantum graphs we
refer to~\cite{kostrykin-schrader:03}.

%
\section{Quantum wave guide model}
\label{sec:qwg.model}
%

\subsection{Branched quantum wave guides}
\label{sec:qwg}
Let $X_\eps$ be a $d$-dimensional manifold. If $X_\eps$ has boundary,
we denote it by $\bd X_\eps$. We assume that $X_\eps$ and $\bd X_\eps$
are disjoint, i.e., $X_\eps$ is the interior of $\clo X_\eps = X_\eps
\cup \bd X_\eps$. In addition, we assume that $X_\eps$ can be
decomposed into open sets $U_\edeps$ and $U_\vxeps$, i.e,
\begin{equation*}
  X_\eps = \bigdisjcup_{e \in E} U_\edeps \disjcup
           \bigdisjcup_{v \in V} U_\vxeps.
\end{equation*}
\begin{notation}
  \label{not:dcup}
  Here and in the sequel, $A=\bigdisjcup_i A_i$ means that $A_i$ are open
  (in $A$), mutually disjoint and the interior of $\bigcup_i \clo A_i$
  equals $A$.
\end{notation}
We have introduced this notion to avoid mentioning boundaries of
dimension $d-1$ which are unimportant in an $\Lsymb_2$-decomposition.
Note that it suffices to consider a chart cover of $X_\eps$ \emph{up
  to a set of measure $0$} 
when dealing
with $\Lsymb_2$-theory.

Denote the metric on $X_\eps$ by $g_\eps$.  We assume that $U_\edeps$
and $U_\vxeps$ are isometric to $(U_e, g_\edeps)$ and $(U_v,
g_\vxeps)$, respectively, where the underlying manifolds are
independent of $\eps>0$. In addition, we assume that $U_e = e \times
F$ where $F$ is a compact $m$-dimensional manifold with $m:=(d-1)$.
The cross section manifold $F$ has boundary depending on whether
$X_\eps$ has a boundary or not.
\begin{notation}
  \label{not:decomp.mfd}
  Here and in the sequel, the subscripts $(\cdot)_\edeps$ and
  $(\cdot)_\vxeps$ (or sometimes only $(\cdot)_e$ and $(\cdot)_v$) denotes the
  restriction of objects living on $X_\eps$ to $U_\edeps$ and
  $U_\vxeps$, respectively. For example, $g_\edeps := g \restr {U_\edeps}$ or
  $u_v:= u \restr{U_v}$. We will switch between different charts
  (e.g., $U_\vxeps$ and $U_e=e \times F \cong (0,\ell_e) \times F$)
  without mentioning. If no confusion can occur, we also omit the
  subscripts.
\end{notation}
\begin{notation}
  \label{not:r.mfd}
  As a Riemannian manifold, $U_e$ carries the metric $g_\edeps$ with
  $\eps=1$. Similarly, $\hat U_\edeps=(U_e, \hat g_\edeps)$ and
  $U_v=(U_v, g_v)$.
\end{notation}
Motivated by our example in \Sec{loop} we assume that the metric
components satisfy
\newcommand{\qwgass}{H$_\eps$}
\newcommand{\eqrefqwg}{\textrm{(\qwgass){}}}
\begin{equation}
  \label{eq:def.met}
  \tag{\qwgass1}
  \begin{array}{r@{\hspace*{0ex}}l@{\hspace*{5ex}}r@{\hspace*{0ex}}l}
         g_\edeps &= (1+\Err(\eps))^2 \de x^2 + \eps^2 h,&
         g_\vxeps &\approx \eps^2 g_v\\
    \hat g_\edeps &=               \de x^2 + \eps^2 h, &
    \hat g_\vxeps &=       \eps^2 g_v
 \end{array}
\end{equation}
where $g_v$ and $h$ are fixed metrics on $U_v$ and $F$, respectively.
For simplicity, we suppose that $\vol_m F=1$. Clearly, we have
\begin{equation}
  \label{eq:met.prod.asym}
  \dd U_\edeps = (1 + \Err(\eps)) \dd \hat U_\edeps
\end{equation}
for the Riemannian densities w.r.t.\ $g_\edeps$ and $\hat g_\edeps$.
To keep the model simple, we also assume that an \emph{exterior} edge
neighbourhood $U_\edeps$ has exact product structure, i.e., that
$g_\edeps=\hat g_\edeps$ for $e \in E_\ext$.
\begin{notation}
  \label{not:approx}
  Here and in the following, $b_\eps = \Err(\eps^\alpha)$ means that $|b_\eps
  \eps^{-\alpha}|$ is bounded by some constant $c>0$ for
  $0<\eps<\eps_0$.  Similarly, $b_\eps \approx \hat b_\eps$ means that
  there exist constants $c_\pm>0$ such that $c_-b_\eps \le \hat b_\eps
  \le c_+ b_\eps$ for all sufficiently small $\eps>0$. The constants
  $c$ and $c_\pm$ are supposed to be independent of $\eps>0$, $z \in
  X_\eps$, $e \in E$ and $v \in V$; e.g., $g_\vxeps \approx \hat
  g_\vxeps$ means that $b_\eps = g_\vxeps(z)(w,w)$ and $\hat b_\eps =
  \hat g_\vxeps(z)(w,w)$ satisfy $b_\eps \approx \hat b_\eps$
  \emph{uniformly} in $\eps>0$, $v \in V$, $z \in U_v$ and $w \in
  T^*_zU_v$.
\end{notation}
Condition~\eqref{eq:def.met} means that on the edge neighbourhood,
the metric $g_\edeps$ differs from $\hat g_\edeps$ only by a small
longitudinal error. On the vertex neighbourhood, we are closed to
the $\eps$-homothetic metric $\hat g_\vxeps$. Note that the
embedded case of~\Sec{loop} is included in this setting. The
estimate $g_\vxeps \approx \eps^2 g_v$ allows us to consider also
\emph{non}-homothetic vertex neighbourhoods $U_\vxeps$ occurring
e.g.\ if the edges are curved up to the vertex,
cf.~\cite[Sec.~3.1]{post:06a}. We can indeed treat a slightly
more general model with off-diagonal terms in the metric (coming
e.g.\ from non-constant radii along the edge neighbourhood) and a
slightly slower scaling at the vertex neighbourhood.  We refer
to~\cite{exner-post:05, post:06a} and keep the simpler model
here, since it already covers the main example, the embedded
quantum graph.

The metric $\hat g_\eps$ on $X_\eps$, close to the original one,
is more adapted to the reduction onto the quantum graph. Note that
$(X_\eps, \hat g_\eps)$ consists of straight cylinders $(U_e, \hat
g_\edeps)$ of radius $\eps$ and \emph{fixed} length $\ell_e$
joined by $\eps$-homothetic vertex neighbourhoods $(U_v, \hat
g_\vxeps)$. The manifold $(X_\eps, \hat g_\eps)$ does \emph{not}
form an $\eps$-neighbourhood of an quantum graph embedded in some
ambient space, since the vertex neighbourhoods cannot be fixed in
the ambient space unless one allows slightly shortened edge
neighbourhoods as we described in the example in \Sec{loop}.
Nevertheless, introducing $\eps$-independent coordinates
simplifies the comparison of the Laplacian on the quantum graph
and the manifold.

In addition, we assume the following uniformity conditions:
\begin{align}
  \label{eq:vol.ew}
  \tag{\qwgass2}
  c_{\vol} := \sup_{v \in V} {\vol_d U_v} < \infty, \qquad \qquad
  \lambda_2 := \inf_{v \in V} \EWN 2 {U_v} > 0,
\end{align}
where $\EWN 2 {U_v}$ denotes the second (first non-zero) Neumann
eigenvalue of $(U_v,g_v)$. In addition, we assume that $X_\eps$ is of
bounded geometry, i.e., we have a global lower bound on the
injectivity radius and the Ricci curvature, namely
\begin{equation}
    \label{eq:curv.bd}
  \tag{\qwgass3}
  r_0(\eps):=\injrad X_\eps > 0, \qquad
  \kappa_0(\eps) :=
   \inf_{\substack{x \in X_\eps\\ v \in T_xX_\eps \setminus\{0\}}}
      \frac{g_\eps \bigl( \Ric(x)v,v \bigr)}{g_\eps(v,v)} > -\infty.
\end{equation}
Both constants will in general depend on $\eps$.  Roughly speaking,
Condition~\eqref{eq:vol.ew} means that $U_v$ remains small (cf.\ the
discussion in~\cite[Rem.~2.7]{post:06a}). The
assumption~\eqref{eq:vol.ew}--\eqref{eq:curv.bd} are trivially
satisfied once the vertex set $V$ is finite.
Assumption~\eqref{eq:curv.bd} still remains true for example if the
set of ``building blocks'', i.e., the sets of isometry classes of
$\{U_v\}_{v \in V}$ and $\{U_e\}_{e \in E}$ are finite. This
assumption is only needed in~\eqref{eq:ell.reg} in order to assure
elliptic regularity.

For further purposes, we need a finer decomposition of $U_v$ into
\begin{equation}
  \label{eq:dec.vx}
  U_v = \bigdisjcup_{e \in E_v} A_\vxed \disjcup U_v^-
\end{equation}
where $A_\vxed \cong (0,\ell_0/2) \times F$ with coordinates $(\check
x, y)$. Note that we have $x\approx \eps \check x$ (if we extend the
coordinate $x$ to $A_\vxed$ and $\check x$ to $U_e$), and therefore
$\de x = \eps \de \check x$. In particular, $g_\vxedeps \approx
\eps^2(\de \check x^2 + h)$ where $g_\vxedeps$ is the restriction of
$g_\eps$ to $A_\vxeps$. Note that this decomposition always exists. If
necessary, we have to remove a small part (of length $\Err(\eps)$) of the
adjacent edge neighbourhood and rescale the coordinates on the
shortened edge neighbourhood in order to obtain again
$\eps$-independent coordinates on the edge neighbourhood.

\begin{notation}
  \label{not:orient}
  We denote $\bd_e U_v$ the boundary part of $U_v$ meeting $\clo U_e$
  and similarly, $\bd_v U_e$ the boundary part meeting $\clo U_v$ (if
  $v \in \bd e$). Similarly, $\bd_e U_v^-$ denotes the common part of
  $\clo U{}_v^-$ and $\clo A_\vxed$.
\end{notation}
\subsection{Magnetic Hamiltonian on the quantum wave guide}
\label{sec:mag.qwg}
We now determine the assumptions on the magnetic and electric
potentials. Here, the magnetic potential is a $1$-form on $X_\eps$ and
$q_\eps$ is a function on $X_\eps$ such that
\begin{equation}
  \label{eq:mag.pot}
  \alpha_\eps \in \Linfty {T^*X_\eps} \qquad \text{and} \qquad
  q_\eps \in \Linfty {X_\eps},
\end{equation}
i.e., $|\alpha_\eps|_{g_\eps}$ and $|q_\eps|$ are essentially bounded
functions on $X_\eps$. As on the quantum graph, we assume for
simplicity that $q_\eps \ge 0$ and that $\alpha_\eps$, $q_\eps$ vanish
on the exterior edge neighbourhoods.  To avoid any difficulties with
the operator domain and elliptic regularity in~\eqref{eq:ell.reg} we
assume that $\alpha_\eps$ is smooth.

In order to compare the magnetic and electric potential with the one
on the quantum graph, we introduce another magnetic and electric
potential $\hat \alpha_\eps$ and $\hat q_\eps$, respectively. The fact
that $\hat \alpha_\eps$ is no longer smooth does not matter since we
use $\hat \alpha_\eps$ only as intermediate step in the verification
of the closeness assumptions in \Sec{closeness.mod}.

Again, motivated by the loop example
in \Sec{loop} we assume that
\begin{equation}
  \label{eq:def.mag}
  \tag{\qwgass4}
  \begin{array}{r@{\hspace*{0ex}}l@{\hspace*{5ex}}r@{\hspace*{0ex}}l}
         \alpha_\edeps &= (a_e+\Err(\eps)) \de x + \eps \omega_\edeps,&
         \alpha_\vxeps &\approx \eps \alpha_v\\
    \hat \alpha_\edeps &=  a_e          \de x, &
    \hat \alpha_\vxeps &= 0
 \end{array}
\end{equation}
where $a_e$ is the magnetic potential on the quantum graph and
where
\begin{equation}
  \label{eq:def.omega}
  \tag{\qwgass5}
  \omega_\edeps = \de_F \vartheta_\edeps = \Err(1), \qquad
  \vartheta_\edeps = \Err(1), \qquad
  \partial_x \vartheta_\edeps = \Err(1).
\end{equation}
In particular, $\omega_\edeps$ is an exact $1$-form on $F$ and
$\alpha_v$ is a fixed $1$-form on $T^*U_v$.

For the electric potential, we assume that
 \begin{equation}
  \label{eq:def.pot}
  \tag{\qwgass6}
  \begin{array}{r@{\hspace*{0ex}}l@{\hspace*{5ex}}r@{\hspace*{0ex}}l}
         q_\edeps &= q_e + \Err(\eps),&
         q_\vxeps &= \Err(1)\\
    \hat q_\edeps &= q_e, &
    \hat q_\vxeps &= 0
 \end{array}
\end{equation}
where $q_e$ is the electric potential on the quantum graph. From
these assumptions, it is clear, that global bounds on the quantum
graph potentials $a$ and $q$ are enough to ensure that
$\alpha_\eps$ and $q_\eps$ are bounded.

We define the magnetic Hamiltonian $H_\eps$ acting in the the
Hilbert space $\HS_\eps := \Lsqr {X_\eps, g_\eps}$ (with the norm
$\norm \cdot$ and inner product $\iprod \cdot \cdot$) via the
quadratic form
\begin{equation}
  \label{eq:h.eps}
  \qf h_\eps (u) :=
  \normsqr {D_\edeps u} + \iprod u {q_\eps u},
\end{equation}
where $D_\edeps := (\de - \im \alpha_\eps)$.  Denote by $\qf d_\eps$
the quadratic form given by $\qf h_\eps$ without field, i.e,
$\alpha_\eps=0$ and $q_\eps=0$. The proof of the following lemma
is straightforward (cf.\ \Lem{h.free}):
\begin{lemma}
  \label{lem:h.eps.free}
  Assume that $\alpha_\eps \in \Linfty {T^*X_\eps}$ and $q_\eps \in
  \Linfty {X_\eps}$, i.e., $|\alpha_\eps|_{g_\eps}$ and $q_\eps$ are
  essentially bounded functions on $X_\eps$. Then $\qf h_\eps$ and
  $\qf d_\eps$ are closed forms on
  \begin{equation}
    \label{eq:dom.h.free.eps}
    \HS_\eps^1 := \Sob {X_\eps} := \bigset {u \in \Lsqr {X_\eps}}
                          {|\de u|_{g_\eps} \in \Lsqr {X_\eps}}
  \end{equation}
  where the derivative is understood in the weak sense.  Furthermore,
  the norms $\norm[1] \cdot := \norm[\qf d_\eps] \cdot$ and $\norm[\qf
  h_\eps] \cdot$ satisfy $\norm[\qf d_\eps] u \approx \norm[\qf h_\eps] u$
  (independently of $\eps$). In particular, the norms are equivalent.
\end{lemma}
We denote by $H_\eps$ and $\Delta_\eps$ the corresponding
operators associated to $\qf h_\eps$ and $\qf d_\eps$. Note that
$\Delta_\eps = \laplacian{X_\eps} \ge 0$ is the usual (Neumann)
Laplacian on $X_\eps$. Since we assumed that $\alpha_\eps$ is smooth
also the operator domains of $H_\eps$ and $\Delta_\eps$ agree, namely
they equal
\begin{equation}
  \label{eq:dom.op.eps}
  \HS^2_\eps := \Sob[2]{X_\eps} :=
  \bigset{u \in \Lsqr {X_\eps}}
     {|\de u|_{g_\eps}, \Delta_\eps u \in \Lsqr{X_\eps}, \,
      \normder u = 0 \text{ on $\bd X_\eps$}}.
\end{equation}
Note that we include the Neumann boundary condition in the definition
of the second order Sobolev space if $\bd X_\eps \ne \emptyset$.

\subsection{Intermediate product model}
\label{sec:prod.mod}
When comparing the magnetic Laplacian $H_\eps$ on the branched
quantum wave guide with the magnetic Laplacian $H$ on the graph,
it will be convenient to use also the magnetic Laplacian $\hat
H_\eps$ defined via the hat-quantities:
\begin{notation}
  Here and in the sequel, the label $\hat \cdot$ refers to the
  product metric $\hat g_\eps$ and the simplified potentials $\hat
  \alpha_\eps$ and $\hat q_\eps$ defined as above. Similarly, a
  Hilbert space defined via $\hat g_\eps$ will carry the label $\hat
  \cdot$, e.g., $\hat \HS_\eps:=\Lsqr {X_\eps, \hat g_\eps}$ with norm
  and inner product $\normhat \cdot$, $\iprodhat \cdot \cdot$, resp.
  The quadratic form $\hat{\qf h}_\eps$ is defined as
  in~\eqref{eq:h.eps} but with $\hat g_\eps$, $\hat \alpha_\eps$ and
  $\hat q_\eps$, instead.
\end{notation}

The main reason why we introduced the intermediate model operator
$\hat H_\eps$ on $\hat \HS_\eps$ is to split the reduction onto
the quantum graph into two steps: In the first step, we discard
the error terms coming from the failure of the metric to be an
exact product as well as from the transverse magnetic and electric
potential terms. Once having established some closeness estimates
on $H_\eps$ and $\hat H_\eps$ in \Lem{asymp.prod}, we will show in
\Sec{closeness} that $H_\eps$ approaches the quantum graph
Hamiltonian $H$ using the intermediate operator $\hat H_\eps$;
this will simplify the estimates used there.

Using our assumptions on the metric and the fields, we have (taking
\Not{decomp.mfd} into account):
\begin{gather}
  \begin{array}{r@{\hspace*{0ex}}l@{\hspace*{5ex}}r@{\hspace*{0ex}}l}
    \normsqr[\edeps] u &= \int_{U_\edeps} |u|^2 \dd U_\edeps,&
    \normsqr[\vxeps] u &= \int_{U_\vxeps} |u|^2 \dd U_\vxeps,\\
    \normsqrhat[\edeps] u &= \eps^m \int_{U_e} |u|^2 \dd F \dd x,&
    \normsqrhat[\vxeps] u &= \eps^d \int_{U_v} |u|^2 \dd U_v,\\
  \end{array}\\
  \begin{split}
      \qf h_\edeps(u) &=
    \begin{array}[t]{r@{\hspace{0pt}}l}
         \int_{U_\edeps} \Bigl[
            g_\edeps^{xx} \bigl|(D_e &+ \Err(\eps)) u|^2 \\
      &+ \dfrac 1 {\eps^2}
          \bigl|(\de_F - \im \eps \omega_\edeps) u\bigr|^2_h +
          q_\edeps |u|^2 \Bigr]  \dd U_\edeps,
    \end{array}\\
    \hat{\qf h}_\edeps(u) &=
       \eps^m \int_{U_e} \Bigl[ |D_e u|^2 +
       \frac 1 {\eps^2} |\de_F u|^2_h
           + q_e |u|^2 \Bigr] \dd F \dd x,\\
  \end{split}\\
  \begin{split}
    \qf h_\vxeps(u) &=
       \int_{U_\vxeps} \bigl[ |(\de-\im \alpha_\vxeps) u|_{g_\vxeps}^2
           + q_\vxeps |u|^2 \bigr] \dd U_\vxeps,\\
    \hat{\qf h}_\vxeps(u) &=
       \eps^{d-2} \int_{U_v} |\de u |_{g_v}^2 \dd U_v
  \end{split}
\end{gather}
where $D_e:=\partial_x - \im a_e$ and $g_\edeps^{xx}:=g_\edeps(\de
x,\de x) = 1+\Err(\eps)$ due to~\eqref{eq:def.met}.  To discard the
transversal magnetic potential $\omega_\edeps$, we need to
introduce an approximate gauge function, namely
\begin{align}
  \label{eq:def.theta}
    \Theta_\edeps(x,y) &:= \e^{\im \vartheta_\edeps(x,y)}\\
    \nonumber
    \Theta_\vxeps(z) &:=
    \begin{cases}
      \chi_\vxed(\check x) + (1-\chi_\vxed(\check x)) \Theta_\edeps(v,y), &
              z=(\check x,y) \in A_\vxed\\
      1, & z \in U^-_v
    \end{cases}
\end{align}
where $\chi_\vxed$ equals $0$ on $\bd_e U_v$ and $1$ on $\bd_e U_v^-$.
The function $\vartheta_\edeps$ was introduced
in~\eqref{eq:def.omega}; we also recall~\eqref{eq:dec.vx} for a
definition of $A_\vxed$, $U^-_v$, $\check x$, and \Not{orient} for the
definition of the boundary $\bd_e U_v$ etc.  In particular, we can
choose $\chi_\vxed$ in such a way that $|\chi'_\vxed| \le 4/\ell_0$
(since the length of $A_\vxed$ is $\ell_0/2$).  Note that the
``gauge'' function $\Theta_\eps$ is unitary only on $U_\edeps$ since
$|\Theta_\edeps|=1$, while on the vertex neighbourhood we have just
$|\Theta_\vxeps|\le 1$.  Note, in addition, that the components
$\Theta_\edeps$ give together a global \emph{Lipschitz-continuous}
function $\Theta_\eps$. We will need this fact in \Sec{closeness}.  A
simple estimate shows that
\begin{equation}
  \label{eq:theta.est}
  \begin{array}{r@{\hspace*{0ex}}l@{\hspace*{5ex}}r@{\hspace*{0ex}}l}
    \norm[\infty] {\Theta_\edeps - 1} &= \Err(\eps), &
    &\de \Theta_\edeps =
       \im \eps(\partial_x \vartheta_\edeps + \omega_\edeps)
              \Theta_\edeps,\\
    \norm[\infty] {\Theta_\vxeps - 1} &= \Err(\eps), &
    &|\de \Theta_\vxeps|_{g_\vxeps} = \Err(1)
  \end{array}
\end{equation}
where e.g.\ $\Err(\eps)=\eps \norm[\infty]{\vartheta_\edeps}$
and~$\Err(1)=4\norm[\infty]{\vartheta_\edeps}/\ell_0 +
\norm[\infty]{|\omega_\edeps|_h}$~(cf.~\eqref{eq:def.omega}).  Now
we are going to provide some estimates which will be used when
comparing the Hamiltonian on the quantum wave guide with the one
on the quantum graph:
\begin{lemma}
  \label{lem:asymp.prod}
  We have
  \begin{gather}
    \label{eq:norm.prod.ed}
      \bigl| \iprodhat[\edeps] u {\hat u} -
               \iprod[\edeps] u {\hat u} \bigr| =
          \Err(\eps) \norm[\edeps] u \normhat[\edeps] {\hat u}\\
    \label{eq:qf.prod.ed}
    \bigl| \hat {\qf h}_\edeps(u, \hat u) -
               \qf h_\edeps(u, \Theta_\edeps \hat u) \bigr| =
          \Err(\eps) \norm[\qf d_\edeps] u
                          \normhat[\hat{\qf d}_\edeps] {\hat u}
            \quad(\text{if }\de_F \hat u=0)\\
    \label{eq:qf.prod.vx}
    \bigl| \qf h_\vxeps(u, \Theta_\vxeps \hat u) \bigr| =
          \Err(1) \norm[\qf d_\vxeps] u
                          \normhat[\hat{\qf d}_\vxeps] {\hat u}
  \end{gather}
  for all functions $u$, $\hat u$ in the appropriate spaces. Here,
  $\Err(\eps)$ and $\Err(1)$ depend only on the error terms $\Err(\eps)$ and
  $\Err(1)$ in~\eqrefgraph\ and~\eqrefqwg.
\end{lemma}
\begin{proof}
  The inner product estimate follows immediately
  from~\eqref{eq:met.prod.asym}.  For the second assertion note that
  \begin{equation*}
    D_\edeps (\Theta_\edeps \hat u) =
      \bigl( \partial_x \hat u - \im (a_e +
       \Err(\eps) - \eps \partial_x \vartheta_\edeps) \hat u\bigr)
        \Theta_\edeps\de x
  \end{equation*}
  where the $y$-component vanishes due to the fact that $\de_F \hat
  u=0$ and that $\de_F \Theta_\edeps=\im \eps \Theta_\edeps
  \omega_\edeps$ cancels the transversal magnetic potential.
  Furthermore, the difference of the $\de x$-components is
  \begin{multline*}
    \conj{D_e u} \, D_e \hat u -
    (1+\Err(\eps)) \conj {(D_e u + \im \Err(\eps)) u} \, (D_e
    (\Theta_\edeps \hat u) + \im \Err(\eps) \Theta_\edeps \hat u) \\=
     \Err(\eps)(\partial_x \conj u + \conj u)
             (\partial_x \hat u + \hat u)
   \end{multline*}
   where $1+\Err(\eps)$ is the error factor in the metric $g_\edeps$ and
   $\Err(\eps)$ in the last line depends only on the errors given in
   assumptions~\eqrefgraph\ and~\eqrefqwg. In addition, the
   $y$-component does not occur.
   The last estimate follows in a similar way using $|\de
   \Theta_\vxeps|_{g_\vxeps} = \Err(1)$ (cf.~\eqref{eq:theta.est}).
\end{proof}
The requirement $\de_F \hat u=0$ in the second estimate is due to the
fact that we used $u$ instead of $\conj \Theta_\edeps u$ in $\hat
h_\edeps$. This is exactly the situation we will need in
\Sec{closeness}.  We will also see that our rough estimate $\Err(1)$
in~\eqref{eq:qf.prod.vx} is already sufficient to ensure that $H_\eps$
approaches the quantum graph Hamiltonian $H$.

%
\section{Complex dilation}
\label{sec:dilation}
%
Next we are going to explain the complex dilation argument. We use
an exterior scaling on the external edges only.

\subsection{Space decomposition}
\label{sec:decomp}

We start with the space decomposition into an interior and exterior
part. Recall that we assumed that each edge neighbourhood of an
external edge $e \in E_\ext$ has exact product structure (i.e.,
$g_\edeps=\hat g_\edeps$) and no field (i.e., $\alpha_\edeps=0$ and
$q_\edeps=0$).
\begin{notation}
  \label{not:decomp}
  Here and in the sequel, the subscript $(\cdot)_\inl$ stands for the
  \emph{internal} component and $(\cdot)_\ext$ for the external
  component of an element in the Hilbert space, respectively,\ for the
  restriction to the subspace $\HS_\inl$ of a quadratic form or an
  operator. We often omit the label $(\cdot)_\inl$ or $(\cdot)_\ext$
  on a function, if it is clear (e.g., we write $\qf h_\inl(f)$
  instead of $\qf h_\inl(f_\inl)$ etc.).
\end{notation}

To avoid difficulties with a cut into an internal and external part at
a vertex, we can introduce artificial vertices of degree $2$ on the
external edges. Note that such vertices do not change the domain of
the graph Hamiltonian since a vertex of degree $2$ with free boundary
conditions means nothing else then continuity of a function and its
derivative at the vertex (cf.~\eqref{eq:kirchhoff}).  Remember that
there is no potential on the external edges.

Without loss of generality we can therefore assume that each boundary
vertex $\bd_- e$ of an external edge $e \in E_\ext$ has degree $2$ and
distance $\ell_0$ from any other vertex in $V$. If this were not the
case for an external edge $e$, just introduce a new boundary vertex at
distance $\ell_0$ from $\bd_-e$ on $e$.

We can also assume that the manifold $X_\eps$ has product structure
near the boundary vertices since we assumed that the edge
neighbourhood $U_\edeps$ has \emph{exact} metric product structure for
external edges $e$. This means in particular, that we do not associate
a vertex neighbourhood to a \emph{boundary} vertex.

We remind the user that we used a different decomposition in
\Sec{complex}. For computational reasons, it is easier to keep the
number of vertices minimal on a quantum graph, but for our purposes,
it is easier to be away from the inner vertices. From an abstract
point of view, of course, both models lead to the same definition of
resonances, cf.~\Lem{dil.ind}.

We denote by $X_{0,\inl}:=(V, E_\inl, \ell)$ the \emph{internal} and
by $X_{0,\ext} := (\Gamma_0, E_\ext,\ell)$ the \emph{external} metric
graph. Note that $X_{0,\ext}$ corresponds to the disjoint union of
$|\Gamma_0|=|E_\ext|$ many half-lines. The boundary vertices
$\Gamma_0$ form the common boundary of $X_{0,\inl}$ and $X_{0,\ext}$.

Similarly, we decompose the manifold $X_\eps$ into
\begin{equation*}
    X_{\eps,\inl} :=
        \bigdisjcup_{e \in E_\inl} U_\edeps \disjcup
            \bigdisjcup_{v \in V} U_\vxeps
        \quad \text{and} \quad
    X_{\eps,\ext} :=
        \bigdisjcup_{e \in E_\ext} U_\edeps
\end{equation*}
(remind \Not{dcup}) and denote the common boundary of $X_{\eps,\inl}$
and $X_{\eps,\ext}$ by $\Gamma_\eps$.  Again, $X_{\eps,\ext}$ consists
of $|E_\ext|$ many disjoint half-infinite cylinders $(0,\infty) \times
F_\eps$.

\begin{notation}
  \label{not:der.ext.ed}
  For a boundary vertex $v=\bd_-e \in \Gamma_0$ with external edge $e
  \in E_\ext$ we set
  \begin{align*}
     f_\inl (v)       &:=  f_e(-0),&
     u_\inl (v,\cdot) &:= u_e(-0,\cdot)\\
     f_\ext (v)       &:= f_e(+0), &
     u_\ext (v,\cdot) &:= u_e(+0,\cdot)\\
     f'_\inl (v)       &:=  f'_e(-0),&
     u'_\inl (v,\cdot) &:= \partial_x u_e(-0,\cdot)\\
     f'_\ext (v)       &:= f'_e(+0), &
     u'_\ext (v,\cdot) &:= \partial_x u_e(+0,\cdot)
  \end{align*}
  where we identify a neighbourhood of $v$ with a neighbourhood of $0
  \in \R$ (positive numbers corresponding to the external part) and
  where $g(\pm 0)$ denotes the left/right limit.  Note that the sign
  convention for $f'_\inl(v)$ differs from the one for internal
  vertices in~\eqref{eq:der.orient}.
\end{notation}
We split the Hilbert space $\HS$ and \ $\HS_\eps$ into two
components, namely we take
\begin{equation}
  \label{eq:decomp.hs}
  \HS = \HS_\inl \oplus \HS_\ext
\end{equation}
and the analogous decomposition for $\HS_\eps$ where
\begin{equation}
  \label{eq:decomp.0.eps}
  \begin{array}{r@{\hspace{0pt}}l@{\qquad}r@{\hspace{0pt}}l}
    \HS_\inl &= \Lsqr {X_{0,\inl}},& \HS_\ext &= \Lsqr {X_{0,\ext}}\\
    \HS_{\eps,\inl} &= \Lsqr {X_{\eps,\inl}},& \HS_{\eps,\ext} &=
    \Lsqr {X_{\eps,\ext}}
  \end{array}
\end{equation}
on the quantum graph and the branched quantum wave guide,
respectively.

\subsection{Dilated operators}
\label{sec:dil.op}
Now we introduce the exterior dilation operator. For $\theta \in
\R$ we define by
\begin{equation*}
 \Phi_e^\theta(x) :=
   \e^\theta x, \qquad x > 0
\end{equation*}
a non-smooth flow on an external edge $e \in E_\ext$.  Clearly,
$\Phi_e^\theta$ extends (by identity) to a (non-smooth) flow on the
graph $X_0$. Similarly,
$\Phi_\edeps^\theta(x,y):=(\Phi_e^\theta(x),y)$ defines a non-smooth
flow on the external edge neighbourhood, again extended to a flow
$\Phi_\eps^\theta$ on $X_\eps$. For a \emph{smooth} version of
exterior dilation we refer to~\cite{hislop-sigal:89,hislop-sigal:96}.
\begin{remark}
  \label{rem:smooth.dil}
  The smooth dilation argument seems to be less technical, at least,
  one does not have to deal with $\theta$-dependent domains (see the
  appendix). The price to pay is a more complicated expression of the
  dilated operator between the interior and exterior part. Since most
  of the technical details are hidden in the abstract criterion, the
  verification of the convergence assumptions for the non-smooth
  dilation is simpler. Moreover, on a graph it is in a sense natural
  to have a ``constant'' scaling at each edge. In addition, both
  dilation arguments leads to the same definition of resonances
  (cf.~\Lem{dil.ind}).
\end{remark}
On an edge $e \in E$ we have then the following group action
\begin{equation}
  \label{eq:flow}
  U^\theta f := (\det D \Phi^\theta)^{1/2} (f \circ \Phi^\theta)
\end{equation}
where
\begin{equation*}
  (\det D \Phi^\theta)^{1/2} =
  \begin{cases}
    1             & \text{on $X_{0,\inl}$,}\\
    \e^{\theta/2} & \text{on $X_{0,\ext}$}
  \end{cases}
\end{equation*}
and similarly for $U_\eps^\theta$.  Clearly, $U^\theta$ and
$U_\eps^\theta$ are $1$-parameter unitary groups with respect to
$\theta \in \R$, acting non-trivially on the \emph{external} part
only.

\begin{notation}
  \label{not:op.dil}
  For a quadratic form $\qf h$ and an operator $H$ in $\HS$ we set
  \begin{equation*}
    \qf h^\theta(f):= \qf h(U^{-\theta} f) \qquad \text{and} \qquad
    H^\theta := U^\theta H U^{-\theta}
  \end{equation*}
  with domains $\dom \qf h^\theta := U^\theta(\dom \qf h)$ and $\dom
  H^\theta := U^\theta(\dom H)$ for \emph{real} $\theta$.
\end{notation}
Clearly, $\qf h^0=\qf h$ and $H^0 = H$.  A simple calculation shows
that for an external edge $e \in E_\ext$ we have
\begin{subequations}
\label{eq:h.dil}
  \begin{equation}
    \label{eq:h.dil.qg}
    \qf h_e^\theta (f) = \e^{-2\theta} \qf h_e(f),\qquad
    (H^\theta f)_e     = -\e^{-2\theta} f_e''\\
  \end{equation}
  on the quantum graph and
  \begin{multline}
      \qf h_\edeps^\theta (u) = \e^{-2\theta}
      \normsqr[\edeps]{\partial_x u} + \frac 1
      {\eps^2}\bignormsqr[\edeps]{|\de_F u|_h},\\
      (H_\eps^\theta u)_e = -\e^{-2\theta}
      \partial_{xx} u_e + \frac 1 {\eps^2} \Delta_F u_e
  \end{multline}
\end{subequations} %
on the manifold. Of course, the action on \emph{internal} edges
remains unchanged. On the quantum graph, the domains are given for a
real $\theta$ by
\begin{subequations}
  \label{eq:def.dil.dom}
  \begin{equation}
    \label{eq:def.dil.qf}
    \HS^{1,\theta} := \dom \qf h^\theta = \bigset{f \in
      \Sob{X_{0,\inl}} \oplus \Sob{X_{0,\ext}}}
    {f_\ext = \e^{\theta/2} f_\inl \text{ on $\Gamma_0$}}
  \end{equation}
  and
  \begin{multline}
    \label{eq:def.dil.op}
    \HS^{2,\theta} := \dom H^\theta = \Bigset{f \in
      \Sob[2]{X_{0,\inl}} \oplus \Sob[2]{X_{0,\ext}}}
    { \\ 
          f_\ext  = \e^{\theta/2} f_\inl, \,
          f'_\ext  = \e^{3\theta/2} f'_\inl
     \text{ on $\Gamma_0$.}}
  \end{multline}
\end{subequations}
Here,
\begin{subequations}
\label{eq:sob.qg}
  \begin{equation}
    \label{eq:sob1.qg.inl}
    \Sob{X_{0,\inl}} := \Cont {X_{0,\inl}} \cap
    \bigoplus_{e \in E_\inl} \Sob e
  \end{equation}
  and
  \begin{equation}
    \label{eq:sob2.qg.inl}
    \Sob[2]{X_{0,\inl}} :=
    \bigset{f \in \Cont {X_{0,\inl}} \cap \bigoplus_{e \in E_\inl} \Sob[2] e}
    {\sum_{v \in E_v} D_e f = 0, \; v \in V}
  \end{equation}
on the internal part and
\begin{equation}
  \label{eq:sob2.qg.ext}
     \Sob{X_{0,\ext}} := \bigoplus_{e \in E_\ext} \Sob e \und
   \Sob[2]{X_{0,\ext}} :=
  \bigoplus_{e \in E_\ext} \Sob[2] e
\end{equation}
\end{subequations} %
on the external part. Note that due to Assumption~\eqref{eq:ext.fin}
we have $\Sob[k]{X_{0,\ext}} \cong \Sob[k]{0,\infty}^{|E_\ext|}$.

On the manifold, we have a very similar definition for
$\HS^{1,\theta}_\eps$ and $\HS^{2,\theta}_\eps$, with first order
Sobolev spaces $\Sob {X_{\eps,\bullet}}$ defined as
in~\eqref{eq:dom.h.free.eps} and second order Sobolev spaces and
second order Sobolev spaces
\begin{equation}
  \label{eq:sob2.mfd}
  \Sob[2]{X_{\eps,\bullet}} :=
  \bigset{ u\restr{X_{\eps,\bullet}}} { u \in \Sob[2]{X_\eps}}
\end{equation}
where $\Sob[2]{X_\eps}$ already includes the Neumann boundary
conditions on $\bd X_\eps$ (if non-empty), i.e., we impose these
boundary conditions only on $\bd X_\eps \cap \clo X_{\eps,\bullet}$,
\emph{not} on $\Gamma_\eps$.

Roughly speaking, the domain of the quadratic form consists of
functions having a jump of magnitude $\e^{\theta/2}$ from the internal
to the external part. The operator domain in addition requires that
the derivative along the common boundary of the internal and external
part has a jump of magnitude $\e^{3\theta/2}$. In particular, even the
quadratic form domain depends on $\theta$.

The expression of $H^\theta$ now serves as a generalization for
$\theta$ in the strip $S_\vartheta = \set{\theta \in \C}{ |\Im \theta|
  < \vartheta/2}$ where $0 \le \vartheta < \pi$. We call $H^\theta$
the \emph{complex dilated} Hamiltonian, and similarly for
$H_\eps^\theta$. We will show in \App{res.est} that
$\{H^\theta\}_\theta$ is a self-adjoint family with spectrum contained
in the common sector $\Sigma_\vartheta$.  In addition, we show that
$R^\theta(z):=(H^\theta-z)^{-1}$ is an analytic family in $\theta$
(for $z$ not in the $\vartheta$-sector $\Sigma_\vartheta=\set{z \in
  \C}{ |\arg z| \le \vartheta}$, cf.\
\Lems{res.complex}{dom.h.complex}). This is a highly non-trivial fact
since $H^\theta$ is neither of type~A nor of type~B, i.e., both
sesquilinear form and operator domain depend on $\theta$ even for real
$\theta$. In other words, the non-smooth exterior scaling as defined
here is a very singular perturbation of the operator $H=H^0$.  The
same statements hold for the complex dilated Hamiltonian
$H_\eps^\theta$ on $\HS_\eps$.

The sesquilinear form $\qf h^\theta$ associated with the operator
$H^\theta$ is defined via
\begin{equation}
  \label{eq:dil.sf.0}
  \qf h^\theta(f,g) := \iprod f {H^\theta g} =
     \qf h_\inl(f_\inl, g_\inl) +
         \e^{-2\theta}\qf h_\ext(f_\ext, g_\ext)
\end{equation}
for $f \in \HS^{1,\conj \theta}$ and $g \in \HS^{2,\theta}$ with
domains as in~\eqref{eq:def.dil.dom}, where
\begin{equation*}
  \qf h_\inl :=
        \bigoplus_{e \in E_\inl} \qf h_e, \qquad \text{and} \qquad
  \qf h_\ext :=
        \bigoplus_{e \in E_\ext} \qf h_e.
\end{equation*}

Similarly, the sesquilinear form $\qf h_\eps^\theta$ associated to
$H_\eps^\theta$ is
\begin{equation}
  \label{eq:dil.sf.eps}
  \qf h^\theta_\eps(u,w) :=
     \qf h_{\eps,\inl}(u_\inl, w_\inl) +
         \e^{-2\theta} \qf h_{\eps,\ext}(u_\ext, w_\ext)
\end{equation}
for $u \in \HS^{1,\conj \theta}$ and $w \in \HS^{2,\theta}$.  We show
in \Lem{sf.bdd} how these sesquilinear forms can be extended to
bounded sesquilinear forms on $\HS^{1,\conj \theta} \times
\HS^{1,\theta}$ and this is actually all we need in order to show the
convergence in the appendices.

\begin{remark}
  \label{rem:dil.qf}
  We naturally have to introduce the sesquilinear forms on
  \emph{mixed} pairs $\HS^{1,\conj \theta} \times \HS^{1,\theta}$ in
  order to formally preserve the analyticity in $\theta$. This is
  exactly the setting we need in order to apply our abstract
  convergence result provided in \App{abstr.crit}.

  The difficulty here is to find a good norm on the natural quadratic
  form domain $\HS^{1,\theta}$. The corresponding expression defined
  via $\qf q^\theta(f):=\iprod f {H^\theta f}$ contains boundary terms
  of the form $\conj f(v) f'(v)$ (on the quantum graph) which are not
  obviously defined on $\HS^{1,\theta}$.

  In addition, it seems to be very difficult to estimate errors in
  terms of the corresponding norm $\norm[\qf q^\theta] \cdot$. There
  has been some confusion on the \emph{quadratic} form domain on
  $\HS^{1,\theta}$ due to the anti-linearity of a sesquilinear form in
  its first argument (cf.\ the Mathematical Reviews entry
  for~\cite{graffi-yajima:83}).

  To avoid these difficulty, we use a simpler norm on $\HS^{1,\theta}$
  related with the unperturbed form $\qf h$ by a simple multiplication
  operator. In this case, we have to assure that the corresponding
  spaces behave like a ``natural'' scale of Hilbert spaces associated
  to $H^\theta$ (cf.~\App{scale}).
\end{remark}

\subsection{Essential and discrete spectrum}

We collect some facts about the family of dilated operators
$\{H^\theta\}_\theta$. Note that we cannot directly apply the
perturbation theory of such operators developped
in~\cite[XIII.10]{reed-simon-1-4} since the form domain of $H_\theta$
(cf.~\eqref{eq:def.dil.qf}) contains \emph{discontinuous} functions
and is therefore not included in the form domain $\HS^1=\Sob{X_\eps}$
of the free operator, even not for real $\theta \ne 0$. In particular,
we cannot directly use the $\HS^{\pm1}$-scale of Hilbert spaces
associated to the free operator.  Nevertheless, most of the
conclusions of~\cite[XIII.10]{reed-simon-1-4} remain true since
$(H^\theta-z)^{-1}$ depends analytically on $\theta \in S_\vartheta$
for $z \notin \Sigma_\vartheta$ as we will see in \App{res.est}.

We first determine the essential spectrum of $H^\theta$ and
$H_\eps^\theta$. Note that the essential spectrum is determined by the
behaviour of $X_\eps$ at infinity. Namely, it does not matter if we
change the operator on a compact set due the invariance of the
essential spectrum under compact perturbations (decomposition
principle).  Recall that we assumed in~\eqref{eq:ext.fin} that we only
have finitely many external edges.
\begin{proposition}
  \label{prop:ess.sp.0}
  The essential spectrum is given by
  \begin{equation*}
    \essspec {H^\theta} \setminus (0, \infty) =
    \e^{2 \theta} [0, \infty).
  \end{equation*}
  If, in addition, $E_\inl$ is also finite (i.e., the internal graph
  $X_{0,\inl}$ is compact), then
  \begin{equation*}
    \essspec {H^\theta} = \e^{-2 \theta} [0, \infty).
  \end{equation*}
\end{proposition}

Similarly, we can prove on the manifold:
\begin{proposition}
\label{prop:ess.sp.eps}
  The essential spectrum is given by
  \begin{equation*}
    \essspec {H_\eps^\theta} \setminus (0, \infty)  =
    \frac 1 {\eps^2} \bigcup_{k \in \N} \EWN k F + \e^{-2 \theta} [0,\infty).
  \end{equation*}
  If, in addition, $E_\inl$ is also finite, then
  \begin{equation*}
    \essspec {H_\eps^\theta} =
    \frac 1 {\eps^2} \bigcup_{k \in \N} \EWN k F + \e^{-2 \theta} [0,\infty).
  \end{equation*}
  In particular, since $\EWN 1 F = 0$, for any bounded set $B \subset
  \C \setminus (0,\infty)$,
  \begin{equation*}
    \essspec {H_\eps^\theta} \cap B =
    \e^{-2\theta}[0,\infty) \cap B
  \end{equation*}
  provided $\eps$ is small enough.
\end{proposition}

Next, we make some general observations on the spectrum of
$H_\eps^\theta$ ($\eps \ge 0$) which are true for both models, the
quantum graph and manifold model:
\begin{proposition}
  \label{prp:disc.spec}
  Assume that $E_\inl$ is finite.
  \begin{enumerate}
  \item The spectrum $\spec{H_\eps^\theta}$ depends only on $\Im
    \theta$ and $\spec{H_\eps^{\conj
        \theta}}=\conj{\spec{H_\eps^\theta}}$.
  \item The discrete spectrum $\disspec{H_\eps^\theta}$ is locally
    constant in $\theta$, i.e., if $0 < \Im \theta_1 \le \Im \theta_2
    < \vartheta/2$ then $\disspec{H_\eps^{\theta_1}} \subset
    \disspec{H_\eps^{\theta_2}}$.
  \item We have $\spec {H_\eps^\theta} \cap [0,\infty) =
    \spec[p]{H_\eps}$ where $\spec[p]{H_\eps}$ denotes the set of
    eigenvalues of $H_\eps$ (which are embedded in the continuous
    spectrum).
  \item The singular continuous spectrum of $H_\eps$ is empty.
  \item There is a subspace $\mathcal A$ satisfying~\eqref{eq:uni.eq}
    such that $\Psi_f(z):=\iprod f {(H_\eps-z)^{-1}f}$ has a
    meromorphic continuation onto the Riemann surface defined by $w
    \to \sqrt w$ if $\eps=0$ resp.~$w \to \sqrt{w-\EWN k F/\eps^2}$,
    $k \in \N$ if $\eps>0$.
   \item $\lambda \in \Sigma_\vartheta$ is a discrete eigenvalue of
     $H_\eps^\theta$ iff there exists $f \in \HS_\eps$ such that the
   meromorphic continuation of $\Psi_f$ has a pole in $\lambda$.
  \end{enumerate}
\end{proposition}
\begin{proof}
  The proof follows closely the proof
  of~\cite[Thm.~XIII.36]{reed-simon-1-4}, so we only comment on the
  differences.  We omit the dependency on $\eps$ here. The basic
  ingredient in the proof is first, the analyticity of the family
  $\{H^\theta\}_\theta$ in the sense that the resolvents are analytic
  in $\theta$, and second, the unitary equivalence
  \begin{equation}
    \label{eq:uni.eq}
    H^{\theta_1+\theta_2} = U^{\theta_1} H^{\theta_2} U^{-\theta_1}
  \end{equation}
  for real $\theta_1$ and complex $\theta_2 \in S_\vartheta$. This
  unitary equivalence holds a priori only for real $\theta_1$ and
  $\theta_2$. But since both sides are analytic in $\theta_2$,
  equality~\eqref{eq:uni.eq} extends therefore to complex $\theta_2
  \in S_\vartheta$. From this and the fact that $\{H^\theta\}_\theta$
  is a self-adjoint family,~(i) follows immediately. In a similar
  way,~(ii) follows noting the fact that an eigenvalue of $H^\theta$
  depends analytically on $\theta$ since $(H^\theta+1)^{-1}$ is
  analytic (cf.~\cite[Thm.~VII.1.8]{kato:66}). In order to prove~(iii)
  and~(iv) as in \cite{reed-simon-1-4}, we need the notion of
  \emph{analytic vectors} with respect to the unitary group $U^\theta$
  (namely w.r.t.\ its self-adjoint generator given by $A:=(x
  \partial_x + \partial_x x)\im/2$ on each external edge). The
  subspace of analytic vectors is defined as
  \begin{equation*}
    \mathcal A := \Bigset {f \in \bigcap_{k \in \N} \dom A^k}
            {\sum_n \frac {t^n}{n!} \norm{A^n f} < \infty}
  \end{equation*}
  for some $t \ge \vartheta/2$.  It then follows that
  \begin{equation}
    \label{eq:an.vec}
    \begin{cases}
      U^\theta(\mathcal A) & \text{is dense in $\HS$ and}\\
      \theta \mapsto U^\theta f       
                           & \text{extends analytically as
        map $S_\vartheta \to \Lsqr{X_\eps}$}
    \end{cases}
  \end{equation}
  for all $f \in \mathcal A$ using~\eqref{eq:uni.eq}
  (cf.~\cite[Ch.~X.6]{reed-simon-1-4}). The analytic extension is then
  \begin{equation*}
    f^\theta := U^\theta f = \sum_n \frac{\theta^n}{n!} (iA)^n f .
  \end{equation*}
  To prove~(v) we just note that a meromorphic continuation of
  $\Psi_f$ is given by
  \begin{equation*}
    \Psi^\theta_f(z)
    = \iprod {f^{\conj \theta}} {(H^\theta-z)^{-1} f^\theta}
  \end{equation*}
  since we have $\Psi^0_f(z)=\Psi^\theta(f)$ a priori only for real
  $\theta$ but by analyticity also for $\theta \in S_\vartheta$.
  For~(vi) we argue as follows: If $g$ is an eigenvector of $H^\theta$
  with eigenvalue $\lambda$ then let $f:=U^{-\theta} g$.  Again, by
  analyticity, we have $U^\theta f = g$ not only for real, but also
  for complex~$\theta$.  In particular, $\Psi_f$ has a pole at
  $\lambda$. On the other side, if $\Psi_f=\Psi_f^\theta$ has a pole
  at $\lambda$, then $\iprod {f^\theta} {\1_{\{\lambda\}}f^\theta} \ne
  0$ and in particular, $f^\theta$ is an eigenvector of $H^\theta$.
\end{proof}

Motivated by~(iii) and~(vi) of the last lemma, we make the following
definition.
\begin{definition}
  \label{def:resonances}
  A \emph{resonance} of $H_\eps$ is a non-real eigenvalue of the
  dilated operator $H_\eps^\theta$ for some $\theta \in S_\vartheta$
  and $0<\vartheta<\pi$.
\end{definition}
Finally, we assure that our definition of resonances does not depend
on where we cut the spaces into an internal and external part (cf.\
also~\cite{helffer-martinez:87}):
\begin{lemma}
  \label{lem:dil.ind}
  The definition of resonances \Def{resonances} does not depend on
  where we cut the graph and the manifold into an external and
  internal part. Furthermore, the definition of resonances is the same
  if we use a smooth flow as in~\cite{hislop-sigal:89}.
\end{lemma}
\begin{proof}
  Denote by $U^\theta$ and $\wt U^\theta$ the exterior dilation
  operators associated to the flow $\Phi^\theta$ and $\wt
  \Phi^\theta$, respectively (cf.~\eqref{eq:flow}), where the flow is
  either a (non-smooth) flow with cut at some point $x_0 \ge 0$ on the
  external edge or smooth.  The main point is to show that there
  exists a subspace $\mathcal A$ which satisfies~\eqref{eq:an.vec} for
  \emph{both} $U^\theta$ \emph{and} $\wt U^\theta$. But since we have $A_{x_0}=((x-x_0)
  \partial_x + \partial_x (x-x_0))\im/2 = A_0$ for the generator, the
  set of analytic vectors of $A_0$ forms such a subspace. Then an
  eigenvalue of the dilated operator with respect to $U^\theta$ or
  $\wt U^\theta$ is a pole of the meromorphic continuation of
  $\Psi_f(z)=\iprod f {(H-z)^{-1}f}$ for some $f \in \mathcal A$, and
  the latter definition is clearly independent of the dilation
  operators.
\end{proof}

%
\section{Closeness of graph and wave-guide model}
\label{sec:closeness.mod}
%

\subsection{Quasi-unitary operators}
\label{sec:quasi.unitary}
We now define quasi-unitary operators mapping from $\HS$ to $\wt \HS$
and vice versa, as well as their analogues on the compatible scales of
order $1$, namely $\HS^{1,\theta}$ and $\wt \HS^{1,\theta}$
(cf.~\Def{quasi-uni} and \Def{closeness}). Here,
\begin{align}
  \label{eq:spaces}
    \HS       &:= \Lsqr {X_0} &
    \wt \HS   &:= \HS_\eps = \Lsqr {X_\eps}
\end{align}
and we define $\HS^{1,\theta}$ and $\wt \HS^{1,\conj \theta}$ as
in~\eqref{eq:def.dil.qf}, but now for \emph{complex} $\theta \in
S_\vartheta$. Using the map
\begin{equation}
  \label{eq:def.t}
  \begin{split}
    \map{T^\theta} \HS \HS, \qquad T^\theta f &:=
               f_\inl \oplus \e^{-\theta/2} f_\ext\\
    \map{\wt T^\theta} \HS \HS, \qquad \wt T^\theta u &:=
               u_\inl \oplus \e^{-\theta/2} u_\ext,
  \end{split}
\end{equation}
we have $\HS^{1,\theta}=T^{-\theta}(\HS^1)$ where $\HS^1=\dom \qf
h = \Sob {X_0}$ is the quadratic form domain of the undilated
Hamiltonian. On $\HS^{1,\theta}$, we use the (complete) norm
\begin{equation}
  \label{eq:norm.1.theta}
  \norm[1,\theta] f := \norm[1]{T^\theta f}
\end{equation}
where $\norm[1]\cdot$ is the norm associated to $\qf d$.  Similarly,
we define $\wt T^\theta$ and a norm on $\wt \HS^{1, \theta}$ via
$\norm[1, \theta] u := \norm[1]{\wt T^\theta u}$ where $\norm[1]
\cdot$ is the norm associated to the free quadratic form $\qf d_\eps$.
We show in \App{res.est}, namely in \LemS{sf.bdd}{emb.1.2.mfd} that we
obtain a scale of order $1$ in the sense of \Def{scale.1}. In
particular, we also show in \App{res.est} that the various constants
$\wt C_i^\theta$ in \Apps{scale}{abstr.crit} associated to the
manifold case are $\eps$-independent.

Let $\map J \HS {\wt \HS}$ be given on the components of $X_\eps$ by
\begin{equation}
  \label{eq:def.j}
  (J_e f)(x,y):= \eps^{-m/2}f_e(x)
     \qquad \text{and} \qquad
  (J_v f) := 0,
\end{equation}
i.e., as an extension independent of the transverse variable. Recall
that $d=m+1 \ge 2$ is the dimension of the manifold $X_\eps$. Next we
define $\map {J^1} {\HS^{1,\theta}} {\wt \HS^{1,\theta}}$ by
\begin{equation}
  \label{eq:j.1}
  \begin{split}
    (J^1_e f)(x,y) &:= \eps^{-m/2} \Theta_\edeps(x,y)f_e(x),\\
    (J^1_v f)(z)   &:= \eps^{-m/2} \Theta_\vxeps(z) f(v)
  \end{split}
\end{equation}
for internal vertices $v \in V \setminus \Gamma_0$ where $\Theta_\eps$
is given in~\eqref{eq:def.theta}. Note that we did not associate a
vertex neighbourhood to the boundary vertices since they have degree
$2$.  Note in addition that the latter operator is well defined: the
function $J^1f$ matches along the different internal components
(recall that $\Theta_\eps$ is a Lipschitz function on $X_\eps$) and
has a jump of relative magnitude $e^{\theta/2}$ from the internal to
the external part.  Finally, $f(v)$ is defined for
$\Sobsymb^1$-functions (cf.~\eqref{eq:point.vx}).

Concerning the mappings in the opposite direction, we first
introduce the following averaging operators
\begin{equation*}
  (N_e u)(x) := \int_F u_e(x,y) \dd F(y) \qquad \text{and} \qquad
  C_v u := \frac 1 {\vol_d U_v} \int_{U_v} u \dd U_v
\end{equation*}
for $u \in \wt \HS = \Lsqr {X_\eps}$. Recall that $\vol_m F=1$.  The
map in the opposite direction $\map {J'} {\wt \HS} \HS$ is given by
\begin{equation}
  \label{eq:j.}
  (J' u)_e (x):= \eps^{m/2} (N_e u)(x), \qquad x \in e.
\end{equation}
Furthermore, we define $\map {J'_1} {\wt \HS^{1,\conj \theta}}
{\HS^{1,\conj \theta}}$ by
\begin{equation}
  \label{eq:j.1.}
  (J'_e{}^{1} u)(x):=
  \eps^{m/2} \Bigl[ N_e u (x) +
       \sum_{v \in \bd e, v \notin \Gamma_0} \rho_{v,e}(x)
              \bigl[ C_v u - N_e u (v) \bigr]
             \Bigr]
\end{equation}
for $x \in e$ on an internal edge $e$ and $J'_e{}^1 u := J'_e u$ for
an external edge $e$.  Here $\rho_{v,e}$ is a smooth cut-off function
such that $\rho_{v,e}(v)=1$ and $\rho_{v,e}(x)=0$ if $d(v,x) \ge
\ell_e/2$.  Moreover, $J'_e{}^{1} u(v)= \eps^{m/2} C_v(u)$ so that
$J'_e{}^{1} u$ fits to a continuous function at each internal vertex
$v \in V \setminus \Gamma_0$ of the the quantum graph. In addition
$J'_e{}^{1} u$ has a jump of magnitude $\e^{\conj \theta/2}$ at the
boundary $\Gamma_0$ since we assumed that at a boundary vertex $v \in
\Gamma_0$ there is no additional vertex neighbourhood. The role of the
conjugation will become clear in \App{abstr.crit}.

Note that the definition of $J$, $J'$ and $J_1'$ are the same as
in the absence of the fields, $\alpha=0$ and $q=0$
(cf.~e.g.~\cite{kuchment-zeng:01, exner-post:05, post:06a}),
while the definition of $J_1$ has an additional phase factor due
to the magnetic potential.

\subsection{Closeness assumptions}
\label{sec:closeness}

Now we are in position to demonstrate that the two Hamiltonians
are close to each other:
\begin{theorem}
  \label{thm:closeness}
  Assume~\eqref{eq:deg.bd}--\eqref{eq:pot.bd} on the quantum graph
  $X_0$ and \eqref{eq:def.met}--\eqref{eq:def.pot} on the manifold
  $X_\eps$. Then the dilated magnetic Hamiltonians $H_\eps^\theta$ and
  $H^\theta$ are $\Err(\eps^{1/2})$-close in the sense of
  \Def{closeness} where the error depends only on the constants and
  errors in the hypotheses (except on $r_0(\eps)$ and
  $\kappa_0(\eps)$), and on $\Re \theta$.
\end{theorem}
\begin{proof}
  We start with Condition~\eqref{eq:j.scale} of the closeness
  assumptions: we estimate
\begin{multline*}
  \normsqr{Jf - J^1f} \\
  = (1+ \Err(\eps)) \sum_{e \in E} 
             \int_{U_e}|\Theta_\edeps - 1|^2 |f_e|^2 \dd F \dd x 
      + \Err(\eps)  \sum_{v \in V}  
             \int_{U_v} |\Theta_\vxeps|^2 \dd U_v |f(v)|^2 \qquad\\
  =     \Err(\eps)  \sum_{e \in E} \normsqr[e]{f_e} 
      + \Err(\eps) c_{\vol} \sum_{v \in V} |f(v)|^2 
  = \Err(\eps) \sum_{e \in E}
          \bigl(\normsqr[e]{f_e} + \normsqr[e]{f_e'} \bigr)
\end{multline*}
where we have used~\eqref{eq:theta.est} and our
assumptions~\eqref{eq:len.bd},~\eqref{eq:def.met},~\eqref{eq:vol.ew}
and~\eqref{eq:def.omega}. In addition, we have used the standard
Sobolev estimate
\begin{equation}
  \label{eq:point.vx}
  |f(0)|^2 \le \frac 4 {\ell_0}
      \bigl( \normsqr[(0,\ell_0/2)] f + \normsqr[(0,\ell_0/2)] {f'} \bigr).
\end{equation}
which implies
\begin{equation}
  \sum_{v \in V} |f(v)|^2 \le
  \frac 4 {\ell_0} \sum_{e \in E}
       \bigl(\normsqr[e]{f_e} + \normsqr[e]{f_e'} \bigr)
\end{equation}
by choosing an edge $e \in E_v$ for each vertex $v$. Clearly, we
can estimate the right-hand side by $\norm[1,\theta] f =
\norm[1]{T^\theta f}$ and we obtain $\norm{Jf - J^1f} =
\Err(\eps^{1/2}) \norm[1,\theta] f$ where $\Err(\eps^{1/2})$ depends now
also on $\Re \theta$.

Next we have
\begin{equation*}
  \normsqr{J' u - J'{}^1 u} =
  \sum_{e \in E} \sum_ {v \in \bd e}
    \eps^m \int_e \rho_{v,e}^2 \dd x \, | C_v u - N_e u(v)|^2
\end{equation*}
since the supports of $\rho_\vxed$, $v \in \bd e$, are disjoint by
construction. As in~\cite[Lemma~5.5]{exner-post:05} (cf.
also~\cite[Lem.~2.10]{post:06a} for the non-compact case) we
can show
\begin{equation*}
  \eps^m |C_v u - N_e u(v)|^2 = \Err(\eps) \normsqr[U_\vxeps] {\de u}
\end{equation*}
for $u \in \Sob {U_\vxeps}$, $v \in \bd e$. Here $\Err(\eps)$ depends
on the errors in~\eqref{eq:def.met}, on $\lambda_2$
in~\eqref{eq:vol.ew} and on $\ell_0$.  Reordering the sum $\sum_{e
\in E} \sum_{v \in \bd e} \normsqr[U_\vxeps] u$, we gain a factor
$d_0$ (the maximal degree of a vertex, cf.~\eqref{eq:deg.bd}). In
particular,
\begin{equation*}
  \normsqr{J'u - J'{}^1u} =
  \Err(\eps) d_0 \sum_{v \in V} \normsqr[\vxeps] {\de u} =
  \Err(\eps) \normsqr[1,\conj \theta] u
\end{equation*}
for $u \in \wt \HS^{1,\conj \theta}$ so that $\norm{J'u - J'{}^1u}
= \Err(\eps^{1/2}) \norm[1,\conj \theta] u$ where again the error
term $\Err(\eps^{1/2})$ depends also on $\Re \theta$.

Assumption~\eqref{eq:j.adj} follows easily
from~\eqref{eq:norm.prod.ed}, i.e.,
\begin{equation*}
  |\iprod {Jf} u - \iprod f {J' u} | \le
  \Err(\eps) \norm f \norm u
\end{equation*}
for $f \in \Lsqr {X_0}$ and $u \in \Lsqr {X_\eps}$.  In the same
way, Assumption~\eqref{eq:j.bdd} follows from
\begin{equation*}
  \normsqr{J f} \le (1+ \Err(\eps)) \normsqr f \qquad \text{and} \qquad
  \normsqr{J' u} \le (1+ \Err(\eps)) \normsqr u.
\end{equation*}

Assumption~\eqref{eq:j.inv} follows from $J'J f = f$ and
\begin{equation*}
  \normsqr{J J' u - u} =
      \sum_{e \in E} \normsqr[U_\edeps] {N_e u - u}
    + \sum_{v \in V} \normsqr[U_\vxeps] u.
\end{equation*}
Now, as in~\cite[Lemmas~3.1,~4.4]{exner-post:05}
(cf.~\cite[Lem.~2.11]{post:06a} for the non-compact case), we
can show
\begin{equation*}
  \normsqr[U_\edeps] {N_e u - u} \le
  \Err(\eps^2) \normsqr[U_\edeps] {\de u}
\end{equation*}
where $\Err(\eps^2)$ depends on the error in~\eqref{eq:def.met} and
on $\EWN 2 F$. Next we have
\begin{equation*}
  \normsqr[U_\vxeps] u = \Err(\eps)
   (\normsqr[U_{\edeps}^+] u + \normsqr[U_{\edeps}^+] {du})
\end{equation*}
(cf.~\cite[Cor.~5.8]{exner-post:05}
or~\cite[Lem.~2.12]{post:06a} for the non-compact case)) with
an error depending on $\ell_0$ and the errors
in~\eqref{eq:def.met} and~\eqref{eq:vol.ew}. Here,
\begin{equation*}
    U_\vxeps^+= U_\vxeps \disjcup \bigdisjcup_{e \in E_v} U_\edeps
\end{equation*}
is the vertex neighbourhood together with its adjacent edge
neighbourhoods.  The last two estimates mean that a function
orthogonal to the constant transversal function or being concentrated
at a vertex neighbourhood cannot be spectrally bounded.  Summing all
these error terms, we obtain $\norm{J J' u - u} = \Err(\eps^{1/2})
\norm[1] u$ for $u \in \HS^1$.

We finally prove~\eqref{eq:j.comm.1} in our model. On each internal
edge, we have the contribution
\begin{equation*}
  \qf h^\theta_e(J_e'{}^1 u, f) - \qf h^\theta_\edeps(u, J_e^{1} f)
  = \qf h^\theta_e(J_e'{}^{1} u, f) -
             \hat{\qf  h}^\theta_\edeps(u, J_\e f) +
            \Err(\eps) \norm[\qf d_\edeps] u \norm[\qf d_e] f,
\end{equation*}
where we used~\eqref{eq:qf.prod.ed}. Note that $J_e^1f = \Theta_\edeps
J_e f$, $\de_F J_e f=0$ and $\norm[\hat{\qf d}_\edeps]{J_e
  f}=\norm[\qf d_e] f$ and that $\Err(\eps)=0$ if $e \in E_\ext$.
Recall that $\normsqr[\qf d_\edeps] u = \normsqr[\edeps] u +
\normsqr[\edeps] {\de u}$ and similarly for the other norms. Now
\begin{equation*}
  \qf h^\theta_e(J_e'{}^1 u, f) -
  \hat{\qf h}^\theta_\edeps(u, J_\e f) =
  \sum_{v \in \bd e, v \notin \Gamma_0} \iprod[\edeps]{D_e \rho_\vxed} f
       \eps^{m/2} \bigl(C_v \conj u - N_e \conj u(v)\bigr)
\end{equation*}
since the longitudinal terms cancel due to the simple form of
$\hat{\qf h^\theta_\edeps}$ and $\de_F J_e f=0$. On external edges we
even have $\qf h^\theta_e(J_e'{}^1 u, f) = \qf h^\theta_\edeps(u, J_\e
f)$ since $U_\edeps$ has exact product structure there.

The vertex contribution is
\begin{equation*}
  \qf h_\vxeps^\theta (u, J_v^1 f) =
  \eps^{-m/2} \qf h_\vxeps^\theta (u, \Theta_\vxeps f(v)) =
  \Err(\eps^{1/2}) \norm[\qf d_\vxeps] u |f(v)|,
\end{equation*}
where we have used~\eqref{eq:qf.prod.vx}. Note that $J_v^1f =
\eps^{-m/2} \Theta_\vxeps f(v)$, $\de f(v) = 0$, and therefore
$\norm[\hat{\qf d}_\vxeps]{f(v)}=\eps^{d/2} |f(v)|$.

Finally, summing up all the error terms, we
obtain~\eqref{eq:j.comm.1} with $\delta=\Err(\eps^{1/2})$, again
depending also on $d_0$ and $\Re \theta$.
\end{proof}

Using the additional information of \Thm{main.app} we can conclude
from \App{abstr.crit} our main result:
\begin{theorem}
  \label{thm:main2}
  Let $0 \le \vartheta < \pi$ and $\theta \in S_\vartheta$, i.e.,
  $|\Im \theta| < \vartheta/2$.  Assume in
  addition~\eqref{eq:deg.bd}--\eqref{eq:pot.bd} on the quantum graph
  $X_0$ and \eqref{eq:def.met}--\eqref{eq:def.pot} on the manifold
  $X_\eps$. If $\lambda(0)$ denotes a resonance of the magnetic
  Hamiltonian $H_0$ with a multiplicity $m>0$ then for a sufficiently
  small $\eps>0$ there exist~$m$ resonances $\lambda_1(\eps), \dots,
  \lambda_m(\eps)$ of $H_\eps$, satisfying $\Im\lambda_j(\eps)<0$ and
  not necessarily mutually different, which all converge to
  $\lambda(0)$ as $\eps\to 0$. The same is true in the case when
  $\lambda(0)$ is an embedded eigenvalue of $H_0$, except that only
  $\Im\lambda_j(\eps)\le 0$ holds in general.
\end{theorem}
Note that if the internal part is compact (i.e., if there are only
finitely many vertices), then the
assumptions~\eqref{eq:deg.bd}--\eqref{eq:len.bd} and
\eqref{eq:vol.ew} are automatically fulfilled.

We can even conclude stronger results from \App{abstr.crit} using the
identification maps $J$ and $J'$ defined in~\eqref{eq:def.j}
and~\eqref{eq:j.}, namely the resolvent convergence and the convergence of the
eigenprojections.
\begin{theorem}
  \label{thm:main.res}
  Under the same assumptions as in the previous theorem, we have
  \begin{subequations}
  \label{eq:main.res}
    \begin{gather}
      \norm{J(H_0^\theta-z)^{-1} - (H_\eps^\theta-z)^{-1}J} 
      = \Err(\eps^{1/2}),\\
      \norm{J(H_0^\theta-z)^{-1}J' - (H_\eps^\theta-z)^{-1}} 
      = \Err(\eps^{1/2})
    \end{gather}
  \end{subequations}
  for $z \notin \Sigma_\vartheta$. The error depends on the same
  quantities as the error in~\Thm{closeness}, and also on $\vartheta$
  and $z$.

  In addition, suppose that $\lambda^\theta(0)$ is a discrete
  eigenvalue of $H_0^\theta$. Let $D$ be an open disc such that $D$
  contains $\lambda$ but no other spectral point of $H_0^\theta$.
  Then~\eqref{eq:main.res} holds when the resolvent is replaced by the
  spectral projection $\1_D(H_0^\theta)$ resp.\ $\1_D(H_\eps^\theta)$.
  If the multiplicity of $\lambda^\theta(0)$ is $1$ with normalised
  eigenfunction $\psi^\theta_0$ (a \emph{resonance} or
  \emph{eigenstate} for $H_0$) then there exists a normalised
  eigenfunction $\psi^\theta_\eps$ (a \emph{resonance} or
  \emph{eigenstate} for $H_\eps$) on the manifold such that
  \begin{equation*}
    \norm{J \psi_0^\theta - \psi_\eps^\theta} = \Err(\eps^{1/2}) \und
    \norm{J' \psi_\eps^\theta - \psi_0^\theta} = \Err(\eps^{1/2}).
  \end{equation*}
\end{theorem}

As a by-product, we also have shown that the spectrum of a magnetic
Hamiltonian on a \emph{non-compact} manifold converges to the
associated non-compact quantum graph Hamiltonian provided our
uniformity assumptions are fulfilled. In particular, we could
approximate fractal spectra such as studied, e.g., in~\cite{bgp:07}
as we have mentioned in the introduction.
\begin{theorem}
  \label{thm:mag.ham}
   Assume~\eqref{eq:deg.bd}--\eqref{eq:pot.bd} on the quantum graph
  $X_0$ and \eqref{eq:def.met}--\eqref{eq:def.pot} on the manifold
  $X_\eps$. Then the spectrum of $H_\eps$ converges to $H_0$ on any
  finite energy interval. The same is true for the essential and
  discrete spectrum.
\end{theorem}
\begin{proof}
  The spectral convergence is a direct consequence of the
  closeness, as it follows from the general theory developed
  in~\cite[Appendix]{post:06a}.
\end{proof}

%
\setcounter{section}{0}
\renewcommand{\thesection}{\Alph{section}}
\section*{Appendices}
%

%
\section{Scale of Hilbert spaces}
\label{app:scale}
%

\subsection{Scale of Hilbert spaces associated with a self-adjoint operator}
\label{sec:scale.sa}

Denote by $\Delta$ a non-negative, self-adjoint operator in the
Hilbert space $\HS$. We sometimes refer to $\Delta$ as the \emph{free}
operator. Throughout the paper we use the convention that $\iprod
\cdot \cdot$ and other sesquilinear forms are \emph{anti-linear} in the
first and linear in the second argument.

A scale of Hilbert spaces can be associated with $\Delta$ as follows: For
fixed $k \ge 0$ we set $\HS^k := \dom \Delta^{k/2}$ equipped with the norm
$\norm[k] u := \norm{(\Delta+1)^{k/2}u}$. For negative powers, we set
$\HS^{-k}:= (\HS^k)^*$ where
\begin{subequations}
  \label{eq:scale.neg}
  \begin{equation}
    (\HS^k)^* :=
         \set{\map \phi {\HS^k} \C}
             {\text{$\phi$ anti-linear and bounded}},
  \end{equation}
  with the norm
  \begin{equation}
    \norm[-k] \phi :=
       \sup_{f \in \HS^k}
             \frac {|\phi(f)|}{\norm[k] f}
  \end{equation}
\end{subequations}
and $\HS$ is embedded in $\HS^{-k}$ via $f \mapsto \iprod \cdot f$. For more
details we refer e.g.\ to~\cite{kps:82}.

\subsection{Scale of Hilbert spaces associated with a self-adjoint
  family of operators}
\label{sec:scale.sect}
Since our dilated operators are no longer self-adjoint, we also need a
scale of Hilbert spaces associated with a particular class of
non-self-adjoint operators, namely \emph{sectorial} operators. Most of
the material on such operators is standard and can be found e.g.\
in~\cite
{kato:66}. We also introduce a scale of order $1$
which is not associated to the natural quadratic form, but easier to
handle in the present application.

Let $\{H^\theta\}_\theta$ with $\theta \in S = \set{w \in \C}{|\Im w|
  < b}$ be a family of closed operators acting in the Hilbert space
$\HS$.
\begin{definition}
  \label{def:sa.fam}
  We say that the family $\{H^\theta\}_\theta$ is
  \emph{self-adjoint}\footnote{In general, $H^\theta$ is self-adjoint
    only for \emph{real} $\theta$.} if $(H^\theta)^*=H^{\conj
    \theta}$.

  The family $\{H^\theta\}$ is called \emph{(spectrally) uniformly
    $\vartheta$-sectorial}\footnote{Usually, an operator is called
    \emph{sectorial}, if $\vartheta < \pi/2$, and if one requires in
    addition that for all $\vartheta_1 \in (\vartheta,\pi/2)$ there is
    a constant $C_0=C_0(\vartheta,\vartheta_1)$ such that
    $\norm{(H^\theta-z)^{-1}} \le C_0/|z|$ for all $z \notin
    S_{\vartheta_1}$. We do not need this fact here.} if $\spec
  {H^\theta}$ is contained in the common sector
  $\Sigma_\vartheta=\set{z \in \C}{ |\arg z| \le \vartheta}$.
\end{definition}
We allow values $0 \le \vartheta < \pi$, although operators with
spectrum \emph{not} contained in the right half-plane are no longer
semi-bounded. The only point we need here is, that $-1$ belongs to the
resolvent set and that we can control the norm of the corresponding
resolvent (denoted by the constant $C_0^\theta$).

From now on we assume that $\{H^\theta\}$ is a self-adjoint, uniformly
$\vartheta$-sectorial family of operators. We start defining the
scales of order $2$, $0$ and $-2$:

Let $\HS^0 := \HS$, $\norm[0] \cdot := \norm \cdot$ and
\begin{equation}
  \label{eq:scale.2}
  \HS^{2,\theta} := \dom H^\theta, \qquad
  \norm[2,\theta] f :=  \norm {(H^\theta+1)f}
\end{equation}
be the spaces of order $0$ and $2$. Since $H^\theta$ is closed and $-1
\notin \spec{H^\theta}$, $\HS^{2,\theta}$ with norm $\norm[2,\theta]
\cdot$ is also a Hilbert space.  The dual space is defined by
\begin{equation}
  \label{eq:scale.-2}
  \HS^{-2,\theta}:= (\HS^{2,\conj \theta})^*
\end{equation}
similarly as in~\eqref{eq:scale.neg}. Note the complex conjugation of
$\theta$ in order to compensate the anti-linearity in the definition of
the dual. In the next two lemmas, we want to assure that $H^\theta$
and its resolvent extend to maps on the scale of order $-2,0,2$:

\begin{lemma}
  \label{lem:scale.incl}
  The embedding $\map \iota \HS {\HS^{-2,\theta}}$, $g \mapsto \iprod
  \cdot g$ is continuous. Furthermore, $\norm{(H^\theta+1)^{-1} g} =
  \norm[-2,\theta]{\iota g}$ for $g \in \HS$, i.e., $\HS^{-2,\theta}$ can be
  considered as the completion of $\HS$ in the norm $\norm[-2,\theta] g :=
  \norm{(H^\theta+1)^{-1}g}$.
\end{lemma}
\begin{proof}
  We have
  \begin{equation}
  \label{eq:norm.-2}
    \norm[2,-\theta] {\iota g} =
    \sup_{f \in \HS^{2,\conj \theta}}
         \frac {|\iprod f g|}{\norm[2,\conj \theta] f} =
    \sup_{h \in \HS}
         \frac {|\iprod h {(H^\theta+1)^{-1}g}|}
                   {\norm h} =
    \norm{(H^\theta+1)^{-1}g}
  \end{equation}
  where $h=(H^{\conj \theta}+1)f$ and the claims follow.
\end{proof}

\begin{lemma}
  \label{lem:scale.2}
  The maps
  \begin{equation}
    \label{eq:op.scale.2}
    \map {(H^\theta+1)} {\HS^{2,\theta}} \HS   \und
    \map {(H^\theta+1)^{-1}} \HS {\HS^{2,\theta}}
  \end{equation}
  are isometries and inverse to each other. Similarly,
  \begin{equation}
    \label{eq:op.scale.-2}
    \map {(H^\theta+1)} \HS {\HS^{-2,\theta}}   \und
    \map {(H^\theta+1)^{-1}} {\HS^{-2,\theta}} \HS
  \end{equation}
  are isometries and inverse to each other. Here $(H^\theta+1)g:=
  \iprod{(H^{\conj \theta}+1)(\cdot)} g$ and $(H^\theta+1)^{-1} \iota
  g := (H^\theta+1)^{-1}g$ extend to an isometry on $\HS^{-2,\theta}$.
  Finally,
  \begin{equation}
    \label{eq:scale.op}
    \map{H^\theta} {\HS^{2,\theta}} \HS    \quad \text{and} \quad
    \map{H^\theta} \HS {\HS^{-2,\theta}}
  \end{equation}
  are bounded maps with norm bounded by $1+C_0^\theta$, where
  $C_0^\theta=\norm{(H^\theta+1)^{-1}}$ in general depends on
  $\theta$.
\end{lemma}
\begin{proof}
  The first two assertions are almost obvious. The last one follows
  from the fact that since $-1 \notin \spec A$, we have $\norm f \le
  C_0^\theta \norm{(H^\theta+1)f}$, and therefore
  \begin{equation*}
    \norm{H^\theta f} \le \norm{(H^\theta+1)f} + \norm f \le
    (1+C_0^\theta) \norm{(H^\theta+1)f} = (1+C_0^\theta)\norm[2,\theta] f.
  \end{equation*}
  Similarly
  \begin{multline*}
    |\iprod{H^{\conj \theta} f} g| \le
    |\iprod {(H^{\conj \theta}+1) f } g| + |\iprod f g| \\ \le
    (\norm[2,\conj \theta] f + \norm f ) \norm g \le
    (1+C_0^\theta)\norm[2,\conj \theta] f  \norm g
  \end{multline*}
  and therefore $\norm[-2,\theta]{H^\theta g} \le (1+C_0^\theta) \norm
  g$.
\end{proof}

So far, we have defined a scale of Hilbert spaces
$\{\HS^{k,\theta}\}_{k,\theta}$, $k=-2,0,2$, associated to the
self-adjoint, uniformly $\vartheta$-sectorial family
$\{H^\theta\}_\theta$, i.e., for $k=0$ and $k=2$, the inclusion map
\begin{equation}
  \label{eq:scale.k.incl}
  \map {\iota} {\HS^{k,\theta}} {\HS^{k-2,\theta}}
\end{equation}
is continuous, $\HS^{k,\theta}$ is dense in $\HS^{k-2,\theta}$ and the
maps
\begin{subequations}
  \label{eq:scale.k}
  \begin{gather}
    \label{eq:scale.k.op}
    \map{H^\theta}{\HS^{k,\theta}}{\HS^{k-2,\theta}},\\
    \label{eq:scale.k.res}
    \map{(H^\theta+1)^{-1}}{\HS^{k-2,\theta}}{\HS^{k,\theta}}
  \end{gather}
  are continuous.
\end{subequations}

Since in our context, the domain $\dom H^\theta$ will depend on the
\emph{complex} parameter $\theta$, the natural quadratic form
associated to $H^\theta$ is not well-adopted to our application
(especially its natural norm). We therefore define the norm on the
Hilbert space of order $1$ in a different way:

\begin{definition}
  \label{def:scale.1}
  Let $\HS^{1,\theta}$ be a linear subspace of $\HS$, and let $\Delta
  \ge 0$ be a self-adjoint, non-negative operator on $\HS$.  We say
  that $\HS^{1,\theta}$ defines a \emph{compatible scale of order $1$
    w.r.t.\ $\Delta$} if the following conditions are fulfilled:
  \begin{enumerate}
  \item There is a family of bounded, invertible operators $\map
    {T^\theta} \HS \HS$, called \emph{compatibility operators} such
    that
    \begin{equation}
      \label{eq:t.ad.inv}
      (T^\theta)^* = T^{\conj \theta} \und
      (T^\theta)^{-1} = T^{- \theta}
    \end{equation}
    for $\theta \in S$. We assume that $\HS^{1,\theta} =
    T^{-\theta}(\HS^1)$ and define a norm
    \begin{equation}
      \label{eq:norm.1}
      \norm[1,\theta] u :=
      \norm[1] {T^\theta u} =
      \norm{(\Delta+1)^{1/2}T^\theta u}
    \end{equation}
    where $\HS^1$ is the element of the Hilbert space scale of order $1$
    associated with $\Delta$.

  \item \label{emb.dense} We assume that $\HS^{2,\theta}$ is a dense
    subspace of $\HS^{1,\theta}$.
  \item \label{emb.1.2.th} We assume that the embedding
    $\HS^{2,\theta} \hookrightarrow \HS^{1,\theta}$ is continuous with
    norm bounded by $C_2^\theta$.
  \item \label{def.sf} Finally, we assume that the \emph{sesquilinear
      form associated to $H^\theta$} defined by $\qf
    h^\theta(f,g):=\iprod f {H^\theta g}$
    is continuous on $\HS^{1,\conj \theta} \times \HS^{1,\theta}$,
    i.e., there exists a constant $C_1^\theta$ such that
    \begin{equation}
      \label{eq:est.sf}
      |\qf h^\theta(f,g)| \le
      C_1^\theta \norm[1,\conj \theta] f   \norm[1,\theta] g
    \end{equation}
    for all $f \in \HS^{1,\conj \theta}$ and $g \in
    \HS^{2,\theta}=\dom H^\theta$.
  \end{enumerate}
\end{definition}
Clearly, by the construction, $\HS^{1,\theta}$ is complete, since
$\HS^1$ is. In addition, since $\HS^{2,\theta}$ is dense in
$\HS^{1,\theta}$, the sesquilinear form $\qf h^\theta$ extends uniquely
to a bounded one on $\HS^{1,\conj \theta} \times \HS^{1,\theta} \to
\C$ which we denote by the same symbol.

We define the dual space as before by
\begin{subequations}
  \label{eq:scale.-1.th}
  \begin{equation}
    \HS^{-1,\theta}:= (\HS^{1,\conj \theta})^*
  \end{equation}
  with the canonical norm $\norm[-1,\theta] \cdot$ as
  in~\eqref{eq:scale.neg}. Note that we can consider $\HS^{-1,\theta}$
  as the completion of $\HS$ in the norm
  \begin{equation}
    \label{eq:norm.-1.th}
    \norm[-1,\theta] u =
    \norm{(\Delta+1)^{-1/2} T^{-\theta}  u}.
  \end{equation}
\end{subequations}
There are simple equivalent characterisations of the last two
conditions~\eqref{emb.1.2.th} and~\eqref{def.sf} following from the
definitions:
\begin{lemma}
  \label{lem:emb.1.2.th}
  Condition~\eqref{emb.1.2.th} is equivalent to the fact that
  \begin{subequations}
    \label{eq:emb.1.2.th}
    \begin{equation}
      \map {(H^\theta+1)^{-1}} \HS {\HS^{1,\theta}}
    \end{equation}
    is norm-bounded by $C_2^\theta$ or equivalently,
    \begin{equation}
      (\Delta+1)^{1/2}T^\theta (H^\theta+1)^{-1}
    \end{equation}
  \end{subequations}
  is a bounded operator in $\HS$ with bound $C_2^\theta$.
\end{lemma}
We also have a sufficient condition:
\begin{lemma}
  \label{lem:res.-1.1}
  Condition~\eqref{emb.1.2.th} follows from the fact that
  \begin{subequations}
    \label{eq:res.1.2.th}
    \begin{equation}
      \map {(H^\theta+1)^{-1}} {\HS^{-1,\theta}}{\HS^{1,\theta}}
    \end{equation}
    is norm-bounded or equivalently,
    \begin{equation}
      (\Delta+1)^{1/2} T^\theta (H^\theta+1)^{-1} T^\theta (\Delta+1)^{1/2}
    \end{equation}
  \end{subequations}
  is a bounded operator in $\HS$.
\end{lemma}
\begin{lemma}
  \label{lem:scale.1}
  The continuity of the sesquilinear form
  \begin{equation}
    \label{eq:def.sf}
    \map {\qf h^\theta}
    {\HS^{1,\conj \theta} \times \HS^{1,\theta}} \C
  \end{equation}
  is equivalent to the fact that
  \begin{subequations}
    \label{eq:h.1.-1.th}
    \begin{equation}
      \map {H^\theta} {\HS^{1,\theta}} {\HS^{-1,\theta}}, \qquad
      g \mapsto \qf h^\theta(\cdot, g)
    \end{equation}
    is norm-bounded by $C_1^\theta$ or equivalently,
    \begin{equation}
      (\Delta+1)^{-1/2} T^{-\theta} H^\theta T^{-\theta} (\Delta+1)^{-1/2}
    \end{equation}
  \end{subequations}
  is a bounded operator in $\HS$ with bound $C_1^\theta$.
\end{lemma}
These observations show that $\{\HS^{k,\theta}\}_{k,\theta}$ behaves
almost like a natural scale of Hilbert spaces; in
particular, $\HS^{k,\theta}$ is dense in $\HS^{k-1,\theta}$: This follows for $k=1$ by the construction of $\HS^{1,\theta}$ and for $k=2$ by \Defenum{scale.1}{emb.dense}. Furthermore, the inclusions
\begin{equation*}
  \map \iota {\HS^{k,\theta}} {\HS^{k-1,\theta}}
\end{equation*}
are continuous for $k=1$ by the construction of $\HS^{1,\theta}$ and
for $k=2$ by \Defenum{scale.1}{emb.1.2.th}. By duality, the same
statements hold for $k=0$ and $k=-1$.  In addition,~\eqref{eq:scale.k}
is valid for $k=0,1,2$ (by \Lem{scale.2} and \Lem{scale.1}) except
that the resolvent is only a continuous map from $\HS$ to
$\HS^{1,\theta}$ (\Lem{emb.1.2.th}).  Therefore, the following
definition is natural:

\begin{definition}
  \label{def:comp.scale}
  We call $\{\HS^{k,\theta}\}_{k,\theta}$, $k=-2,-1,0,1,2$, a
  \emph{compatible} scale if $\HS^{1,\theta}$ is a compatible scale of
  order $1$ (\Def{scale.1}) and $\{\HS^{k,\theta}\}_{k,\theta}$,
  $k=-2,0,2$, is a scale in the sense
  of~\eqref{eq:scale.k.incl}--\eqref{eq:scale.k}.
\end{definition}

%
\section{An abstract convergence criteria for non-selfadjoint operators}
\label{app:abstr.crit}
%

In this section we are going to prove the resolvent convergence for
self-adjoint, uniformly $\vartheta$-sectorial families of (closed)
operators $\{H^\theta\}_\theta$ and $\{\wt H^\theta\}_\theta$ acting
in $\HS$ and $\wt \HS$, respectively, for all $\theta$ in the strip
$S_\vartheta$ (i.e. $|\Im \theta| < \vartheta/2$).
\begin{notation}
  \label{not:norm.op}
  We will use the obvious notation $\norm [k \to m] A$ for the norm of
  the operator $\map A {\HS^k} {\HS^m}$ where $\HS^k$ is an element of
  the scale w.r.t.\ the self-adjoint operator $\Delta \ge 0$.
  Similarly, we write $\norm[k,\theta \to m,\theta] A$ for the norm of
  the operator $\map A {\HS^{k,\theta}} {\HS^{m,\theta}}$ where
  $\{\HS^{k,\theta}\}_{k,\theta}$, $k=-2, \dots, 2$, is a compatible
  scale associated to the operator $H^\theta$
  (cf.~\Defs{scale.1}{comp.scale}).

  Furthermore, we employ the analogous \emph{tilded} notation for the
  respective objects acting in the Hilbert space $\wt \HS$, namely the
  self-adjoint operator $\wt \Delta \ge 0$ with the scale $\{\wt
  \HS^k\}_k$ and the operator $\wt H^\theta$ giving rise to the scale
  $\{\wt \HS^{k,\theta}\}_{k,\theta}$.
\end{notation}

Next we introduce the notion of \emph{quasi-unitarity} up to an
error $\delta>0$. In our application, $\delta=\delta(\eps)$ where
$\eps$ is the parameter appearing in the operators \emph{and}
domains and \ Hilbert spaces. We prefer to formulate the results
below without mentioning explicitly the parameter $\eps$.

\begin{definition}
  \label{def:quasi-uni}
  Suppose that we have linear operators
  \begin{equation}
  \label{eq:quasi-uni}
    \map J      \HS    {\wt \HS} \qquad \text{and} \qquad
    \map {J'}   {\wt \HS} \HS.
  \end{equation}
  We say that $J$ and $J$ are \emph{$\delta$-quasi-unitary} w.r.t.\
  the operators $\Delta$ and $\wt \Delta$ iff the following conditions
  hold for $\delta>0$:
  \begin{gather}
    \label{eq:j.adj}
    \norm {J-{J'}^*} \le \delta,\\
    \label{eq:j.inv}
    \norm[1\to0] {\1 - J' J} \le \delta, \qquad
    \norm[1\to0] {\1 - J J'}  \le \delta,\\
    \label{eq:j.bdd}
    \norm J     \le 2, \qquad \qquad
    \norm {J'}  \le 2,
  \end{gather}
  where $\norm[1\to0] A = \norm{A(\Delta+1)^{-1/2}}$ is the norm of
  $\map A {\HS^1} \HS$ and the analogous norm is used on $\wt \HS$.
\end{definition}

This allows us to specify what we mean by closeness of operators
$H^\theta$ and $\wt H^\theta$:
\begin{definition}
  \label{def:closeness}
  We say that the operators $H^\theta$ and $\wt H^\theta$ in $\HS$ and
  $\wt \HS$, respectively, are \emph{$\delta$-close} w.r.t.\ the
  $\delta$-quasi-unitary operators $J$ and $J'$ or briefly,
  \emph{$\delta$-close} if there exist compatible scales
  $\HS^{1,\theta}$ and $\wt \HS^{1,\theta}$ of order $1$ associated to
  $H^\theta$ and $\wt H^\theta$ with compatibility operators
  $T^\theta$ and $\wt T^\theta$ in the sense of \Def{scale.1}, and if
  there exist operators\footnote{The operators $J^1$ and $J'{}^1$ need
    not to be bounded.}
  \begin{equation*}
    \map{J^1=J^{1,\theta}} {\HS^{1,\theta}}{\wt \HS^{1,\theta}}
        \quad\text{and}\quad
    \map{J'{}^1=J'^{1,\conj \theta}}
          {\wt \HS^{1,\conj \theta}}{\HS^{1, \conj \theta}}
  \end{equation*}
  such that
  \begin{equation}
    \label{eq:j.scale}
    \norm[1,\theta \to 0]{J^1 - J} \le \delta
        \qquad \text{and} \qquad
    \norm[1, \conj \theta \to 0]{J'{}^1 - J'} \le \delta.
  \end{equation}
  and
  \begin{gather}
    \label{eq:j.comm}
    \norm[1,\theta \to -1,\theta]
    {(J'{}^1)^* H^\theta - \wt H^\theta J^1} \le \delta, \\
    \label{eq:j.t}
    \wt T^\theta J = J T^\theta, \qquad\qquad
    T^\theta J' = J' \wt T^\theta,
  \end{gather}
  where $\norm[1,\theta \to 0] A = \norm{A
    T^{-\theta}(\Delta+1)^{-1/2}}$ and similarly on $\wt \HS$ and
  where
  \begin{equation*}
       \norm[1,\theta \to -1,\theta] V =
       \norm{(\wt \Delta+1)^{-1/2} \wt T^{-\theta} V
               T^{-\theta}(\Delta+1)^{-1/2}}.
\end{equation*}
\end{definition}
\begin{remark}
  \label{rem:closeness}
  \indent
  \begin{enumerate}
  \item We do not exclude that $\delta$ depends on $\theta$.
  \item Note that $H^\theta$ in~\eqref{eq:j.comm} is a bounded
    operator as map $\HS^{1,\theta}$ to $\HS^{-1,\theta}$
    (cf.~\eqref{eq:h.1.-1.th}) and similarly for $\wt H^\theta$.
  \item \label{j.comm} Denote the associated sesquilinear forms to
    $H^\theta$ and $\wt H^\theta$ by $\qf h^\theta$ and $\wt{\qf
      h}^\theta$, respectively (cf.~\Defenum{scale.1}{def.sf}).
    Then~\eqref{eq:j.comm} is equivalent to
    \begin{equation}
      \label{eq:j.comm.1}
      \tag{\ref{eq:j.comm}'}
      \bigl| \qf h^\theta (J'{}^1 u, f ) -
      \wt {\qf h}^\theta (u, J^1 f ) \bigr|
      \le
      \delta \norm[1,\conj \theta] u \norm[1,\theta] f
    \end{equation}
    for $u \in \wt \HS^{1,\conj \theta}$ and $f \in \HS^{1,\theta}$.
    In fact, we will see in the proof of \Thm{res} (the only point,
    where $J^1$ and $J'{}^1$ enter), that it is enough to
    have~\eqref{eq:j.comm.1} only for $f$ and $u$ in the
    \emph{operator} domains, i.e., $f \in \HS^{2,\theta}$ and $u \in
    \wt \HS^{2,\conj \theta}$.  Since $\HS^{2,\theta}$ is dense in
    $\HS^{1,\theta}$ and similarly on $\wt \HS$ by
    \Defenum{scale.1}{emb.dense}, this implies of
    course~\eqref{eq:j.comm}.
  \end{enumerate}
\end{remark}

An immediate consequence is the following:
\begin{lemma}
  \label{lem:j.theta}
  With the previous notation  we have
  \begin{gather}
    \label{eq:j.inv.theta}
    \norm{(J'J - \1)f} \le C^\theta_3 \delta \norm[2,\theta] f\\
    \label{eq:j.iso.theta}
    \bigl| \normsqr{Jf} - \normsqr f \bigr| \le
    C^\theta_4 \delta \normsqr[2,\theta] f
  \end{gather}
  for $f \in \HS^{2,\theta}$ and similarly on $\wt \HS$.
\end{lemma}
\begin{proof}
  We estimate
  \begin{equation*}
    \norm{(J'J - \1)f} \le
    \norm[1 \to 0]{(J'J - \1) T^{-\theta}} \norm[1] {T^\theta f} \le
    \norm{T^{-\theta}} \delta C_2^\theta \norm[2,\theta] f =:
    C^\theta_3 \delta \norm[2,\theta] f
  \end{equation*}
  using~\eqref{eq:j.inv},~\eqref{eq:j.t}
  and \Defenum{scale.1}{emb.1.2.th}. Similarly,
  \begin{multline*}
    \bigl| \normsqr{Jf} - \normsqr f \bigr| =
    \bigl| \iprod {(J^*J - \1) f} f \bigr| \\ \le
    \bigl| \iprod {T^{-\theta} (J^*-J')J  T^\theta f}
            {T^{-\theta} T^\theta f} \bigr| +
    \bigl| \iprod {T^{-\theta}(J'J - \1) T^\theta f}
               {T^{-\theta} T^\theta f} \bigr| \\ \le
    3 \normsqr{T^{-\theta}} \delta \normsqr[1]{T^\theta f} \le
    3 \normsqr{T^{-\theta}} (C_2^\theta)^2 \delta \normsqr[2,\theta] f =:
    C^\theta_4 \delta \normsqr[2,\theta] f
  \end{multline*}
  using again assumptions in~\Def{scale.1}, \Def{quasi-uni}
  and~\Def{closeness}.  The estimates on $\wt \HS$ follow similarly.
\end{proof}

We can now state the convergence of the resolvents:
\begin{theorem}
  \label{thm:res}
  Assume that the families $(H_\theta)$ and $(\wt H_\theta)$ are
  $\delta$-close w.r.t.\ the quasi-unitary operators $J$ and $J'$,
  then
  \begin{equation}
  \label{eq:res}
      \norm {\wt R^\theta J - J R^\theta} \le
      C^\theta_5 \delta,
  \end{equation}
  where $R^\theta:=(H^\theta +1)^{-1}$, $\wt R^\theta := (\wt H^\theta
  +1)^{-1}$ and $C^\theta_5 := (1+C_0^\theta +C_2^\theta) (1+\wt
  C_0^{\conj \theta} + \wt C_2^{\conj \theta})$.
\end{theorem}
\begin{proof}
  We write
  \begin{equation*}
    \wt R^\theta J - J R^\theta =
    \wt R^\theta \bigl[ J H^\theta - \wt H^\theta J \bigr]  R^\theta
  \end{equation*}
  where the operator in the bracket maps from $\HS^{2,\theta}$ to
  $\HS^{-2,\theta}=(\HS^{2,\conj \theta})^*$. This operator can be
  decomposed into
  \begin{multline*}
    J H^\theta - \wt H^\theta J \\=
    (J-J'^*) H^\theta + (J'-J'^1)^* H^\theta +
    \bigl((J'{}^1)^*H^\theta - \wt H^\theta J^1\bigr) +
    \wt H^\theta (J^1 - J)
  \end{multline*}
  where
  \begin{equation}
    \label{eq:j1.adj}
    \map{(J'{}^1)^*}{\HS^{-1, \theta}=(\HS^{1,\conj \theta})^*}
                {\wt \HS^{-1, \theta}=(\wt \HS^{1, \conj \theta})^*}.
  \end{equation}
  Now $H^\theta$ is bounded as a map from $\HS^{2,\theta}$ to $\HS$,
  as well as $\wt H^\theta$ is bounded as map from $\HS$ to
  $\HS^{-2,\theta}$ with the bounds $C_0^\theta+1$ and\ $\wt C_0^\theta+1$,
  respectively, cf.~\Lem{scale.2}. Next, the inclusion $\HS^{2,\theta}
  \hookrightarrow \HS^{1,\theta}$ is bounded with bound $C_2^\theta$,
  and similarly in the space $\wt \HS$
  (cf.~\Defenum{scale.1}{emb.1.2.th}). Finally, we can sum up all
  the error terms to arrive at the given bound.
\end{proof}

Denote by $\rho(H)$ the \emph{resolvent set} of $H$. A simple argument
allows us to deal with all $z$ in $\rho(H^\theta)$ \emph{and}
$\rho(\wt H^\theta)$:
\begin{theorem}
  \label{thm:res.z}
  Suppose that $z_0, z \in \rho(H^\theta) \cap \rho(\wt H^\theta)$, then
  \begin{equation*}
    \norm {V(z)} \le C^\theta_5(z) \norm {V(z_0)}
  \end{equation*}
  where $V(z):=\wt R^\theta(z) J - J R^\theta(z)$ and
  $R^\theta(z):=(H^\theta - z)^{-1}$ for $z \in \rho(H^\theta)$, and
  similarly for $\wt H^\theta$. In particular,
  \begin{equation}
    \label{eq:res.z}
    \norm {V(z)} \le C^\theta_6(z) \delta
  \end{equation}
  under the assumptions of \Thm{res}. The constants $C^\theta_5(z)$ and
  $C^\theta_6(z)$ depend continuously on $z$.
\end{theorem}
\begin{proof}
  Setting for brevity $R:=R^\theta$ and $\wt R := \wt R^\theta$, we have
  \begin{multline*}
    V(z) =
    V(z_0) + (z-z_0)\bigl(\wt R(z) \wt R(z_0)J - J R(z) R(z_0)\bigr) \\ =
    V(z_0) + (z-z_0)\bigl(\wt R(z) V(z_0) + V(z) R(z_0)\bigr)
  \end{multline*}
  where we have used the second resolvent identity. Reordering the terms
  we get
  \begin{equation*}
    V(z) \bigl[\1 - (z-z_0) R(z_0)\bigr] =
    \bigl[\1 + (z-z_0) \wt R(z) \bigr] V(z_0)
  \end{equation*}
  Since $\1 + (z-z_0) R(z)$ is the inverse of $\1 - (z-z_0) R(z_0)$,
  we obtain
  \begin{equation}
    \label{eq:res.z2}
    V(z) =
       \bigl[\1 + (z-z_0) \wt R(z) \bigr] V(z_0)
       \bigl[\1 + (z-z_0) R(z)\bigr]
  \end{equation}
  and the estimate follows with
  \begin{equation}
    \label{eq:const.c1}
    C^\theta_5(z):= \Bigl(1 + \frac{|z-z_0|}{\wt d(z)} \Bigr)
                    \Bigl(1 + \frac{|z-z_0|}{d(z)}     \Bigr)
  \end{equation}
  where $d(z):=\norm{R(z)}^{-1}$ and similarly for $\wt d(z)$.
  Estimate~\eqref{eq:res.z} follow immediately from \eqref{eq:res}
  with $C^\theta_6(z) := C^\theta_5 C^\theta_5(z)$.
\end{proof}

We can now easily extend the convergence results to a suitable
class of holomorphic functions of the operators:
\begin{theorem}
  \label{thm:hol.calc}
  Suppose that $\phi$ is a holomorphic functions in a neighbourhood of
  a simply connected domain $D \subset \C$ such that \sloppy $D$ is
  disjoint from $\spec {H^\theta}$ and $\spec {\wt H^\theta}$ for
  $H^\theta$ and $\wt H^\theta$ being $\delta$-close and $\delta$
  small enough. Suppose in addition that $\phi \in \Lp[1] {\bd D,
    C^\theta_5(z)\dd|z|}$ (cf.~\eqref{eq:const.c1}). Then
  \begin{equation}
    \label{eq:eq:hol.calc}
    \norm{\phi(\wt H^\theta)J - J \phi(H^\theta)} \le C^\theta_7 \delta
  \end{equation}
  where the constant depends only on $\theta$ and $\phi$. The
  integrability condition on $\phi$ is in particular satisfied if the
  curve is compact.
\end{theorem}
\begin{proof}
  Since $D$ is contained in the resolvent set of both operators and
  due to our integrability assumption on $\phi$, the holomorphic
  spectral calculus applies,
  \begin{equation*}
    \phi(H^\theta) =
    \frac 1 {2 \pi \im} \oint_{\bd D}
                      \frac {\phi(z)}{z - H^\theta} \dd z,
  \end{equation*}
  and a similar claim is valid for $\wt H^\theta$. This implies
  \begin{equation*}
    J \phi(H^\theta) - \phi (\wt H^\theta) J =
    -\frac 1 {2 \pi \im} \oint_{\bd D}
                      \bigr(J R^\theta(z) - \wt R^\theta(z) J \bigr)
                  \phi(z) \dd z.
  \end{equation*}
  and therefore,
  \begin{equation*}
    \norm{J \phi(H^\theta) - \phi(\wt H^\theta) J} \le
         \frac {\delta} {2\pi} \int_{\wt \gamma}
               C_6^\theta(z)|\phi(z)| \dd |z| =:
         C^\theta_7 \delta
  \end{equation*}
  Since $C^\theta_6(z)=C^\theta_5 C^\theta_5(z)$ depends continuously
  on $z$, the right-hand side is in particular finite if $\bd D$ is
  compact.
\end{proof}


Now we are able to demonstrate the main result of this section namely
the convergence of eigenprojections and eigenvalues. For the discrete
spectrum of $H^\theta$ it is not necessary to consider the
\emph{whole} spectrum of $\wt H^\theta$, we only need to make sure
that we are away from its \emph{essential} spectrum.
\begin{theorem}
  \label{thm:ew}
  Suppose that $\lambda$ is a discrete eigenvalue of $H^\theta$ with
  multiplicity $m>0$. Let $D \subset \rho(H^\theta)$ be an open disc
  such that $D$ contains $\lambda$ but no
  other spectral point of $H^\theta$. If $\clo D \cap \essspec {\wt
    H^\theta} = \emptyset$ for $\wt H^\theta$ being $\delta$-close to
  $H^\theta$, then
  \begin{equation*}
    \norm{J \1_{\{\lambda\}}(H^\theta) -
            \1_D (\wt H^\theta) J} \le C^\theta_7 \delta
  \end{equation*}
  where $C^\theta_7$ depends only on $\theta$ and $D$.

  In particular, if $m$ denotes the multiplicity of $\lambda$, then
  there exist $m$ discrete eigenvalues $\wt \lambda_1, \dots, \wt
  \lambda_m$ (not necessarily mutually distinct) in the discrete spectrum of
  $\wt H^\theta$ such that
  \begin{equation*}
    | \wt \lambda_j - \lambda | \le \eta(\delta), \qquad
    j = 1, \dots, m,
  \end{equation*}
  where $\eta(\delta) \to 0$ as $\delta \to 0$.
\end{theorem}
\begin{proof}
  We choose a sequence $\wt H_n:= \wt H_n^\theta$ which is
  $\delta_n$-close to $H = H^\theta$ where $\delta_n \to 0$. Since
  $\clo D \cap \bigcup_n \essspec {\wt H_n} = \emptyset$ and
  $\bigcup_n \disspec {\wt H_n}$ is countable, there exists a closed
  curve $\wt \gamma$ in $D$, disjoint from $\bigcup_n \spec {\wt H_n}$
  enclosing $\lambda$ but no other spectral point of $H$. Denote by
  $\wt D$ the enclosed region in $\C$, i,e., $\bd \wt D \subset
  \rho(H) \cap \rho(\wt H_n)$ is parametrised by $\wt \gamma$ and $\wt
  D \cap \spec H = \{\lambda\}$. Then we can apply the previous
  theorem and obtain
  \begin{equation}
    \label{eq:proj}
    \norm{J \1_{\{\lambda\}}(H) - \1_{\wt D} (\wt H_n) J} =
    \norm{J \1_{\wt D}(H) - \1_{\wt D} (\wt H_n) J} \le
         C^\theta_7 \delta_n
  \end{equation}
  where $C^\theta_7$ is finite and depend only on $\theta$ and $D$.

  For the eigenvalue convergence we first denote by
  $P=\1_D(H)=\1_{\{\lambda\}}(H)$ and $\wt P=\1_D(\wt H)$ the
  corresponding spectral projections. We start proving that $\dim
  P(\HS) = \dim \wt P(\wt \HS)$.  Note first that $P(\HS) \subset
  \HS^{2,\theta}$ for $f=Pf \in P(\HS)$ since $\norm[2,\theta]
  f=|\lambda+1|\norm f$. Then we estimate
  \begin{equation}
    \label{eq:proj.lower}
    \norm{\wt PJf} \ge \norm{Jf}-\norm{\wt P J - J P} \norm f \ge
    \bigl(1 - \sqrt {C^\theta_4 \delta} \,
                 |\lambda+1| - C^\theta_7 \delta \bigr)
    \norm f
  \end{equation}
  using~\eqref{eq:j.iso.theta} and~\eqref{eq:proj}.  If $\delta$ is
  small enough, the right-hand side is still positive. In particular,
  $\wt P J$ is injective on $P(\HS)$ so that $(\wt P J)(P(\HS))$ has
  at least the dimension of $P(\HS)$, i.e., $\dim P(\HS) \le \dim \wt
  P(\wt \HS)$.  The opposite inequality follows similarly.

  Now it is almost obvious that in every neighbourhood $\wt D$ of
  $\lambda$ satisfying the above assumption there are $m$ (not
  necessarily mutually distinct) eigenvalues $\wt \lambda_j$ of $\wt H^\theta$
  provided $\delta$ is small enough (cf.~\cite[Ch.~II.5.1]{kato:66}).
\end{proof}

In the case of one-dimensional projections we can even show the
convergence of the corresponding eigenvectors. Note that
generically, the eigenvalues are simple (cf.~\cite{uhlenbeck:76}):
\begin{theorem}
  \label{thm:eigenvectors}
  Suppose that $\psi$ is a normalized eigenvector of $H^\theta$ with
  eigenvalue $\lambda$ of multiplicity $1$ and that $\lambda \notin
  \essspec {\wt H^\theta}$. Then there exist an eigenvalue $\wt
  \lambda$ of $\wt H^\theta$ of multiplicity $1$ arbitrary close to
  $\lambda$ and a unique eigenvector $\wt \psi$ (up to a unitary
  scalar factor) and constants $C^\theta_8$, $\wt C^\theta_8>0$
  depending only on $\theta$ and $\lambda$ such that
  \begin{align*}
    \norm{J \psi - \wt \psi} &\le C^\theta_8 \delta, &
    \norm{J' \wt \psi - \psi} &\le \wt C^\theta_8\delta
  \end{align*}
  provided $H^\theta$ and $\wt H^\theta$ are $\delta$-close and
  $\delta>0$ is small enough.
\end{theorem}
\begin{proof}
  The first assertion follows from the previous theorem. Denote the
  corresponding eigenprojections by $P$ and \ $\wt P$, respectively. For the
  eigenvector convergence, note that
  \begin{equation*}
    \wt \psi = \frac 1 {\iprod {\wt P J \psi} {J\psi}} \, \wt P J \psi
  \end{equation*}
  since $\wt P$ is a one-dimensional projection. Note in addition that
  \begin{equation}
    \label{eq:proj.lower2}
    \iprod {\wt P J \psi} {J \psi} =
    \normsqr {\wt P J \psi} \ge
    \frac 1 4 \normsqr \psi =
    \frac 1 4, \qquad 0 < \delta < \delta_0
  \end{equation}
  for some $\delta_0>0$ due to~\eqref{eq:proj.lower}. Now,
  \begin{multline*}
    \norm{J \psi - \wt \psi} =
    \Bignorm{J P \psi -
          \frac 1 {\iprod {\wt P J \psi} {J \psi}} \, \wt P J \psi} \\ \le
    \norm{(JP - \wt PJ)\psi} +
       \bigg| 1 - \frac 1  {\iprod {\wt P J \psi} {J \psi}} \bigg|
       \norm{\wt P J \psi} \\ \le
    C^\theta_7 \delta +
        8 \bigr| \iprod {(\wt P J - J P)\psi} {J\psi} +
           \normsqr{J \psi} - \normsqr \psi \bigr| \\\le
    (17 C^\theta_7 +8C^\theta_4 |1+\lambda|^2) \delta =:
    C^\theta_8 \delta
  \end{multline*}
  since $\psi=P\psi$ and $\norm \psi=1$ using the previous
  theorem,~\eqref{eq:j.bdd},~\eqref{eq:j.iso.theta}
  and~\eqref{eq:proj.lower2}.  The second estimate follows immediately
  from
  \begin{equation*}
    \norm{J' \wt \psi - \psi} \le
    \norm{J'(\wt \psi - J \psi)} + \norm{(J'J-\1)\psi} \le
    \bigl(2 C^\theta_8 + C^\theta_3 |1+\lambda| \bigr) \delta =:
     \wt C^\theta_8 \delta
  \end{equation*}
  using~\eqref{eq:j.inv.theta}
  All the estimates are valid for $0<\delta<\delta_0$.
\end{proof}

%
\section{Analyticity and a resolvent estimate}
\label{app:res.est}
%

Here we sketch the proof of analyticity of the complex dilated
Hamiltonian $H^\theta$ as given in~\Sec{dilation}. We follow
closely the proof given in~\cite{cdks:87}. We repeat the arguments
here, since we are not aware of a proof in the quantum graph case.
In addition, we need a stronger assertion, namely an explicit
control of the norm of the resolvent $R^\theta:=(H^\theta+1)^{-1}$
as a map from $\HS$ to $\HS^{1,\theta}$ (cf.~\Lem{emb.1.2.th}).
On the quantum graph, it is enough to show that the operator is
bounded, but on the manifold, we need a uniform control of the
constant with respect to the shrinking parameter $\eps$. Since the
proof of the analyticity and the resolvent estimate is basically
the same, we state it in an abstract way for both models at the
same time. The main idea in showing the analyticity is to compare
$H^\theta$ with the decoupled operator $H^{\theta,\Dir}$ where the
decoupling is achieved via an additional Dirichlet condition at
the boundary between the interior and exterior part.

We first need some notation.  Assume that the Hilbert space splits
into an interior and exterior part, namely $\HS=\HS_\inl \oplus
\HS_\ext$ (cf.~\Sec{decomp} and~\eqref{eq:decomp.0.eps}).
\begin{notation}
  \label{not:dil}
  We constantly use the subscripts $(\cdot)_\inl$ and $(\cdot)_\ext$
  for the \emph{interior} and \emph{exterior part}, respectively.
  Similarly, quadratic forms, operators and functions with these
  subscripts are understood in the obvious way. In addition, $\bullet$
  stands either for ``$\inl$'' or
  ``$\ext$''. 
\end{notation}



\subsection*{Decomposition and quadratic forms}

We will make common use of minimal and maximal quadratic form domains
which corresponds to Neumann and Dirichlet boundary conditions for the
associated operators.  Note that the classical Neumann boundary
conditions appear only in the domain of the associated operator (for
details, see e.g.~\cite{reed-simon-1-4}).

Suppose that $\qf h$ is a quadratic form of the magnetic Hamiltonian
on $\HS$ (either on the quantum graph or the manifold).  Denote by
$\HS_\bullet=\Lsqr{X_\bullet}$ the corresponding subspace of $\HS$ for
$X_\bullet=X_\inl$ or $X_\ext$.

The quadratic form $\qf h^\Neu_\bullet=\qf h_\bullet$ with domain
$\HS^{1,\Neu}_\bullet=\HS^1_\bullet$ associated to the Neumann
operator on $X_\bullet$ consists of those functions $u \in
\HS_\bullet$ such that $\qf h_\bullet(u)$ is defined and finite. In
particular, we have
\begin{equation}
  \label{eq:qf}
  \HS^1_\bullet = \Sob{X_{0,\bullet}} \und
  \HS^1_\bullet = \Sob{X_{\eps,\bullet}}
\end{equation}
where the Sobolev spaces $\Sob{X_{\eps,\bullet}}$ are defined
in~\eqref{eq:dom.h.free} and~\eqref{eq:dom.h.free.eps} on the quantum
graph and the manifold, respectively.

We often omit the superscript $(\cdot)^\Neu$ on the Neumann quadratic
form and its domain (when the boundary does not separate the domain
into separate parts), since Neumann boundary conditions mean no
restriction on the quadratic form domain.

We assume that the quadratic forms can be written as
\begin{subequations}
  \label{eq:qf.dec}
  \begin{align}
    \label{eq:qf.inl}
    \qf h_\inl(u) &= \normsqr{\partial_\inl (\chi u)} +
    \qf h^\orth_\inl (\chi u) + \qf h^\rest_\inl(u)\\
    \label{eq:qf.ext}
    \qf h_\ext(u) &= \normsqr{\partial_\ext u } + \qf h^\orth_\ext (u)
  \end{align}
\end{subequations}
for $u = u_\inl \oplus u_\ext \in \HS^{1,\Neu}$ where
$\partial_\bullet u=\partial_x u$ is the derivative w.r.t.\ the
coordinate $x$ (oriented towards infinity on the external edge) and. In
addition, $\chi$ is assumed to be a smooth cut-off function such that
$\chi=1$ near the common boundary and equals $0$ away from it.
Furthermore, we assume that $\qf h^\rest_\inl(u)=0$ for functions with
support near the boundary. To be more concrete, we give the
expressions in our examples: On the manifold we have
\begin{subequations}
\label{eq:qf.dec.ex}
  \begin{align}
    \label{eq:qf.dec.mfd}
    \qf h^\orth_{\eps,\bullet}(u) =
    \frac 1 {\eps^2} \int_{X_{\eps,\bullet}} |\de_F u|_h^2 \dd X_{\eps}
    &\und
    \qf h^\rest_{\eps,\inl}(u) = \normsqr[X_{\eps,\inl}] {\de((1-\chi)u)},\\
\intertext{where we can choose $\chi$ independently of $\eps$ in the manifold
  case due to our decomposition \emph{away} from the internal
  vertices: Namely, $\Gamma_\eps$ has distance $\ell_0$ from any
  internal vertex due to our assumptions in \Sec{decomp}.
  On the quantum graph, we simply have}
    \label{eq:qf.dec.qg}
    \qf h^\orth_{0,\bullet}(f) = 0
    &\und
    \qf h^\rest_{0,\inl}(f) = \normsqr[X_{0,\inl}]{(1-\chi)f)'}
  \end{align}
\end{subequations}

The quadratic form $\qf h^\Dir_\bullet$ with domain $\HS^{1,\Dir}$
associated to the Dirichlet operator on $X_\bullet$ is defined as
restriction of $\qf h_\bullet$ on the subset of functions in
$\HS^{1,\Neu}_\bullet$ vanishing on the common boundary $\Gamma$ of
$X_\bullet$ and $X \setminus X_\bullet$.

\begin{notation}
  \label{not:dir.neu}
  The superscripts $(\cdot)^\Neu$ and $(\cdot)^\Dir$ will always refer
  to \emph{Neumann} and \emph{Dirichlet boundary conditions} on the
  common boundary $\Gamma$ of the interior and exterior part,
  respectively.
\end{notation}
We also need the corresponding forms on the whole space, namely
\begin{subequations}
  \label{eq:qf.whole}
  \begin{align}
    \label{eq:qf.dom.whole}
    \HS^{1,\Neu} := \HS^{1,\Neu}_\inl \oplus \HS^{1,\Neu}_\ext &\und
    \HS^{1,\Dir} := \HS^{1,\Dir}_\inl \oplus \HS^{1,\Dir}_\ext\\
  \intertext{together with their natural quadratic forms}
    \qf h^\Neu := \qf h^\Neu_\inl \oplus \qf h^\Neu_\ext &\und
    \qf h^\Dir := \qf h^\Dir_\inl \oplus \qf h^\Dir_\ext.
  \end{align}
\end{subequations}

\begin{notation}
\label{not:qf}
  For a non-negative quadratic form (i.e., $\qf h(u) \ge 0$ for $u \in
  \dom \qf h$) we define the associated natural norm as
  \begin{equation}
    \label{eq:def.qf.norm}
    \norm[1] u := \bigl( \normsqr u + \qf h(u) \bigr)^{1/2}.
  \end{equation}
  We refer to $\HS^1=\dom \qf h$ with norm $\norm[1] \cdot$ as
  \emph{space of order $1$}. We use similar notation for the various
  quadratic forms defined in this section.
\end{notation}
In our application, all the quadratic forms defined here, are closed,
so that the corresponding scales of order $1$ are indeed Hilbert
spaces.

\subsection*{Boundary maps}

We also need the boundary maps in order to express the various
boundary conditions. It will be convenient to use an
$\eps$-independent space in the manifold case:
\begin{notation}
  \label{not:restr.map}
  Let $\map {S_\bullet} {\HS^1_\bullet} \HSaux$ be the \emph{the
    restriction} or \emph{boundary map} onto the boundary $\Gamma=\bd
  X_\bullet$ given by
  \begin{align}
    \label{eq:trace.1}
     u \longmapsto \{u(v)\}_{v \in \Gamma_0} &\und
     u \longmapsto \eps^{m/2} u \restr {\Gamma_1}\\
     \intertext{where}
    \label{eq:bd.space}
    \HSaux=\lsqr{\Gamma_0} \cong \C^{|\Gamma_0|} &\und
    \HSaux=\Lsqr{\Gamma_1},
  \end{align}
  in the quantum graph and manifold cases, respectively.  Here,
  $\Gamma_1$ is the rescaled boundary $\Gamma_\eps$ with $\eps=1$.
  Note that the number of boundary vertices equals the number of
  external edges which we assumed to be finite in~\eqref{eq:ext.fin}.
\end{notation}

In the manifold case, we also have a scale on the boundary Hilbert
space $\HSaux=\Lsqr {\Gamma_1}$, namely $\HSaux^k=\Sob[k]{\Gamma_1}$.
In particular, we can define the dual $\HSaux^{-1/2}$ of
$\HSaux^{1/2}$ with respect to the pairing $\map
{(\cdot,\cdot)_\HSaux} {\HSaux^{-k} \times \HSaux^k} \C$. In addition
to the boundary map $S_\bullet=S^1_\bullet$ we need a similar map of
order $0$, namely
\begin{equation}
  \label{eq:trace.scale0}
  \map {S^0_\bullet} {\HS_\bullet} {\HSaux^{-1/2}}, \quad
  u \longmapsto \eps^{m/2} u \restr{\Gamma_1}.
\end{equation}

Note that on the quantum graph case, there is no such scale since
$\HSaux=\lsqr {\Gamma_0}\cong \C^{|\Gamma_0|}$. This means in
particular, that $\HSaux^*=\HSaux$ and $S_\bullet^*$ is a map from
$\HSaux$ into $\HS_\bullet^{-1}$.

We have to make sure that $\norm{S_\bullet}$ and $\norm{S^0_\bullet}$
do not depend on $\eps$ in the manifold case:
\begin{lemma}
\label{lem:norm.trace}
  The norm of the restriction maps
  \begin{gather}
    \label{eq:norm.trace.1}
      \map {S_\bullet=S^1_\bullet}
           {\HS^{1,\Neu}_\bullet = \Sob{X_{\eps,\bullet}}}
      {\HSaux^{1/2}=\Sob[1/2]{\Gamma_1}}, \qquad
      u \longmapsto \eps^{m/2} u \restr {\Gamma_1}\\
    \label{eq:norm.trace.0}
      \map {S^0_\bullet}{\HS=\Lsqr{X_{\eps,\bullet}}}
      {\HSaux^{-1/2}=\Sob[-1/2]{\Gamma_1}}, \qquad
      u \longmapsto \eps^{m/2} u \restr {\Gamma_1}
\end{gather}
are bounded independently of $\eps$.
\end{lemma}
\begin{proof}
  \sloppy We have $\norm {S_\bullet^1u} = \eps^{m/2}
  \norm{\tau_\bullet^1 \iota^1_\eps u}$ where $\map
  {\tau_\bullet^1}{\Sob{X_{1,\bullet}}}{\Sob[1/2]{\Gamma_1}}$ is the
  trace map, $\Sob{X_{1,\bullet}}$ is the Sobolev space
  $\Sob{X_{\eps,\bullet}}$ with $\eps=1$ fixed and $\map
  {\iota^1_\eps}{\Sob{X_{\eps,\bullet}}}{\Sob{X_{1,\bullet}}}$.  Now,
  $\norm{\iota^1_\eps}=\eps^{-m/2}$, so that $\norm{S_\bullet^1} \le
  \norm{\tau_\bullet^1}$. Clearly, the latter norm is independent of
  $\eps$. Similarly, $S^0_\bullet$ is the composition of
  $\tau^0_\bullet$ and $\iota^0_\eps$ with
  $\map{\tau^0_\bullet}{\Lsqr{X_{1,\bullet}}}{\Sob[-1/2]{\Gamma_1}}$
  and
  $\map{\iota^0_\eps}{\Lsqr{X_{\eps,\bullet}}}{\Lsqr{X_{1,\bullet}}}$.
  Again, $\norm{\iota^0_\eps}=\eps^{-m/2}$ and the result follows.
\end{proof}

\subsection*{Coupled quadratic forms}

With the help of the boundary maps, we can express the Dirichlet
quadratic form domain as
\begin{equation}
  \label{eq:df.dom.dir}
  \HS^{1,\Dir}_\bullet = \ker S_\bullet \subset \HS^{1,\Neu}_\bullet.
\end{equation}

We define the \emph{undilated} Hamiltonian via its quadratic form
$\qf h$ on
\begin{subequations}
\label{eq:qf.0}
  \begin{gather}
    \label{eq:qf.0.dom}
    \HS^1 :=
      \set{u \in \HS^{1,\Neu}} {S_\ext u = S_\ext u} =
      \ker (-S_\inl + S_\ext)\\
  \intertext{with form given by}
  \label{eq:qf.0.def}
  \qf h(u) := \qf h_\inl (u) + \qf h_\ext(u),
  \end{gather}
\end{subequations}
where
\begin{equation*}
  \label{eq:def.sum.op}
  (-S_\inl + S_\ext)u := -S_\inl u_\inl + S_\ext u_\ext
\end{equation*}
for $u=u_\inl \oplus u_\ext$. We will often omit the subscripts
$u=u_\inl$ etc.\ if they are clear from the context.

Similarly, we define the \emph{dilated} quadratic form $\qf h^\theta$,
for the moment for \emph{real} $\theta$ only, on the space
\begin{equation}
  \label{eq:qf.dom.th}
  \HS^{1,\theta} := \set{u \in \HS^{1,\Neu}} {S_\ext u =
    \e^{-\theta/2} S_\ext u} =
  \ker (-S_\inl + \e^{-\theta/2} S_\ext)
\end{equation}
and we set
\begin{subequations}
\label{eq:qf.th}
  \begin{gather}
    \label{eq:qf.th.def}
    \qf h^\theta (u)= \qf h_\inl(u) + \qf h_\ext^\theta(u),\\
    \label{eq:qf.th.ext}
    \qf h_\ext^\theta (u) =
       \e^{-2\theta} \normsqr{\partial_\ext u } + \qf h^\orth_\ext (u)
  \end{gather}
\end{subequations}
(cf.~\eqref{eq:qf.dec}).  Note that the dilated form $\qf h^\theta$
agrees with the free form $\qf h$ if $\theta=0$.

The various quadratic form domains satisfy the following inclusions,
also called \emph{Dirichlet-Neumann bracketing}, namely,
\begin{equation}
  \label{eq:dir.neu.brack}
  \HS^{1,\Dir}:=\HS^{1,\Dir}_\inl \oplus \HS^{1,\Dir}_\ext \subset
  \HS^{1,\theta} \subset
  \HS^{1,\Neu}_\inl \oplus \HS^{1,\Neu}_\ext =: \HS^{1,\Neu}
\end{equation}
If we equip the spaces with their canonical quadratic form norm as in
\Not{qf}, these inclusions are also bounded and induce bounded maps on
the corresponding dual spaces (cf.~\eqref{eq:scale.neg}), e.g.\
\begin{equation}
  \label{eq:scale.incl}
    \HS^{1,\Dir} \stackrel {\iota_{1,\theta}} \longrightarrow
    \HS^{1,\theta} \und
    \HS^{-1,\theta} \stackrel {\iota_{-1,\theta}}
          \longrightarrow \HS^{-1,\Dir}
  \end{equation}
  where $\iota_{-1,\theta}=(\iota_{1,\conj \theta})^*$.

The following estimate follows immediately from~\eqref{eq:qf.dec}:
\begin{lemma}
  \label{lem:norm.long}
  We have
  \begin{equation}
    \label{eq:norm.long}
    \normsqr{\partial_\bullet u} \le \qf h_\bullet(u)
  \end{equation}
  for functions $u \in \HS^1_\bullet$ (with support close to the
  boundary if $\bullet=\inl$). In particular, the
  operator\footnote{Strictly speaking, $\partial_\inl$ is defined only
    on the subset of functions with support close to the boundary.}
  $\map {\partial_\bullet} {\HS^1_\bullet} {\HS_\bullet}$ has a norm
  bounded by $1$.
\end{lemma}

\subsection*{Associated operators and Sobolev spaces of second order}

We denote the Dirichlet operator on $X_\bullet$ corresponding to the
Dirichlet quadratic form $\qf h^\Dir_\bullet$ by $H^\Dir_\bullet$ with
domain $\HS^{2,\Dir}_\bullet$.

If we are on the exterior part, we also need the dilated version,
namely we denote by $H^{\theta,\Dir}_\ext$ the operator associated to
the form $\qf h^{\theta,\Dir}_\ext$ which is the restriction of $\qf
h^\theta_\ext$ (cf.~\eqref{eq:qf.th.ext}) onto $\HS^{1,\Dir}$. Note
that the domains of $H^\Dir_\ext$ and $H^{\theta,\Dir}_\ext$ agree and
that the operators agree for $\theta=0$.  The \emph{decoupled}
operator $H^{\theta,\Dir}$ is then the direct sum, namely
$H^{\theta,\Dir}=H^\Dir_\inl \oplus H^{\theta,\Dir}_\ext$.

Before defining the \emph{coupled} dilated operators we introduce
minimal and maximal (non-selfadjoint) operators with respect to the
common boundary $\Gamma$ of the internal and external part.

Let $H^{\theta,{\min}}_\bullet$ be the restriction of
$H^{\theta;\Dir}_\bullet$ to
\begin{equation}
  \label{eq:def.d.part}
  \mathcal D^2_\bullet := \set{u \in \HS^{2,\Dir}_\bullet}
  {S_\bullet \partial u = 0}
\end{equation}
(i.e., the intersection of the Dirichlet and Neumann operator domain)
and set
\begin{equation}
  \label{eq:def.d}
  \mathcal D^2 :=
  \mathcal D^2_\inl \oplus \mathcal D^2_\ext
    \und
  H^{\theta,\min} :=
  H^{\min}_\inl \oplus H^{\theta,\min}_\ext.
\end{equation}
The corresponding maximal operators are defined by
\begin{equation}
  \label{eq:def.h.max}
  H^{\theta,{\max}}_\bullet := (H^{\theta,{\min}}_\bullet)^*, \qquad
  H^{\theta,{\max}} :=
  H^{\max}_\inl \oplus H^{\theta, {\max}}_\ext
\end{equation}
with domains
\begin{equation}
  \label{eq:def.w}
  \mathcal W^2_\bullet := \dom H^{\max}_\bullet, \qquad
    \mathcal W^2 := \mathcal W^2_\inl \oplus \mathcal W^2_\ext.
\end{equation}
independent of $\theta$ also for $\bullet=\ext$.

Since we assumed that the magnetic potential on the manifold
$\alpha_\eps$ is smooth and that $a$ on the graph is smooth inside
each edge, we can characterise $\mathcal W^2_\bullet$ via the Sobolev
spaces already introduced earlier, namely $\mathcal W^2_\bullet =
\Sob[2]{X_{\eps,\bullet}}$ for $\eps \ge 0$ (see~\eqref{eq:sob.qg} for
$\eps=0$ and~\eqref{eq:sob2.mfd} for $\eps>0$).

We now define the \emph{dilated} operator $H^\theta$ as the operator
associated to the quadratic form $\qf h^\theta$ for \emph{real}
$\theta$. Its domain is given by
\begin{multline}
  \label{eq:op.dom.dil}
  \HS^{2,\theta} :=
  \set{u \in \mathcal W^2}
    {S_\ext u = \e^{-\theta/2} S_\ext u, \;
     S_\ext \partial_\inl u = \e^{-3\theta/2} S_\ext \partial_\ext u }
   \\
    = \ker (-S_\inl + \e^{-\theta/2} S_\ext) \cap \ker [(-S_\inl
    + \e^{-3\theta/2} S_\ext) \partial]
\end{multline}
where $\partial := \partial_\inl \oplus \partial_\ext$ .  In our
application, $H^\theta$ acts formally on exterior edges as
in~\eqref{eq:h.dil}. As before, we denote by $H=H^0$ and
$\HS^2=\HS^{2,0}$ the undilated operator and domain, respectively.

We will need in \Lem{dom.h.complex} the following facts from elliptic
regularity. In the manifold case, we have the \emph{continuous}
embeddings
\begin{equation}
  \label{eq:ell.reg}
  \map{\iota_{\mathrm {ell}}}{\HS^2}{\mathcal W^2}
  \quad \text{and} \quad
  \map{\iota_{\mathrm {ell}}}{\HS^{2,\Dir}}{\mathcal W^2}
\end{equation}
where the space $\mathcal W^2$ is endowed with a suitable Sobolev norm.
On the quantum graph, such estimates are almost trivial (under
Assumption~\eqref{eq:len.bd}) and on the manifold, we refer
e.g.~to~\cite[Prop.~3]{aubin:76}. Note that on the manifold, we need
to consider this embedding only for \emph{fixed} $\eps$; we do not
need a global constant for all $\eps>0$. In general, $\iota_{\mathrm
  {ell}}$ has a finite norm \emph{depending on $\eps$}, since $X_\eps$
is of bounded geometry by Assumption~\eqref{eq:curv.bd} with constants
depending on $\eps$ and since we imposed no bounds on (general)
derivatives of the magnetic vector potential $\alpha_\eps$.

\subsection*{Analyticity}
The first aim in this section is to show that the family
$\{H^\theta\}_\theta$ with domain $\HS^{2,\theta}$ can be extended
analytically into the complex strip $S_\vartheta$ (i.e., $|\Im \theta|
< \vartheta/2$) and has spectrum contained in $\Sigma_\vartheta$
(i.e., each $z$ in the spectrum satisfies $|\arg z| \le \vartheta$).
Analyticity here means, that the resolvent
\begin{equation}
  \label{eq:def.res.dil}
  R^\theta(z) := (H^\theta - z)^{-1} 
\end{equation}
for $z \notin \Sigma_\theta$ depends analytically on $\theta$ as
operator in $\HS$.

To this aim we introduce the decoupled dilated operator
\begin{equation}
   \label{eq:def.dec.dil}
  H^{\theta,\Dir} := H^\Dir_\inl \oplus H^{\theta,\Dir}_\ext,
\end{equation}
where $H^\Dir_\inl$ and $H^{\theta,\Dir}_\ext$ denotes the
operator with \emph{Dirichlet} boundary conditions at the boundary
$\Gamma$ of $X_\inl$ and $X_\ext$.  We are now able to state the
first lemma on analytic dependence.
\begin{lemma}
  \label{lem:dec.analytic}
  The decoupled dilated Hamiltonian $\{H^{\theta,\Dir}\}_\theta$
  extends to an analytic family of type~A into the strip $\theta \in
  S_\vartheta = \set {\theta \in \C} {2|\Im \theta| < \vartheta}$. In
  addition, $\spec{H^{\theta,\Dir}}$ is contained in the sector
  $\Sigma_\vartheta = \set{ z \in \C}{|\arg z|\le \vartheta}$ and
  therefore a self-adjoint, uniformly $\vartheta$-sectorial family in
  the sense of \Def{sa.fam}
\end{lemma}
\begin{proof}
  The proof is almost obvious due to the explicit expression of the
  operators and the fact that
  \begin{equation*}
    \spec{H^{\theta,\Dir}} =
    \spec{H^\Dir_\inl} \cup \spec{H^{\theta,\Dir}_\ext} \subset
    \Sigma_\vartheta
  \end{equation*}
  since $H^{\theta,\Dir}$ has numerical range in the sector
  $\Sigma_\vartheta$ by~\eqref{eq:qf.th}.  Note that the domain of
  $H^{\theta,\Dir}$ is independent of $\theta$ due to the decoupling.
\end{proof}

Now we are going to extend the definition of the coupled operators
$H^\theta$ for real $\theta$ to the complex strip $S_\vartheta$. We
follow closely~\cite{cdks:87}.  We want to compare the resolvent
$R^\theta(z):=(H^\theta-z)^{-1}$ with the decoupled resolvent
$R^{\theta,\Dir}(z):=(H^{\theta,\Dir}(z))^{-1}$. To do so, want to
express the difference $R^\theta(z) - R^{\theta,\Dir}(z)$ in terms of
an explicit sequence of bounded and analytic operators, for the moment
for \emph{real} $\theta$ only. Since this expression will serve as
generalisation for \emph{complex} $\theta$, we formulate it already
for the complex case in order to formally respect analyticity.

Denote by
\begin{equation}
  \label{eq:res.scale}
  \map{\hat R(z)} {\HS^{-1}}{\HS^1}
\end{equation}
the undilated resolvent $(H-z)^{-1}$ of $H$ as an operator in the
natural scale of Hilbert spaces $\HS^k$ associated to the self-adjoint
operator $H$ (cf.~\Sec{scale.sa}). Since on $\HS^1$, the boundary
values on the internal and external part agree by~\eqref{eq:qf.0}, we
can define a bounded map
\begin{equation}
  \label{eq:def.s}
       \map S {\HS^1} {\HSaux}, \quad f \mapsto S_\inl f =
       S_\ext f
\end{equation}
with dual $\map{S^*} \HSaux {\HS^{-1}}$.

The following arguments for the quantum graph and the manifold differ
slightly due to the fact that the boundary space $\HSaux$ allows a
natural scale of Sobolev spaces only on the manifold.

We start on the quantum graph and define a bounded operator
\begin{equation}
    \label{def:b.theta.qg}
    B^\theta (z) :=
       S^{1,\theta} \partial R^{\theta,\Dir}(z) \colon
       \HS \stackrel{R^{\theta, \Dir}(z)} \longrightarrow
       \HS^{2,\Dir} \stackrel{\iota_{\mathrm{ell}}} \hookrightarrow
       \mathcal W^2 \stackrel{\partial} \longrightarrow
       \HS^{1,\Neu} \stackrel{S^{1,\theta}} \longrightarrow \HSaux
\end{equation}
for $\theta \in S_\vartheta$ and $z \notin \Sigma_\vartheta$ where
\begin{gather*}
  \map{S^{1,\theta}}{\HS^{1,\Neu}}{\HSaux \cong \C^{|\Gamma_0|}}, \qquad
  f \longmapsto -S_\inl f + \e^{-3\theta/2} S_\ext f,\\
  \map{\partial := \partial_\inl \oplus \partial_\ext}
              {\mathcal W^2} {\HS^{1,N}}.
\end{gather*}
For further purposes, we need to express the adjoint operator as a
solution operator.
\begin{lemma}
  \label{lem:adj.b}
  On the quantum graph, the adjoint $\map{(B^{\conj \theta}(\conj
    z))^*} {\HSaux} \HS$ of $B^\theta(z)$ is given as follows: If
  $f=(B^{\conj \theta}(\conj z))^*F$ with $F \in \HSaux$, then $f \in
  \mathcal W^2$ and $f$ is the unique solution of the Dirichlet
  problem
   \begin{equation}
     \label{eq:b.adj}
     (H^{\theta,\max}-z) f = 0, \quad
     f_\inl(v)=F(v), \quad
     f_\ext(v)=\e^{\theta/2} F(v)
   \end{equation}
   for all boundary vertices $v \in \Gamma_0$. In particular, $f$
   satisfies all inner boundary conditions and the jump condition
   along $\Gamma_0$.
\end{lemma}
\begin{proof}
  Let $\wt g \in \HS$, $F \in \HSaux$ and denote $g:= (H^{\conj
    \theta,\Dir}-z)^{-1} \wt g$. Then
  \begin{equation*}
    \iprod[\HS] {\wt g} f =
    \bigl( (B^{\conj \theta}(z))^*\wt g, F\bigr)_\HSaux =
    \sum_{v \in \Gamma_0}
        \bigl(-\conj g'_\inl(v)+\e^{-3\theta/2}
                  \conj g'_\ext(v)\bigr) F(v).
  \end{equation*}
  In particular, $0=\iprod {\wt g} f = \iprod {(H^{\conj
      \theta,\Dir}-\conj z) g} f$ for functions $\wt g$ with support
  away from the boundary vertices. Choosing $\wt g \in \Cci e$, and
  since we assumed that the potentials $a_e$ and $q_e$ are smooth
  inside an internal edge $e$ we conclude that $f_e$ is smooth as
  solution of the ODE $(-\partial_x + \im a_e)(\partial - \im a_e)f_e
  + q_e f_e = zf_e$ and $-f''_e=z f_e$ on external edges. To conclude
  that $f$ also satisfies all inner boundary conditions we use the
  arguments of~\cite[Lem.~2.2]{kostrykin-schrader:99}. In particular,
  we conclude that $f \in \mathcal W^2$ and $(H^{\theta,\max}-z)f=0$.
  Now, using general functions $\wt g$, we have
  \begin{equation*}
    \iprod {\wt g} f =
    \iprod {(H^{\conj \theta,\Dir}-\conj z) g} f =
    \sum_{v \in \Gamma_0}
        \bigl(-\conj g'_\inl(v) f_\inl(v) +\e^{-2\theta}
                  \conj g'_\ext(v) f_\ext(v)\bigr)
  \end{equation*}
  since $g$ vanishes on $\Gamma_0$. It follows that $F(v)=f_\inl(v)$
  and $f_\ext(v)=\e^{\theta/2}f_\inl(v)$ for boundary vertices $v \in
  \Gamma_0$.
\end{proof}

\begin{lemma}
  \label{lem:adj.b.fac}
  We can factorize the adjoint map on the quantum graph by the bounded
  maps
   \begin{equation}
     \label{eq:b.adj.fac}
     (B^{\conj \theta}(\conj z))^* =
       \bigl(\id_{\mathcal W^2} -
          (H^{\theta,\Dir}-z)^{-1} (H^{\theta,\max}-z)\bigr)
       E^\theta \colon  \HSaux \longrightarrow \mathcal W^2,
   \end{equation}
   where
   \begin{equation*}
     \map {E^\theta} \HSaux {\mathcal W^2}
   \end{equation*}
   is a bounded extension map such that $\wt f:=E^\theta F$ is
   constant near $\Gamma_0$ and $F(v)=\wt f_\ext(v)=\e^{\theta/2}\wt
   f_\inl(v)$ for boundary vertices $v \in \Gamma_0$. In particular,
   \begin{equation}
     \label{eq:b.adj.w2}
     \map {(B^{\conj \theta}(\conj z))^*} \HSaux {\mathcal W^2}
   \end{equation}
   is a bounded map and depends analytically on $\theta$ (and $z$).
   Finally, the dual of~\eqref{eq:b.adj.w2}, namely
   \begin{equation}
     \label{eq:b.w2}
     \map {(B^{\conj \theta}(\conj z))^{**}} {\mathcal W^{-2}} \HSaux ,
   \end{equation}
   is bounded and an extension of $\map {B^{\conj \theta}(\conj z)}
   \HS \HSaux$.
\end{lemma}
The extension map can for example be defined as
\begin{equation*}
  E^\theta F(x) := \chi^\theta(x)F(v)
\end{equation*}
near the boundary vertex $v$ where $\chi^\theta$ is a smooth map with
compact support and derivatives bounded in terms of $1/\ell_0$
(cf.~\eqref{eq:len.bd}) such that $\chi^\theta_\inl=1$ and
$\chi^\theta_\ext=\e^{\theta/2}$ near $v$.
\begin{proof}
  A direct calculation shows that~\eqref{eq:b.adj.fac} defines the
  (unique) solution of the Dirichlet problem (see
  e.g.~\cite[Lem.~D.1]{hislop-post:pre06}). The boundedness follows
  since the maps in the factorization~\eqref{eq:b.adj.fac} are
  bounded.  The analyticity is a consequence of the explicit form
  of~\eqref{eq:b.adj.fac} and~\eqref{eq:h.dil.qg}. Note that no space in
  the factorization depend on $\theta$.
\end{proof}

\begin{lemma}
  \label{lem:krein.qg}
  For $\theta \in S_\vartheta$ and $z \notin \Sigma_\vartheta$, the map
  \begin{equation}
    \label{eq:w.scale0}
    \map{W^\theta(z) := \bigl(B^{\conj \theta}(\conj z)\bigr)^*
         S \wh R(z) S^* B^\theta(z)} \HS \HS
  \end{equation}
  on the quantum graph is bounded and analytic. Furthermore, it is
  also bounded and analytic considered as map $\wt W^\theta(z)$ from
  $\HS$ into $\mathcal W^2$.

  For \emph{real} $\theta$, we have
  \begin{equation}
    \label{eq:diff.res}
    W^\theta(z) = R^\theta(z) - R^{\theta,\Dir}(z).
  \end{equation}
\end{lemma}
\begin{proof}
  The boundedness and analyticity follows from the preceding two
  lemmas.  The proof of~\eqref{eq:diff.res} is essentially the same as
  in~\cite[Lem.~A.2]{cdks:87} and basically an application of Greens
  formula for \emph{real} $\theta$.
\end{proof}

On the manifold, we have a similar assertion:
\begin{lemma}
  \label{lem:krein.qwg}
  For $\theta \in S_\vartheta$ and $z \notin \Sigma_\vartheta$, the
  map $W^\theta(z)$ from $\HS$ into $\HS$ defined as
  in~\eqref{eq:w.scale0} but now with
  \begin{gather}
    \label{def:b.theta.qwg}
    B^\theta (z) :=
       S^{0,\theta} \partial R^{\theta,\Dir}(z) \colon
       \HS \stackrel{R^{\theta, \Dir}(z)} \longrightarrow
       \HS^{1,\Dir} \stackrel{\partial} \longrightarrow
       \HS \stackrel{S^{0,\theta}} \longrightarrow \HSaux^{-1/2},\\
    \nonumber
    \map{(B^{\conj \theta}(\conj z))^*} {\HSaux^{1/2}} \HS\\
    \nonumber
    \map{S^{0,\theta}} \HS {\HSaux^{-1/2}}, \quad
      f \longmapsto -S^0_\inl f + \e^{-3\theta/2} S^0_\ext f,\\
    \nonumber
    \map{\partial :=
        \partial_\inl \oplus \partial_\ext} {\HS^{1,N}} \HS,\\
    \nonumber
       \map S {\HS^1} {\HSaux^{1/2}}, \quad f \mapsto S_\inl f =
       S_\ext f, \qquad \map{S^*} {\HSaux^{-1/2}} {\HS^{-1}}.
  \end{gather}
  on the quantum graph is bounded and analytic.  In addition,
  $\norm{W^\theta(z)}$ is bounded w.r.t.~$\eps$.
  
  For \emph{real} $\theta$ and $z \notin \Sigma_\vartheta$, we have
  again~\eqref{eq:diff.res}.
\end{lemma}
\begin{proof}
  Again, the boundedness and analyticity follows from the explicit
  representation. The $\eps$-independence of the norm
  follows from \Lems{norm.trace}{norm.long}. The last assertion is
  again similar to the proof of~\cite[Lem.~A.2]{cdks:87}.
\end{proof}

As in~\cite{cdks:87}, we now define the operator $R^\theta(z)$ also
for \emph{complex} $\theta \in S_\vartheta$ via the
formula~\eqref{eq:diff.res}, i.e.,
\begin{equation}
  \label{eq:def.res.complex}
  \map{R^\theta (z) := W^\theta(z) + R^{\theta,\Dir}(z)} \HS \HS.
\end{equation}
In particular, we have:
\begin{lemma}
  \label{lem:res.complex}
  \indent
  \begin{enumerate}
  \item For $z \notin \Sigma_\vartheta$, the family of $R^\theta(z)$
    is analytic in $\theta \in S_\vartheta$.
  \item The operators $R^\theta(z)$ satisfy the resolvent equation for
    $z \notin \Sigma_\vartheta$.
  \item The kernel of $R^\theta(z)$ is trivial.
  \end{enumerate}
  In particular, $R^\theta(z)$ is the resolvent of an operator
  \begin{equation}
    \label{def:h.theta.complex}
    H^\theta u := (R^\theta)^{-1}u - u, \qquad
    u \in \dom H^\theta := \HS^{2,\theta} :=  R^\theta (\HS)
  \end{equation}
  where $R^\theta:=R^\theta(-1)$ and the family $\{H^\theta\}_\theta$
  is self-adjoint with spectrum contained in the sector
  $\Sigma_\vartheta$. Finally, the norm of $R^\theta$ as operator on
  $\HS=\Lsqr{X_\eps}$ is independent of $\eps$ in the manifold case.
\end{lemma}
\begin{proof}
  (i)~The first assertion follows immediately from~\Lem{dec.analytic}
  and the explicit formula for $W^\theta(z)$ given
  in~\eqref{eq:w.scale0} (now for complex $\theta$). (ii)~The
  resolvent equation is fulfilled for real $\theta$, since then, the
  operator is the resolvent of a (self-adjoint) operator. Due to
  analyticity, the resolvent equation remains true for all $\theta \in
  S_\vartheta$. (iii)~To prove the third assertion, we claim that
  \begin{equation}
    \label{eq:h.lem}
    \bigiprod {(H^{\conj \theta, \Dir} - \conj z) \phi} {W^\theta (z) v} = 0
  \end{equation}
  for all $\phi \in \mathcal D^2$, for all $v \in \HS$, all $z \in
  \compl{(\Sigma_\vartheta)}$ and all $\theta \in S_\vartheta$ where
  $\mathcal D^2$ was defined in~\eqref{eq:def.d}.  The claim can
  easily be seen from
  \begin{equation*}
    \bigiprod[\HS] {(H^{\conj \theta, \Dir} - \conj z)\phi} {W^\theta(z) v} =
    \bigl( {B^{\conj \theta}(\conj z)
              (H^{\conj \theta, \Dir} - \conj z)\phi},
            {S^1 R(z) (S^1)^* B^\theta(z) v} \bigr)_\HSaux
  \end{equation*}
  using~\eqref{eq:w.scale0} for complex $\theta$, where
  $(\cdot,\cdot)_\HSaux$ is the pairing $\HSaux \times \HSaux$ in the
  quantum graph case and $\HSaux^{-1/2} \times \HSaux^{1/2}$ in the
  manifold case. Now the left-hand side of the last inner product
  vanishes since $B^{\conj \theta}(\conj z) (H^{\conj \theta, \Dir} -
  \conj z)\phi = (-S^0_\inl + \e^{-3\theta/2}S^0_\ext)\phi =0$ for
  $\phi \in \mathcal D^2$ and therefore,~\eqref{eq:h.lem} is
  fulfilled.

  Suppose finally, that $R^\theta(z) v=0$. Then we have
  \begin{equation*}
    0 =
   \bigiprod {(H^{\conj \theta, \Dir} - \conj z) \phi} {R^\theta(z) v}=
   \bigiprod {(H^{\conj \theta, \Dir} - \conj z) \phi}
               {R^{\theta, \Dir}(z) v}
  \end{equation*}
  for $\phi \in \mathcal D^2$ due to~\eqref{eq:h.lem} . Since $(H^{\conj
    \theta, \Dir} - \conj z)(\mathcal D^2)$ is dense it follows that
  $R^{\theta, \Dir}(z) v=0$ and therefore $v=0$ since the latter
  operator is injective as resolvent.

  To conclude we observe that~(ii) and~(iii) imply that $R^\theta(z)$
  is a resolvent (cf.~\cite[p.~428]{kato:66}) for all $z \notin
  \Sigma_\vartheta$, i.e., $\spec{H^\theta} \subset \Sigma_\vartheta$.
  In addition, the family $\{H^\theta\}_\theta$ is self-adjoint since
  $(W^\theta)^*=W^{\conj \theta}$ and $(R^{\theta,\Dir})^*=R^{\conj
    \theta, \Dir}$. Finally, $\norm{R^\theta} \le
  \norm{R^{\theta,\Dir}} + \norm{W^\theta}$ is bounded independently
  of $\eps$ by the spectral calculus and the preceding lemma.
\end{proof}

We finally characterize the domain of $H^\theta$.
\begin{lemma}
  \label{lem:dom.h.complex}
  For \emph{complex} $\theta \in S_\vartheta$, the domain of
  $H^\theta$ is given by $\HS^{2,\theta}$ as in~\eqref{eq:op.dom.dil}
  and $H^\theta u = H^{\max} u$ where $H^{\max}$ is defined
  in~\eqref{eq:def.h.max}.
\end{lemma}
\begin{proof}
  Let $u = R^\theta v \in \dom H^\theta$ then
  \begin{equation*}
    \bigiprod{(H^{\conj \theta,{\min}}+ 1)\phi} u =
    \bigiprod{(H^{\conj \theta, \Dir} + 1)\phi} {R^{\theta,\Dir}v} =
    \iprod \phi v
   \end{equation*}
   for all $\phi \in \mathcal D^2$. In particular, due to the definition
   of the adjoint operator, we have $u \in \dom H^{\theta,{\max}}$ and
   $H^{\theta,{\max}} u +u = v$. In particular, $H^{\theta,{\max}} u =
   (R^\theta)^{-1} u - u$ so that finally, $H^\theta u =
   H^{\theta,{\max}} u$ using the
   definition~\eqref{def:h.theta.complex}.

   To show that $u$ belongs to the set defined on the right-hand side
   of~\eqref{eq:op.dom.dil} we will first show that $R^\theta =
   R^{\theta,\Dir} + W^ \theta$ defines a bounded and analytic map
   from $\HS$ into $\mathcal W^2$ denoted by $\wt R^\theta$.  For
   $R^{\theta,\Dir}$ this follows from the sequence of maps
   \begin{equation}
     \label{eq:res.dir.incl}
     \wt R^{\theta,\Dir} \colon
     \HS  \stackrel{R^{\theta,\Dir}} \longrightarrow
     \HS^{2,\Dir} \stackrel {\iota_{\mathrm{ell}}} \longrightarrow
     \mathcal W^2
   \end{equation}
   on the quantum graph and the manifold (cf.~\eqref{eq:ell.reg}) and
   \Lem{dec.analytic}).

   The fact that $\wt W^\theta$ defines a bounded and analytic map
   from $\HS=\Lsqr {X_0}$ into $\mathcal W^2$ in the quantum graph
   case was already shown in \Lem{krein.qg} (setting $z=-1$).  On the
   manifold, we decompose $W^\theta$ into the sequence of bounded maps
   (denoted again by $\wt W^\theta$ when considered as map $\HS$ into
   $\mathcal W^2$)
   \begin{equation*}
     \HS            \stackrel{B^{0,\theta}} \longrightarrow
     \HSaux^{1/2}   \stackrel{(S^0)^*} \longrightarrow
     \HS            \stackrel R \longrightarrow
     \HS^2          \stackrel {\iota_{\mathrm{ell}}} \longrightarrow
     \mathcal W^2   \stackrel{S^2} \longrightarrow
     \HSaux^{3/2}   \stackrel{(B^{-2,\conj \theta})^*} \longrightarrow
     \mathcal W^2
   \end{equation*}
   where $\map{S^k} {\HS^k}{\HSaux^{k-1/2}}$ are the usual trace maps
   (note that $S^{-1}$ is \emph{not} an inverse of $S$) and
   $B^{0,\theta}$ resp.\ $B^{-2,\conj \theta}$ are defined as
   \begin{gather*}
     B^{0,\theta} \colon
     \HS            \stackrel{R^{\theta,\Dir}} \longrightarrow
     \HS^{2,\Dir}   \stackrel {\partial} \longrightarrow
     \HS^{1,\Neu}          \stackrel {S^1} \longrightarrow
     \HSaux^{1/2}\\
     B^{-2,\conj \theta} \colon
     \mathcal W^{-2}\stackrel{(\iota_{\mathrm{ell}})^*} \longrightarrow
     \HS^{-2,\Dir}  \stackrel{R^{\theta,\Dir}} \longrightarrow
     \HS            \stackrel {\partial} \longrightarrow
     \HS^{-1,\Neu}       \stackrel {S^{-1}} \longrightarrow
     \HSaux^{-3/2}.
   \end{gather*}
   Now all these maps are bounded and also analytic.

   Since now in both cases, $\wt R^\theta$ and $\wt W^\theta + \wt
   R^{\theta,\Dir}$ are analytic, and since they agree for \emph{real}
   $\theta$ by \Lem{krein.qg} and \Lem{krein.qwg}, they agree for
   \emph{all} $\theta \in S_\vartheta$. Finally, since $\wt W^\theta +
   \wt R^{\theta,\Dir}$ is bounded, the same is true for $\wt
   R^\theta$.

   To finish the proof that $u$ belongs to the set defined on the
   right-hand side of~\eqref{eq:op.dom.dil}, we note that
   \begin{equation*}
     \bigl[(-S_\inl) \oplus \e^{-\theta/2} S_\ext \bigr] \wt R^\theta
             \quad \text{and} \quad
     \bigl[(-S_\inl \partial_\inl) \oplus
          \e^{-3\theta/2} S_\ext \partial_\ext \bigr] \wt R^\theta
   \end{equation*}
   are bounded and analytic as operators in $\HS$ since the operators
   in the brackets are bounded and analytic from $\mathcal W^2$ to
   $\HS$. These operators vanish for real $\theta$ due
   to~\eqref{eq:op.dom.dil} and vanish therefore for all $\theta \in
   S_\vartheta$.

   For the opposite inclusion, we have to check that a function $u$
   belonging to the set on the right-hand side
   of~\eqref{eq:op.dom.dil} is of the form $u=R^\theta v$. A
   straightforward calculation using similar arguments as in
   \Lems{krein.qg}{krein.qwg} shows that $v:= H^{\theta,{\max}} u +u$
   is the right candidate.
\end{proof}

Finally, we have shown that $H^\theta$ with the above domain
$\HS^{2,\theta}$ is a self-adjoint, analytic family of operators with
spectrum in the sector $\Sigma_\vartheta$ either on the quantum graph
as well as on the manifold.

\subsection*{Compatible scales}

To conclude this section, we have to check that there is a compatible
scale of order $1$ w.r.t.\ the operator $\Delta \ge 0$ in the sense of
\Def{scale.1}.

We define the compatibility operators $\{T^\theta\}_\theta$ as
\begin{equation}
  \label{eq:def.t.}
  \map{T^\theta} \HS \HS, \qquad
  T^\theta u := u_\inl \oplus \e^{-\theta/2} u_\ext.
\end{equation}
Clearly, these operators are bounded, invertible and
satisfy~\eqref{eq:t.ad.inv}.  As in the abstract setting of
\App{scale} we define
\begin{equation}
  \label{eq:1.theta}
  \HS^{1,\theta} := T^{-\theta} (\HS^1), \qquad
  \norm[1,\theta] u := \norm[1]{T^\theta u} =
       \norm{(\Delta+1)^{1/2}T^\theta u}
\end{equation}
for \emph{complex} $\theta \in S_\vartheta$ and $\HS^{-1,\theta} :=
(\HS^{1,\conj \theta})^*$. Note that we could also use the undilated
\emph{magnetic} Hamiltonian $H$ instead of the Laplacian $\Delta$
in~\eqref{eq:1.theta}, since their quadratic forms are equivalent
(cf.~\Lems{h.free}{h.eps.free}). In addition, the
definition~\eqref{eq:1.theta} agrees with the one given
in~\eqref{eq:qf.dom.th} (a priori only for \emph{real} $\theta$).

The density assumption \Defenum{scale.1}{emb.dense}, namely the
density of $\HS^{2,\theta}$ in $\HS^{1,\theta}$ (or equivalently, that
$T^\theta(\HS^{2,\theta}) \subset T^\theta(\HS^{1,\theta})=\HS^1$ is
dense in $\HS^1$) follows by standard arguments . We omit the proof
since we do not need the density in order to apply the results of
\App{abstr.crit}: The only point where the density enters is the
unique continuation of $\qf h^\theta$ to $\HS^{1,\conj \theta} \times
\HS^{1,\theta}$ (cf.~\Defenum{scale.1}{def.sf}), but we only need the
sesquilinear form on $\HS^{1,\conj \theta} \times \HS^{2,\theta}$
(resp.\ in the dual form $\wt \HS^{2,\conj \theta} \times \wt
\HS^{1,\theta}$) (cf.~\Remenum{closeness}{j.comm}).

Next, we  have to show that the sesquilinear form $\qf h^\theta$
associated to the operator $H^\theta$ satisfies
\Defenum{scale.1}{def.sf}:
\begin{lemma}
  \label{lem:sf.bdd}
    The sesquilinear form $\qf h^\theta(f,g):=\iprod f {H^\theta g}$, $f \in
    \HS^{1,\conj \theta}$, $g \in \HS^{2,\theta}$ extends to a
    bounded, sesquilinear form
    \begin{equation}
      \label{eq:sf.bdd}
      \map {\qf h^\theta}
            {\HS^{1,\conj \theta} \times \HS^{1,\theta}} \C
    \end{equation}
    with bound $C_1^\theta$ depending only on $\Re \theta$ and the
    bounds on the potentials $\norm[\infty] a$ and $\norm[\infty] q$.
\end{lemma}
\begin{proof}
  Partial integration shows that
  \begin{equation*}
    \iprod f {H^\theta g} =
    \qf h_\inl(f,g) + \qf h^\theta_\ext (f,g)
  \end{equation*}
  for $f \in \HS^{1,\conj \theta}$ and $g \in \HS^{2,\theta}$; in
  particular, there are no boundary terms. On the internal part, we
  have $\qf h_\inl(f,g) \le (\qf h_\inl(f) \, \qf h_\inl(g))^{1/2}$
  and $\qf h_\inl(f) \le \normsqr[\qf h_\inl] f \approx \normsqr[\qf
  d_\inl] f =\normsqr[1,\theta]{f_\inl}$ using
  \Lems{h.free}{h.eps.free}. On the external part, we estimate $|\qf
  h_\ext^\theta(f,g)| \le 2 \cosh (\Re \theta) \norm[1,\conj \theta]
  {f_\ext} \norm[1,\theta] {g_\ext}$ and the claim follows.
\end{proof}

Finally, we have to check the embedding assumption
\Defenum{scale.1}{emb.1.2.th} or the equivalent resolvent
estimate~\eqref{eq:emb.1.2.th}. We can prove even more, namely the
stronger resolvent estimate~\eqref{eq:res.1.2.th}. Again, the proofs
differ slightly in the quantum graph and manifold case.

\begin{lemma}
  \label{lem:emb.1.2.qg}
  \sloppy There exists a constant $C_2^\theta$ depending only on $\Re
  \theta$, $\norm[\infty] a$ and $\norm[\infty] q$ such that the
  resolvent $R^\theta$ extends to a bounded map $\map {\wh
    R^\theta}{\HS^{-1,\theta}}{\HS^{1,\theta}}$ with norm bounded by
  $C_2^\theta$. In particular, we have
  \begin{equation}
    \label{eq:emb.1.2.th.qg}
    \normsqr[1,\theta] f \le C_2^\theta \normsqr[2,\theta] f
  \end{equation}
  for all $f \in \HS^{2,\theta}$ on the quantum graph.
\end{lemma}
\begin{proof}
  Similar as in the proof of \Lem{dom.h.complex} we show that each
  summand on the RHS of $R^\theta=W^\theta + R^{\theta,\Dir}$ extends
  individually to bounded maps $\HS^{-1,\theta} \longrightarrow
  \HS^{1,\theta}$. We start with $W^\theta$ and note that
  \begin{equation}
    \label{eq:w.scale1}
    \map{\wh W^\theta :=
      (B^{\conj \theta})^*
        S \wh R S^* B^\theta} {\HS^{-1,\theta}} {\HS^{1,\theta}}
  \end{equation}
  is bounded where we consider $B^\theta$ as bounded map $\HSaux
  \longrightarrow \mathcal W^{-2}$ together with its dual (cf.
  \Lem{adj.b.fac}). In that lemma we have also seen that the adjoint
  $(B^{\conj \theta})^*$ maps into $\HS^{1,\theta}$. Since the
  inclusion $\mathcal W^2 \cap \HS^{1,\theta}$ (with $\mathcal
  W^2$-norm) into $\HS^{1,\theta}$ is continuous, $\wh W^\theta$ is
  bounded.  Furthermore, $\wh W^\theta$ and $W^\theta$ coincide on the
  dense set $\HS$ and $\wh W^\theta$ is the unique extension of
  $W^\theta$ onto $\HS^{-1,\theta}$.

  Similarly, $\iota_{1,\theta} \wh R^{\theta,\Dir} \iota_{-1,\theta}$
  is bounded and agrees with $R^{\theta,\Dir}$ on $\HS$: Here, the
  inclusion map $\iota_{1,\theta}$ defined in~\eqref{eq:scale.incl} is
  bounded (see \Lem{h.free}) and $\map{\wh
    R^{\theta,\Dir}}{\HS^{-1,\Dir}}{\HS^{1,\Dir}}$. Note that the norm
  of $\iota_{1,\theta}$ depends on $\Re \theta$ and on $a$ resp.~$q$
  since $H^{\theta,\Dir}$ is the decoupled \emph{magnetic} Hamiltonian
  and the norm on $\HS^{1,\theta}$ (cf.~\eqref{eq:1.theta}) is defined
  with the \emph{free} Hamiltonian.  Finally we have seen that
  $R^\theta$ extends to a bounded map $\map{\wh
    R^\theta}{\HS^{-1,\theta}}{\HS^{1,\theta}}$ as desired.
\end{proof}

Recall that due to~\eqref{eq:def.mag} and~\eqref{eq:def.pot}, the
magnetic potential $\alpha_\eps$ and the electric potential $q_\eps$
are bounded independently of the squeezing parameter $\eps$.
\begin{lemma}
  \label{lem:emb.1.2.mfd}
  \sloppy There exists a constant $\wt C_2^\theta$ depending only on
  $\Re \theta$, $\norm[\infty] {\alpha_\eps}$ and $\norm[\infty]
  {q_\eps}$, and \emph{not} on $\eps$, such that the resolvent
  $R^\theta$ extends to a bounded map $\map {\wh
    R^\theta}{\HS^{-1,\theta}}{\HS^{1,\theta}}$ with norm bounded by
  $\wt C_2^\theta$. In particular, we have
  \begin{equation}
    \label{eq:emb.1.2.th.mfd}
    \normsqr[1,\theta] u \le \wt C_2^\theta \normsqr[2,\theta] u
  \end{equation}
  for all $u \in \HS^{2,\theta}$ on the manifold.
\end{lemma}
\begin{proof}
  Similar as in the previous proof, we show first that $\map{\wh
    W^\theta}{\HS^{-1,\Dir}}{\HS^{1,\Dir}}$ defined as
  in~\eqref{eq:w.scale1}, where now $B^\theta=B^{-1,\theta}$ with
  \begin{equation}
    \label{eq:def.b.th.-1}
    B^{-1,\theta} :=
       S^{0,\theta} \partial R^{\theta,\Dir} \colon
       \HS^{-1,\Dir} \stackrel{R^{\theta, \Dir}} \longrightarrow
       \HS^{1,\Dir} \stackrel{\partial} \longrightarrow
       \HS \stackrel{S^{0,\theta}} \longrightarrow \HSaux^{-1/2},
  \end{equation}
  is bounded: From \Lems{norm.trace}{norm.long} we see that the norm
  of $B^{-1,\theta}$ is bounded independently of $\eps$, and
  therefore, the same is true for the norm of $\wh W^\theta$. It can
  easily be seen by the very definition that $\norm{\wh
    R^{\theta,\Dir}}=1$. As before, $\wh W^\theta$ and $\wh
  R^{\theta,\Dir}$ are extensions of the corresponding operators on
  $\HS$. Finally, the norm of the inclusion map $\iota_{1,\theta}$
  depends on $\norm[\infty]{\alpha_\eps}$ and $\norm[\infty]{q_\eps}$,
  but can be bounded \emph{independently} on $\eps$ by our
  assumptions.
\end{proof}

Summarizing the results of \Lems{res.complex}{dom.h.complex} and
\Lems{emb.1.2.qg}{emb.1.2.mfd} we have shown the following theorem:
\begin{theorem}
  \label{thm:main.app}
  The family $\{H^\theta\}_{\theta \in S_\vartheta}$ is a
  self-adjoint, analytic family of operators with domain given
  by~\eqref{eq:op.dom.dil}. In addition, $\{\HS^{1,\theta}\}_\theta$
  is a compatible scale of order $1$ with respect to the free operator
  $\Delta=\laplacian{X_\eps}$ in both the quantum graph and manifold
  case.  Finally, the constant $C_0^\theta=\norm{(H^\theta+1)^{-1}}$
  in \Lem{scale.2} and the constant $C_2^\theta$ in \Def{scale.1} do
  not depend on $\eps$ in the manifold case. In particular, the
  results of \App{scale} and \App{abstr.crit} apply.

\end{theorem}
\subsection*{Acknowledgments}

The research was supported in part by the Czech Academy of
Sciences and Ministry of Education, Youth and Sports within the
projects A100480501 and LC06002, and by the ESF project ``Spectral
Theory and Partial Differential Equations (SPECT)''. The second
author was partly supported by the DFG through the
Grant~Po~1034/1-1, and appreciates the hospitality extended to him
in the Isaac Newton Institute (Cambridge) during the programme
``Analysis on Graphs and its Applications'' where this work was
finished.

\newcommand{\etalchar}[1]{$^{#1}$}

\end{document}